\documentclass[preprint]{aastex63}
\usepackage{CJK}
\usepackage{comment}

\newcommand{\hcop}{HCO$^+$}
\newcommand{\hocp}{HOC$^+$}
\newcommand{\msol}{M$_{\odot}$}
\newcommand{\enum}[2]{${#1}\times 10^{#2}$}
\newcommand{\httcop}{H$^{13}$CO$^{+}$}
\newcommand{\ps}{s$^{-1}$}
\newcommand{\psc}{cm$^{-2}$}
\newcommand{\pcc}{cm$^{-3}$}

\submitjournal{ApJ}

\shorttitle{ALCHEMI: Radicals in NGC\,253}
\shortauthors{Harada et al.}
\graphicspath{{./}{figs/}}
\begin{document}

\title{Starburst Energy Feedback Seen Through \hcop/HOC$^+$ Emission in NGC\,253 from ALCHEMI}

\correspondingauthor{Nanase Harada}
\email{nanase.harada@nao.ac.jp}

\author[0000-0002-6824-6627]{Nanase Harada}
\affiliation{National Astronomical Observatory of Japan, 2-21-1 Osawa, Mitaka, Tokyo 181-8588, Japan}
\affiliation{Institute of Astronomy and Astrophysics, Academia Sinica, 11F of AS/NTU
Astronomy-Mathematics Building, No.1, Sec. 4, Roosevelt Rd, Taipei 10617, Taiwan}
\affiliation{Department of Astronomy, School of Science, The Graduate University for Advanced Studies (SOKENDAI), 2-21-1 Osawa, Mitaka, Tokyo, 181-1855 Japan}

% Core authors. PIs and data reduction team. These authors appear on the first page of every ALCHEMI article

\author[0000-0001-9281-2919]{Sergio Mart\'in}
\affiliation{European Southern Observatory, Alonso de C\'ordova, 3107, Vitacura, Santiago 763-0355, Chile}
\affiliation{Joint ALMA Observatory, Alonso de C\'ordova, 3107, Vitacura, Santiago 763-0355, Chile}
%\email{Sergio.Martin@eso.org}
%
\author[0000-0003-1183-9293]{Jeffrey G.~Mangum}
\affiliation{National Radio Astronomy Observatory, 520 Edgemont Road,
  Charlottesville, VA  22903-2475, USA}
%\email{jmangum@nrao.edu}

\author[0000-0001-5187-2288]{Kazushi Sakamoto}
\affiliation{Institute of Astronomy and Astrophysics, Academia Sinica, 11F of AS/NTU
Astronomy-Mathematics Building, No.1, Sec. 4, Roosevelt Rd, Taipei 10617, Taiwan}
%\email{ksakamoto@asiaa.sinica.edu.tw}

\author[0000-0002-9931-1313]{Sebastien Muller}
\affiliation{Department of Space, Earth and Environment, Chalmers University of Technology, Onsala Space Observatory, SE-439 92 Onsala, Sweden}
%\email{mullers@chalmers.se}

\author[0000-0001-8153-1986]{Kunihiko Tanaka}
\affil{Department of Physics, Faculty of Science and Technology, Keio University, 3-14-1 Hiyoshi, Yokohama, Kanagawa 223--8522 Japan}
%\email{ktanaka@phys.keio.ac.jp}

\author[0000-0002-6939-0372]{Kouichiro Nakanishi}
\affiliation{National Astronomical Observatory of Japan, 2-21-1 Osawa, Mitaka, Tokyo 181-8588, Japan}
\affiliation{Department of Astronomy, School of Science, The Graduate University for Advanced Studies (SOKENDAI), 2-21-1 Osawa, Mitaka, Tokyo, 181-1855 Japan}
%\email{nakanisi.k@nao.ac.jp}

\author[0000-0002-7758-8717]{Rub\'en~Herrero-Illana}
\affiliation{European Southern Observatory, Alonso de C\'ordova 3107, Vitacura, Casilla 19001, Santiago de Chile, Chile}
\affiliation{Institute of Space Sciences (ICE, CSIC), Campus UAB, Carrer de Magrans, E-08193 Barcelona, Spain}
%\email{rherrero@eso.org}

\author[0000-0002-1413-1963]{Yuki Yoshimura}
\affiliation{Institute of Astronomy, Graduate School of Science,
The University of Tokyo, 2-21-1 Osawa, Mitaka, Tokyo 181-0015, Japan}
%\email{yyoshimura@ioa.s.u-tokyo.ac.jp}

\author[0000-0002-4355-6485]{Stefanie~M\"uhle}
\affiliation{Argelander-Institut f\"ur Astronomie, Universit\"at Bonn, Auf dem H\"ugel 71, D-53121 Bonn, Germany}(*)
%\email{muehle@astro.uni-bonn.de}

% Other authors.  first group, alphabetical
\author[0000-0002-1316-1343]{Rebeca Aladro}
\affiliation{Max-Planck-Institut f\"ur Radioastronomie, Auf dem H\"ugel 69, 53121 Bonn, Germany}
%\email{aladro@mpifr-bonn.mpg.de}

\author[0000-0001-8064-6394]{Laura Colzi}
\affiliation{Centro de Astrobiolog\'ia (CSIC-INTA), Ctra. de Ajalvir Km. 4, Torrej\'on de Ardoz, 28850 Madrid, Spain}
\affiliation{INAF-Osservatorio Astrofisico di Arcetri, Largo Enrico Fermi 5, 50125, Florence, Italy}

\author[0000-0002-2887-5859]{V\'ictor M. Rivilla}
\affiliation{Centro de Astrobiolog\'ia (CSIC-INTA), Ctra. de Ajalvir Km. 4, Torrej\'on de Ardoz, 28850 Madrid, Spain}
\affiliation{INAF-Osservatorio Astrofisico di Arcetri, Largo Enrico Fermi 5, 50125, Florence, Italy}

%%Other authors second group, alphabetical

\author[0000-0002-5828-7660]{Susanne Aalto}
\affiliation{Department of Space, Earth and Environment, Chalmers University of Technology, Onsala Space Observatory, SE-439 92 Onsala, Sweden}
%\email{saalto@chalmers.se}

\author[0000-0002-2333-5474]{Erica Behrens}
\affiliation{Department of Astronomy, University of Virginia, P.~O.~Box 400325, 530 McCormick Road, Charlottesville, VA 22904-4325}

\author[0000-0002-7495-4005]{Christian Henkel}
\affiliation{Max-Planck-Institut f\"ur Radioastronomie, Auf dem H\"ugel
  69, 53121 Bonn, Germany}
\affiliation{Astronomy Department, Faculty of Science, King Abdulaziz
  University, P.~O.~Box 80203, Jeddah 21589, Saudi Arabia}
%\email{chenkel@mpifr-bonn.mpg.de}

\author[0000-0003-4025-1552]{Jonathan Holdship}
\affiliation{Leiden Observatory, Leiden University, PO Box 9513, NL-2300 RA Leiden, The Netherlands}
\affiliation{Department of Physics and Astronomy, University College London, Gower Street, London WC1E 6BT}
%\email{jrh@star.ucl.ac.uk}

\author[0000-0003-3537-4849]{P. K. Humire}
\affiliation{Max-Planck-Institut f\"ur Radioastronomie, Auf dem H\"ugel 69, 53121 Bonn, Germany}
%\email{phumire@mpifr-bonn.mpg.de} 

\author[0000-0001-9436-9471]{David S.~Meier}
\affiliation{New Mexico Institute of Mining and Technology, 801 Leroy Place, Socorro, NM 87801, USA}
\affiliation{National Radio Astronomy Observatory, PO Box O, 1003 Lopezville Road, Socorro, NM 87801, USA}
%\email{david.meier@nmt.edu}

\author[0000-0003-0563-067X]{Yuri Nishimura}
\affiliation{Institute of Astronomy, Graduate School of Science,
The University of Tokyo, 2-21-1 Osawa, Mitaka, Tokyo 181-0015, Japan}
\affiliation{ALMA Project, National Astronomical Observatory of Japan, 2-21-1, Osawa, Mitaka, Tokyo 181-8588, Japan}
%\email{yuri@ioa.s.u-tokyo.ac.jp}

\author[0000-0001-5434-5942]{Paul P.~van der Werf}
\affiliation{Leiden Observatory, Leiden University, PO Box 9513, NL-2300 RA Leiden, The Netherlands}
    
\author[0000-0001-8504-8844]{Serena Viti}
\affiliation{Leiden Observatory, Leiden University, PO Box 9513, NL-2300 RA Leiden, The Netherlands}
\affiliation{Department of Physics and Astronomy, University College London, Gower Street, London WC1E 6BT}
\begin{abstract}
%background, context
%Star formation may be quenched by feedback from starburst events as a consequence of
%the energy injected from radiative heating, mechanical shocks, turbulence caused by massive stars or supernovae. 
%Astrochemistry can be used as a tool to study such feedback on the surrounding interstellar medium
%as it is a probe of the ionization and heating from UV photons or cosmic rays produced by massive stars or supernovae.
%aim
%We study these effects of feedback from radiation or high-energy particles on the chemical composition in 
%the central molecular zone (CMZ) of the local starburst galaxy NGC\,253.
%method
%We make use of multiple transitions of \hcop~and its metastable isomer \hocp~ 
%from the ALMA large program ALCHEMI.
Molecular abundances are sensitive to UV-photon flux and cosmic-ray ionization rate. In starburst environments, the effects of high-energy photons and particles are expected to be stronger.
We examine these astrochemical signatures through multiple transitions of \hcop~and its metastable isomer \hocp~in the center of the starburst galaxy NGC 253 using data from the ALMA large program ALCHEMI.
The distribution of the HOC$^+$(1-0) integrated intensity shows its association with ``superbubbles", cavities created 
either by supernovae or expanding HII regions.
The observed HCO$^+$/HOC$^+$ abundance ratios are $\sim 10-150$,
and the fractional abundance of \hocp~relative to H$_2$ is $\sim 1.5\times 10^{-11} - 6\times 10^{-10}$, which implies that
the HOC$^+$ abundance in the center of NGC\,253 is significantly higher than in quiescent spiral-arm dark clouds in the Galaxy and the Galactic center clouds. 
Comparison with chemical models implies either an interstellar radiation field of $G_0\gtrsim 10^3$ if the maximum visual extinction is $\gtrsim 5$,
or a cosmic-ray ionization rate of $\zeta \gtrsim 10^{-14}$ s$^{-1}$ (3-4 orders of magnitude higher than that within clouds in the Galactic spiral-arms)
to reproduce the observed results.
From the difference in formation routes of \hocp, we propose that a low-excitation line of \hocp~traces cosmic-ray dominated regions, while high-excitation lines trace photodissociation regions. 
Our results suggest that the interstellar medium in the center of NGC 253 is significantly affected by energy input from UV-photons and cosmic rays, sources of energy feedback.

%Although the observed abundances and abundance ratios alone do not differentiate whether they are due to UV photons or cosmic rays,
%we suggest that the excitation of HOC$^+$ is higher in photodissociation regions (PDRs) than in cosmic-ray dominated regions because efficient formation of HOC$^+$ in PDRs involves a reaction that becomes faster with high temperatures and high densities. 
%conclusion
\end{abstract}

\keywords{galaxies: individual(NGC\,253) --- galaxies: ISM --- astrochemistry --- galaxies: abundances --- galaxies: starburst --- radio lines: galaxies}


\section{Introduction} \label{sec:intro}
Starburst galaxies are great laboratories to study
feedback mechanisms because they possess the most vigorous star formation.
Starburst activities inject vast amounts of energy into the surrounding interstellar medium (ISM) \citep[e.g., ][]{2012MNRAS.421.3522H,2013ApJ...776....1K}. 
For example, massive stars (M$\gtrsim 8$ \msol) produce large amounts of UV photons, which can heat and ionize the gas \citep{1985ApJ...291..722T}. 
When these massive stars end their lives, they explode as supernovae producing high cosmic-ray fluxes.
Another form of starburst feedback is the injection of mechanical energy into the ISM through turbulence or shocks.
These energy sources are also major components of the ISM pressure \citep[thermal, turbulence, magnetic, and cosmic-ray pressure;][]{2005ism..book.....L}, which could hinder future star formation,
although it could help star formation locally. 
Since the molecular gas component is the fuel for future
star formation, it has been the subject
of numerous studies to investigate the effect of such feedback. The most direct observable 
effect of feedback is the imprint of the kinematic
signature on the gas properties (e.g., 
in the form of shocks or outflows),
which has been studied with transitions of 
carbon monoxide or a few other transitions with
relatively bright emission \citep[e.g., ][]{2019MNRAS.483.4586F}.
Other properties of molecular gas can be best studied with chemistry.

%Connection between the chemistry and radiation, molecular ISM
Astrochemistry has been widely studied in Galactic star-forming regions in order to constrain 
the physical properties of the molecular gas.
For example, DCO$^+$/HCO$^+$ and N$_2$D$^+$/N$_2$H$^+$ abundance ratios have been used
as indicators of ionization degree which contributes to the balance between the formation and destruction rates of these species in the steady state \citep[e.g.,][]{2002P&SS...50.1133C}.
Astrochemical studies towards external galaxies have been conducted for decades, especially with single-dish telescopes \citep[e.g., ][]{1988A&A...201L..23H,1991A&A...245..457M,2006ApJS..164..450M,2015AA...579A.101A}.
However, the chances to explore the potential of astrochemistry in external galaxies with higher angular resolution were only opened up quite recently due to the development of sensitive interferometers, such as ALMA (the Atacama Large Millimeter/submillimeter Array) and NOEMA (the NOrthern Extended Millimeter Array).\footnote{An important exception is work by \citet{2005ApJ...618..259M}, which presented spatially-resolved extragalactic astrochemistry already with Owens Valley Millimeter Array.}
%extragalactic astrochemistry 
%cite: Meier et al., Viti et al., Martin et al., Harada et al., 
Since then, spatially-resolved studies of astrochemistry have been proven to be possible in a few extragalactic sources (e.g., in the active galactic nuclei hosts NGC 1068 and NGC 1097, the starburst galaxies NGC 253, M83, and NGC 3256) \citep[e.g.,][]{2014PASJ...66...75T,2014A&A...570A..28V,2015A&A...573A.116M,2015ApJ...801...63M,2015PASJ...67....8N,2018ApJ...855...49H,2019ApJ...884..100H}.

%NGC 253, starburst, winds, superbubbles, feedback mechanisms
%Sakamoto+06,11, Bolatto+13, Leroy+18, Krieger+19, 
%TH85, Ulvestad97, 
NGC\,253 is one of the closest starburst galaxies at a distance of 3.5 Mpc \citep[][$1''$=17 pc]{2005MNRAS.361..330R}. It hosts a total star formation rate of 5~M$_{\odot}$\,yr$^{-1}$ of which 2~M$_{\odot}$\,yr$^{-1}$ is concentrated within the central few hundred parsecs \citep{leroy_alma_2015}. This type of central structure in galactic centers are referred to as a central molecular zone (CMZ) \citep{1996ARA&A..34..645M,sakamoto_star-forming_2011}. Strong radio/sub-mm continuum sources suggest multiple star clusters in the CMZ \citep{turner_1_1985,ulvestad_vla_1997,leroy_forming_2018}.
Super hot cores, early stages of super star clusters, are reported by \citet{2020MNRAS.491.4573R}.
There are multiple signs of feedback in the CMZ of NGC\,253. 
The kinetic temperature of the molecular gas has been measured to be very high \citep[$T\gtrsim 300$~K for 10-pc scales;][]{mangum_fire_2019}.
Bubble-like cavities in molecular gas have been found in multiple locations. These cavities are called superbubbles, indicating either expanding HII regions or hypernovae \citep{sakamoto_molecular_2006,sakamoto_star-forming_2011,bolatto_suppression_2013,krieger_molecular_2019}.
There are also outflows observed in X-rays \citep{strickland_chandra_2000,strickland_chandra_2002}, H$\alpha$ \citep{westmoquette_spatially_2011}, and molecules \citep{1985ApJ...299..312T,bolatto_suppression_2013,2017ApJ...835..265W,krieger_molecular_2019,2021ApJ...912....4L}.

% ALCHEMI survey
The chemistry in NGC\,253 has been found to be fairly rich already in single-dish 
observations \citep{2006ApJS..164..450M,2015AA...579A.101A}, and even more so in ALMA observations \citep{2015ApJ...801...63M,ando_diverse_2017,2020ApJ...897..176K}. 
To further explore this rich chemistry, a large spectral scan of the central molecular zone of NGC253 has been performed with the ALMA large program ALCHEMI (ALMA Comprehensive High-resolution Extragalactic Molecular inventory; Mart\'in et al. submitted), covering nearly all ALMA Bands from 3 to 7 (except a few frequency ranges blocked by telluric lines).

% Alternative text covering last paragraph: To further explore this rich chemistry, a large spectral scan of the central molecular zone of NGC253 has been performed with the ALMA large program ALCHEMI (ALMA Comprehensive High-resolution Extragalactic Molecular inventory), covering nearly all ALMA bands 3 to 7 (except few frequency ranges blocked by telluric lines).  

%specific molecules HOC+, CO+, HCO, etc. - cite a few literatures
This study makes use of the ALCHEMI survey data to examine the effects of UV photons and cosmic rays through the abundances of the reactive molecular ion HOC$^+$.
HOC$^+$ is the metastable isomer of HCO$^+$ with an energy difference of 16,600 K \citep{1984ApJ...279..322D}. 
Unlike HCO$^+$, which tends to be more widely distributed, HOC$^+$ has high fractional abundances only in regions strongly irradiated by UV photons, X-rays, or cosmic-ray particles.
Therefore, the \hcop/\hocp~ratio varies significantly among sources (Table \ref{tab:lit}).
For example, the HCO$^+$/HOC$^+$ ratio is high
in quiescent dense dark clouds, with ratios $\gtrsim 1000$ \citep{1997ApJ...481..800A}.
On the other hand, the HCO$^+$/HOC$^+$ ratios are lower in photodissociation regions (PDRs). 
These PDRs include Galactic diffuse clouds \citep[\hcop/\hocp $= 70-120$;][]{2004AA...428..117L}, and the 
Orion dense PDR \citep[\hcop/\hocp = 145-180;][]{2017AA...601L...9G}. 
Starburst galaxies are also expected to contain 
significant numbers of PDRs.  Low \hcop/\hocp~ratios were found in the starburst galaxy M82 \citep[\hcop/\hocp = 44-136; ][]{2008AA...492..675F,2015AA...579A.101A} and in single-dish observations of NGC\,253 \citep[][\hcop/\hocp = 63-80 and 30, respectively]{2009ApJ...706.1323M, 2015AA...579A.101A}.
An \hcop/\hocp~ratio of 55 was also found in the z=0.89 molecular absorber toward PKS1830-211 \citep{2011AA...535A.103M}.
In the Galactic center clouds of the Sgr~B2 region,
where we expect effects of UV-photons, X-rays, or cosmic rays, the \hcop/\hocp~ratio was found to be 300-1500 \citep{2020ApJ...895...57A}.
Model calculations show that HOC$^+$ is also enhanced by cosmic rays \citep{2018ApJ...868...40A}.
It has also been suggested that \hocp~can be efficiently produced in X-ray dominated regions (XDRs) \citep{2007ApJ...664L..23S}. In the central region surrounding the AGN in NGC\,1068, a low ratio of \hcop/\hocp$=30-80$ was found, possibly due to XDRs \citep{2004AA...419..897U}.
In the AGN-driven outflows of Mrk~231 and Mrk~273, very low values of the HCO$^+$/HOC$^+(3-2)$ intensity ratios, $\sim 10$, were observed by \citet{2015AA...574A..85A} and \citet{2018AA...617A..20A}, respectively.
A possible reason for such a low ratio is an outflow in the case of Mrk~231, as the cavity created by an outflow allows radiation or particles to travel further. There was no obvious cause in  Mrk~273.
We intend to compare the spatially-resolved CMZ of NGC 253 with these environments in terms of the \hocp~abundance in order to elucidate the starburst feedback to the ISM. 
%cite: Gerin et al., Goicochea+, Apponi, Aalto, Aladro, 
%%%%%%%%%%%%%%%%%%%%%%%%%%%%%%%%%%%%%%%%%%%%%%%%%%%%%%%%%%%%%%%%%%%%%%%

%%%%%%%%%%%%%%%%%%%%%%%%%%%%%%%%%%%%%%%%%%%%%%%%%%%%%%%%%%%%%%%%%%%%%%%%%%%%%%%%%%
\begin{deluxetable*}{lccc}
%\tablenum{1}
\tablecaption{Literature values of [\hcop]/[\hocp] and [\hocp]/[$H_2$] \label{tab:lit}}
\tablewidth{0pt}
\tablehead{
\colhead{Region} & \colhead{[\hcop]/[\hocp]} & \colhead{[\hocp]/[H$_2$]} &\colhead{Reference}
}
%\decimalcolnumbers
\startdata
\multicolumn{4}{c}{Extragalactic sources}\\
\hline
NGC 253 & 30-80 &\enum{(1.7-2.4)}{-10}&(1), (2)\\
M82 & 44-136 & \nodata &(3)\\
NGC 1068 &30-80 &\enum{4}{-9} &(4)\\
Mrk 231 &5-10$^{a}$ &\nodata &(5)\\
Mrk 273 &5$^{a}$ &\nodata &(6)\\
PKS1830-211 &55 & \enum{1.6}{-10} &(7)\\
\hline
\multicolumn{4}{c}{Galactic PDRs}\\
\hline
Horsehead Nebula & 75-200 &\enum{(0.4-0.8)}{-11} &(8)\\
Orion Bar &145-180 & \enum{3}{-11} &(9),(10)\\
NGC 7023 &50-120 &\enum{7}{-12} &(10)\\
%S140 & & \\
%NGC 2023 & & \\
%M17SW &&\\
\hline
\multicolumn{4}{c}{Galactic center clouds}\\
\hline
Sgr B2 &337-1541 &\enum{(0.5-2.8)}{-11} &(11)\\
\hline
\multicolumn{4}{c}{Galactic dense clouds}\\
\hline
DR21(OH) & 2600 & \enum{1}{-12} &(12)\\
G34.3 & 4000 & \enum{2}{-13} &(12)\\
L134N &$>4500$ &$<$\enum{7}{-13} &(12)\\
NGC 2024 &900&\enum{2}{-12}&(12)\\
%NGC 7538 &3500&\enum{2}{-12}&\\
%Orion (3N, 1E) &2000&\enum{5}{-13}&\\
Orion KL &2100&\enum{2}{-13}&(12)\\
W3 (OH) &6000&\enum{8}{-13}&(12)\\
W51M &1300&\enum{5}{-13}&(12)\\
\hline
\multicolumn{4}{c}{Galactic diffuse clouds}\\
\hline
B0355+508 &72&\enum{6.1}{-11}&(13)\\
B0415+379 &117&\enum{2.6}{-11}&(13)\\
B0528+134 &$>30$&$<$\enum{1.6}{-10}&(13)\\
B1730-130 &$>18$&$<$\enum{4.3}{-10}&(13)\\
B2200+420 &70&\enum{6.5}{-11}&(13)\\
\enddata
\tablecomments{$^a$ Abundance ratios are derived with the assumption of optically thin media.
(1) \citet{2009ApJ...706.1323M} (2)\citet{2015AA...579A.101A} (3) \citet{2008AA...492..675F}
(4) \citet{2004AA...419..897U} (5) \citet{2015AA...574A..85A} (6) \citet{2018AA...617A..20A} 
(7) \citet{2011AA...535A.103M}
(8) \citet{2009AA...498..771G} (9) \citet{2017AA...601L...9G} (10) \citet{2003AA...406..899F} (11) \citet{2020ApJ...895...57A} (12)\citet{1997ApJ...481..800A}  
(13) \citet{2004AA...428..117L}}
\end{deluxetable*}

%organization of this paper.
In this paper, we analyze high-resolution observations of the emission from the molecular ions \hcop~and \hocp, and make use of chemical models to explain the observed abundances.
This paper is organized as follows.
Section \ref{sec:obs} explains the data analysis methods. 
We present our results of velocity-integrated intensity images in Section \ref{sec:mom0} and
spectral shape in Section \ref{sec:spec}.
The derived column densities are discussed in Section \ref{sec:col}.
In Section \ref{sec:model}, we present our chemical model calculations that are to be compared with our observations. 
We discuss the implications of our results in Section \ref{sec:disc}, and we summarize our work in Section \ref{sec:sum}.

\section{Observations and data analysis}\label{sec:obs}
Details of the ALMA observations, data reduction procedures, and data products are presented in the survey description paper of ALCHEMI (Mart\'in et al. submitted).
Here we summarize the details of the ALCHEMI observations relevant to this paper. 
The ALCHEMI spectrally scanned mosaic covers a region of $50'' \times 20''$ (850 $\times$ 340 pc) in spatial extent, comprising the CMZ within NGC\,253. The phase center of the observations is $\alpha=00^h47^m33.26^s$, $\delta=-25^\circ17'17.7''$ (ICRS).
Data cubes are uniformly convolved to an angular resolution of 1\farcs6 corresponding to 27\,pc.
Continuum emission was subtracted in the image plane,
on a per-pixel basis, using the publicly available software STATCONT \citep{2018A&A...609A.101S}.
From the continuum-subtracted cubes, we extracted spectral channels covering a velocity range of $\pm 500\,$km s$^{-1}$ centered on our transitions of interest to create individual spectral transition cubes with a spectral resolution of $\Delta v= 10\,$km\,\ps. 
The primary beam correction was applied to these cubes as a part of the ALCHEMI imaging process (Mart\'in et al. submitted).
To obtain as high signal-to-noise ratios (S/N) as possible and to prevent neighboring transitions close in frequency from  contaminating our velocity-integrated images,
we masked regions of line cubes on a per-channel basis where 
emission is not expected, based on spectral line cubes of CO and HCN. There reference data cubes were
generated with no primary beam correction, using the same grid
with the same pixel coordinates and channel velocities.
Then, we smoothed them to twice coarser angular resolution, and we masked out the regions without either a 5$\sigma$-detection in CO(1-0) or a 2$\sigma$ detection in HCN(1-0). 
Although a mask using CO(1-0) should be sufficient, HCN(1-0) was used because there is a certain region in the cube where HCN(1-0) emission had a higher 
signal-to-noise ratio than CO(1-0).
This mask was applied to the primary-beam corrected line cubes of molecular transitions that are presented in this paper (\hcop, \httcop, and \hocp).
Although $5\sigma$ used for the mask from CO(1-0) seems rather high for a cutoff if we are applying it to transitions with a similar intensity as CO(1-0), the transitions we analyze in this paper have
a brightness temperature at least a factor of several lower than CO(1-0), and the use of a high cutoff in CO should not lead to significant amounts of missing \hocp~and \hcop~emission in our analysis. This was confirmed by a visual inspection comparing masked and unmasked \hcop~and \hocp~cubes.
This procedure is different from the commonly-used approach where 2- to 3-$\sigma$ cutoff is determined from the transition of which the image is made. We employ the above approach so that the low-level emission of species with weak emission
can effectively be collected, and that neighboring lines can be excluded from image cubes. 
This approach should not create any bias because the CO(1-0) transition is much more
extended and stronger than \hcop, \httcop, and \hocp. 
The RMS values per spectral channel of the transitions used in this article are listed in Table \ref{tab:rms}.

\begin{deluxetable}{cccccccc} 
\tablecolumns{8} 
\tablewidth{0pc} 
\tabletypesize{\scriptsize} 
\tablecaption{Spectroscopic properties of the observed transitions of HOC$^+$, HCO$^+$, and H$^{13}$CO$^+$, and RMS noise values of the corresponding image cubes}
\label{tab:rms}
\tablehead{\colhead{Species} & \colhead{Transition} &\colhead{Rest frequency$^{(a)}$} &\colhead{$E_{\rm up}$$^{(b)}$} &\colhead{$A_{\rm ul}$$^{(c)}$} &\colhead{RMS} &\colhead{RMS} &\colhead{$\sigma_{typ}$}\\
\colhead{} &\colhead{} & \colhead{(GHz)} &\colhead{(K)} &\colhead{(\ps)} &\colhead{(mJy beam$^{-1}$)} &\colhead{(mK)} &\colhead{(K\,km\,\ps)}}
\startdata 
HOC$^{+}$ & J=1-0 &  89.487414 &  4.3 & $2.1 \times 10^{-5}$   &0.20 & 12. &0.75\\
          & J=3-2 & 268.451094 & 25.8 & $7.4 \times 10^{-4}$  &1.2  & 7.6 &0.48\\
          & J=4-3 & 357.921987 & 42.9 & $1.8 \times 10^{-3}$  &2.2   &8.2 &0.52\\
HCO$^{+}$ & J=1-0 &  89.1885247 &  4.3 & $4.2 \times 10^{-5}$ &0.23  &14. &0.87\\
          & J=3-2 & 267.5576259 & 25.7 & $1.5 \times 10^{-3}$   &1.4   &9.7 &0.61\\
          & J=4-3 & 356.7342230 & 42.8 & $3.6 \times 10^{-3}$  &2.4   &9.0 &0.57\\
H$^{13}$CO$^+$ & J=1-0 &  86.7542884 &  4.2 & $3.9 \times 10^{-5}$ &0.12  & 7.9 &0.50\\
          & J=3-2 & 260.2553390 & 25.0 & $1.3 \times 10^{-3}$  &1.0   &7.3 &0.46\\
          & J=4-3 & 346.9983440 & 41.6 & $3.3 \times 10^{-3}$ &2.0   &8.2 &0.52\\
\enddata 
\tablecomments{RMS values of a single channel with $\Delta v=10$\,km\,\ps are shown.
$(a)$ Frequency taken from the CDMS 
\citep[https://cdms.astro.uni-koeln.de;][]{2001AA370L49M,2005JMoSt.742..215M}; $(b)$ Upper level energy of the transition; $(c)$ $A_{ul}$: Einstein coefficient of spontaneous emission.} 
\end{deluxetable}

%%%%%%%%%%%%%%%%%%%%%
%\begin{table}[ht!]
%\caption{Spectroscopic properties of the transitions of HOC$^+$, HCO$^+$, and H$^{13}$CO$^+$ covered in ALCHEMI %data.}
%\label{tab:specpro}
%\begin{center} \begin{tabular}{cccc}
%%\hline \hline
%Transition & Rest frequency $^{(a)}$& $E_{up}$ $^{(b)}$& $A_{ul}$ $^{(c)}$\\
%     & (GHz)   & (K)     & (s$^{-1}$) \\
%\hline
%HOC$^+$ (1-0) &  89.487414 &  4.3 & $2.1 \times 10^{-5}$ \\
%%HOC$^+$ (2-1) & 178.972051 & 12.9 & $2.0 \times 10^{-4}$ \\
%HOC$^+$ (3-2) & 268.451094 & 25.8 & $7.4 \times 10^{-4}$ \\
%HOC$^+$ (4-3) & 357.921987 & 42.9 & $1.8 \times 10^{-3}$ \\
%\hline
%HCO$^+$ (1-0) &  89.1885247 &  4.3 & $4.2 \times 10^{-5}$ \\
%%HCO$^+$ (2-1) & 178.3750563 & 12.8 & $4.0 \times 10^{-4}$ \\
%HCO$^+$ (3-2) & 267.5576259 & 25.7 & $1.5 \times 10^{-3}$ \\
%HCO$^+$ (4-3) & 356.7342230 & 42.8 & $3.6 \times 10^{-3}$ \\
%\hline
%H$^{13}$CO$^+$ (1-0) &  86.7542884 &  4.2 & $3.9 \times 10^{-5}$ \\
%%H$^{13}$CO$^+$ (2-1) & 173.5067003 & 12.5 & $3.7 \times 10^{-4}$ \\
%H$^{13}$CO$^+$ (3-2) & 260.2553390 & 25.0 & $1.3 \times 10^{-3}$ \\
%H$^{13}$CO$^+$ (4-3) & 346.9983440 & 41.6 & $3.3 \times 10^{-3}$ \\
%\hline
%\end{tabular} \end{center}
%\tablecomments{$(a)$ Frequency taken from the CDMS\footnote{https://cdms.astro.uni-koeln.de/} %\citep{2001A&A...370L..49M,2005JMoSt.742..215M}; $(b)$ Upper level energy of the transition; $(c)$ $A_{ul}$: Einstein %coefficient of spontaneous emission.}
%\end{table}
%%%%%%%%%%%%%%%%%%%%%%%%%%%%%%%%%%%%%%%%%%%%%%
\section{Velocity-Integrated Intensity Images and Ratios}\label{sec:mom0}

\subsection{Velocity-Integrated Images}\label{sec:velint}
\begin{figure}[h]
\centering{
\includegraphics[width=1.\textwidth]{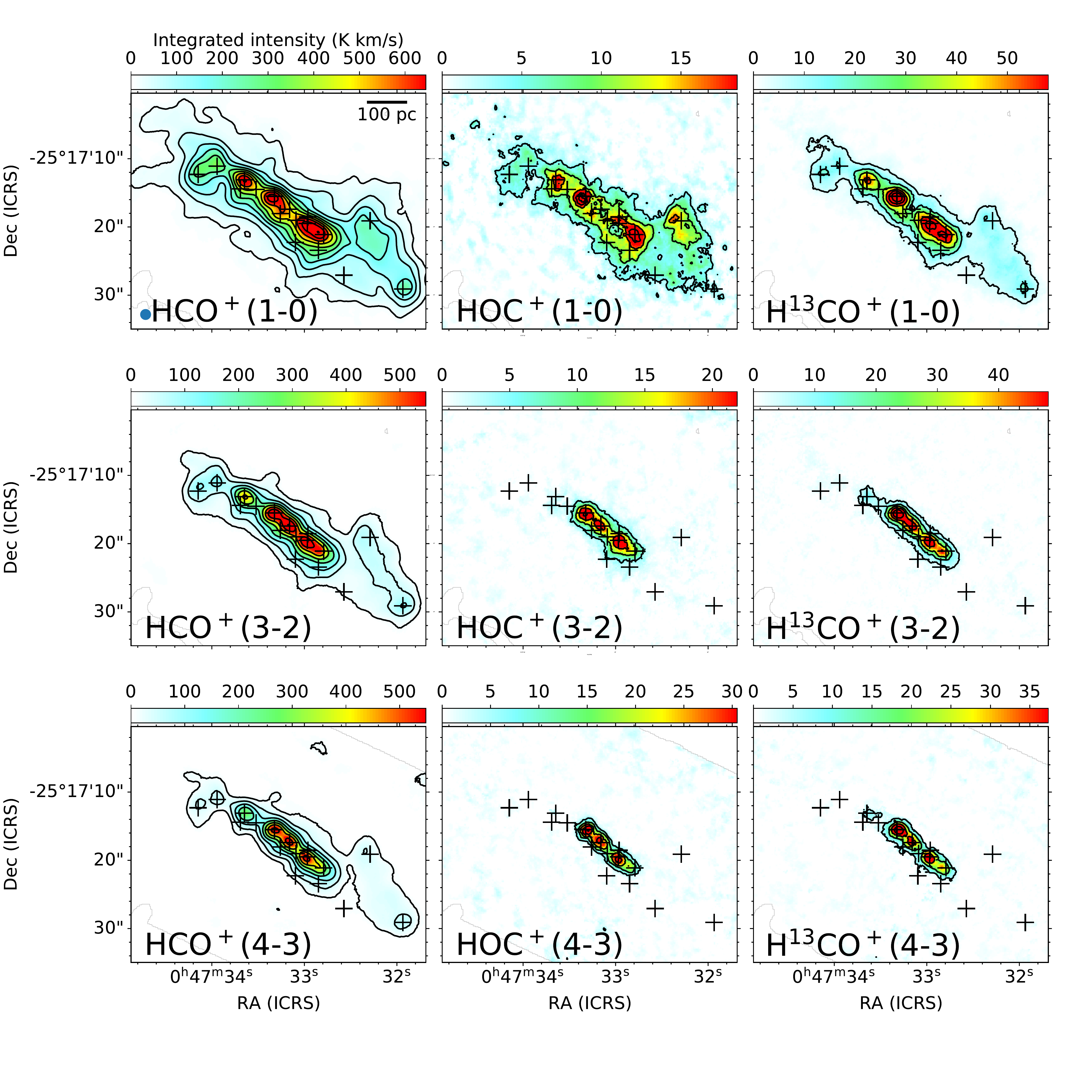}
}
%\caption{Velocity-integrated intensity images of the $J=1-0$, $3-2$, and $4-3$ transitions of HCO$^+$, H$^{13}$CO$^+$, and HOC$^+$.
%The beam size of 1.6$''$ is shown at the bottom left corner of the first panel as a blue circle. 
%Contour levels ($N= 1,2,3...$)are HCO$^+$(1-0): 20$n^{1.7}\sigma_{\rm typ}$ with $\sigma_{\rm typ}$ being a typical error defined in Appendix \ref{sec:app_rms} ($\sigma_{\rm typ}=0.87$ K km s$^{-1}$), HCO$^+$ (3-2): 20$n^{2.0}\sigma_{\rm typ}$ with $\sigma_{\rm typ} = 0.61$ K km s$^{-1}$, HCO$^+$ (4-3): $20n^{2.0}\sigma_{\rm typ}$ with $\sigma_{\rm typ} = 0.57$ K km s$^{-1}$, 
%HOC$^+$(1-0): $5n^{1.3}\sigma_{\rm typ}$ with $\sigma_{\rm typ} = 0.75$ K km s$^{-1}$, HOC$^+$(3-2): $10n^{1.3}\sigma_{\rm typ}$ with $\sigma_{\rm typ} = 0.48$ K km s$^{-1}$, HOC$^+$(4-3): $10n^{1.3}\sigma_{\rm typ}$ with $\sigma_{\rm typ} = 0.52$ K km s$^{-1}$,
%H$^{13}$CO$^+$(1-0): $10n^{1.7}\sigma_{\rm typ}$ with $\sigma_{\rm typ}= 0.50$ K km s$^{-1}$,  H$^{13}$CO$^+$(3-2): $10n^{1.5}\sigma_{\rm typ}$ with $\sigma_{\rm typ} = 0.46$ K km s$^{-1}$, H$^{13}$CO$^+$ (4-3): $10n^{1.5}\sigma_{\rm typ}$ with $\sigma_{\rm typ}=0.52$ K km s$^{-1}$. Grey crosses are positions analyzed 
%in this paper (See Section \ref{sec:velint} and Table \ref{tab:pos_coord}). \label{fig:mom0}}
\caption{Velocity-integrated intensity images of the $J=1-0$, $3-2$, and $4-3$ transitions of HCO$^+$, H$^{13}$CO$^+$, and HOC$^+$.
The beam size of 1.6$''$ is shown at the bottom left corner of the first panel as a blue circle. 
Contour levels ($N= 1,2,3...$) are drawn for multiples of typical errors $\sigma_{\rm typ}$ listed in 
Table~\ref{tab:rms} for transitions HCO$^+$(1-0): 20$n^{1.7}\sigma_{\rm typ}$, HCO$^+$ (3-2): 20$n^{2.0}\sigma_{\rm typ}$, HCO$^+$ (4-3): $20n^{2.0}\sigma_{\rm typ}$, 
HOC$^+$(1-0): $5n^{1.3}\sigma_{\rm typ}$, HOC$^+$(3-2): $10n^{1.3}\sigma_{\rm typ}$, HOC$^+$(4-3): $10n^{1.3}\sigma_{\rm typ}$,
H$^{13}$CO$^+$(1-0): $10n^{1.7}\sigma_{\rm typ}$,  H$^{13}$CO$^+$(3-2): $10n^{1.5}\sigma_{\rm typ}$, H$^{13}$CO$^+$ (4-3): $10n^{1.5}\sigma_{\rm typ}$. Grey crosses are positions analyzed 
in this paper (See Section~\ref{sec:velint} and Table~\ref{tab:pos_coord}). \label{fig:mom0}}
\end{figure}

Figure \ref{fig:mom0} shows velocity-integrated 
intensity images of HCO$^+$, H$^{13}$CO$^+$, and HOC$^+$ in the $J=1-0$, $3-2$ and $4-3$ transitions. 
Spectroscopic properties of these transitions are 
listed in Table \ref{tab:rms}.
We did not include $J=2-1$ transitions due to the poor atmospheric transmission close to the telluric water absorption near 183~GHz, although they are included in the Band 5 follow-up of our survey. 
The RMS noise values of these images are not uniform throughout the images because we used the mask per channel (see Section \ref{sec:obs}), and the number of channels used for integration is different for each pixel.
We discuss how we estimated the RMS noise in these images in Appendix \ref{sec:app_rms}. 
Despite the non-uniform RMS values within the image, we can define $\sigma_{\rm typ}$, typical error values for each image, assuming that the number of channels integrated is 40.
Contours in moment 0 images (Figure \ref{fig:mom0}) are displayed with multiples of $\sigma_{\rm typ}$.

In this paper, we analyzed positions we named M\# or A\#, where \# is a number.
M1-10 are positions taken from \citet{2015ApJ...801...63M} corresponding to their clumps \#1-10,
and additional positions, A1-8, are also chosen for analysis.
Out of these clumps, we did not include M1 due to the low S/N of the emission (position not shown with a grey cross in Figure \ref{fig:mom0}),
and M6 for the presence of an absorption feature (see Section \ref{sec:spec}) in our analysis.

The additional points are chosen for the following reasons. The position A1 is close to another superbubble identified by \citet{krieger_molecular_2019},
A2 is placed at the base of the outflow (southwest streamer), A3, A4, and A5 are at 
the edge of the superbubble surrounding M5, and other positions are chosen somewhere in between the molecular clumps. The position A6 is close to clump 5 in \citet{leroy_alma_2015}. The positions A7 and A8 are also on the shell of superbubbles,
different from the ones surrounding M5. Additional information of these positions is shown in Table \ref{tab:pos_coord}.

\begin{figure*}[h]
\centering{
\includegraphics[width=0.7\textwidth]{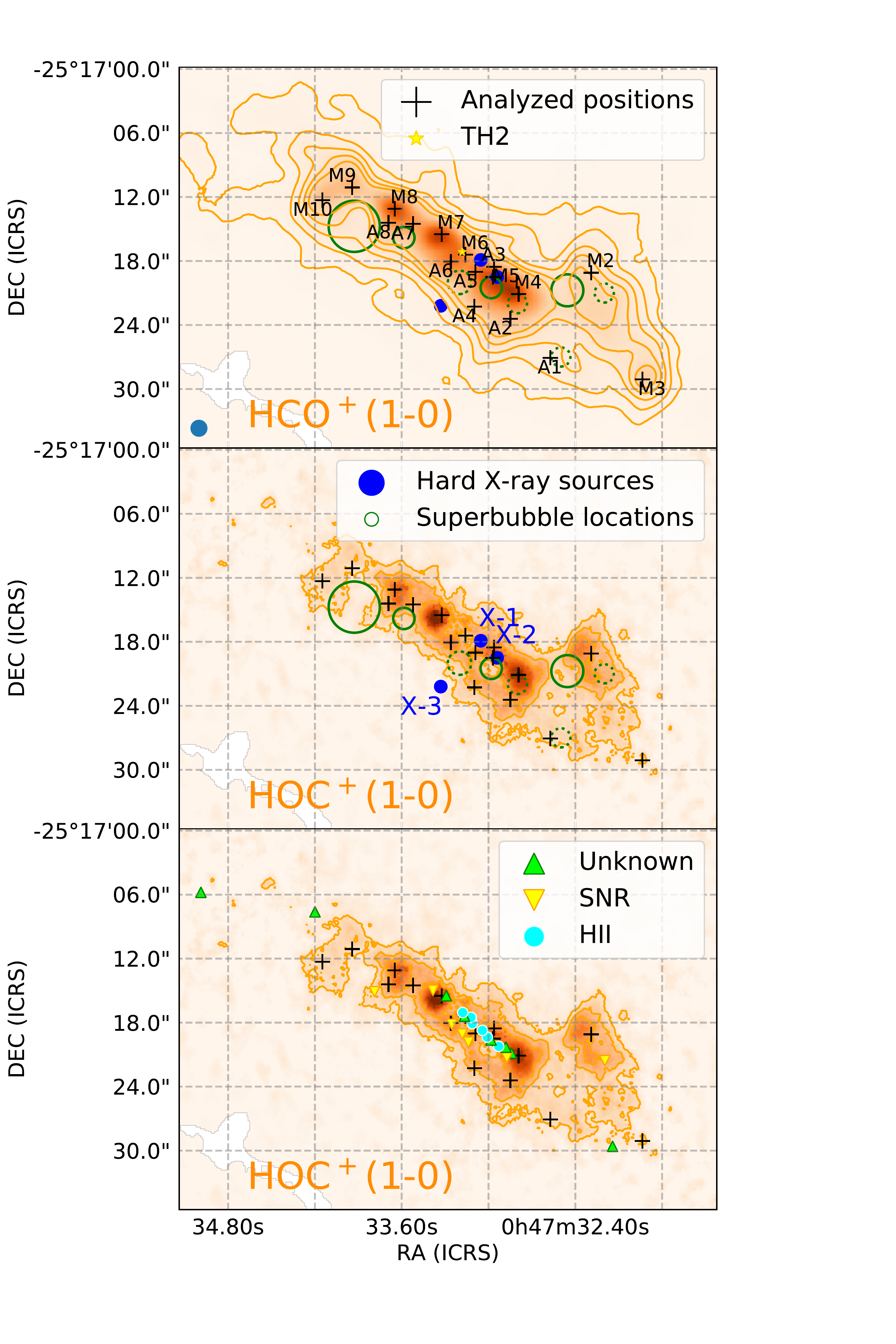}
}
\caption{Positions that are analyzed in this paper, hard X-ray sources, and superbubble positions
are overlaid on (top) HCO$^+$(1-0) and (middle) HOC$^+$(1-0) moment 0 images. Superbubble locations 
are shown with green solid circles for ones that were identified in our data,
while those that we do not identify in our data, but were reported in \citet{krieger_molecular_2019}
are shown with dotted green circles with only approximate sizes and locations. Labels of analyzed positions are shown in the top figure while hard X-ray sources are labeled in the middle figure for the legibility.
(Bottom) Radio sources from \citet{ulvestad_vla_1997} 
are also plotted over \hocp(1-0) according to 
their spectral type (supernova remnants, HII regions, or unknown).
\label{fig:pos_mom0} }
\end{figure*}

To aid the discussion of the distribution of emission,
we plotted the positions analyzed in this paper,  along with the positions of hard X-ray sources \citep{2010ApJ...716.1166M}, molecular superbubbles \citep{sakamoto_molecular_2006,krieger_molecular_2019}, and radio sources \citep{ulvestad_vla_1997} on the moment 0 images of HCO$^+$ and HOC$^+$ (1-0) (Figure \ref{fig:pos_mom0}). 
We make a distinction between superbubbles identified in our data and ones that are not identified 
in our data, but are reported by \citet{krieger_molecular_2019} because the ones taken from the literature only have approximate positions (solid and dotted green circles in Figure \ref{fig:pos_mom0}).

%%%%%%%%%%%%%%%%%%%%%
%%%%%%%%%%%%%%%%
%%%%%%%%%%%%%%%%%%%%%
%%%%%%%%%%%%%%%%%%%%%
%%%%%%%%%%%%%%%%%%%%%%%%%%%%%%%%%%%%%%%%%%%%%%%%%%%%%%%%%%%%%%%%%%%%%%%%
The distribution of emission varies with species and transitions.
The integrated intensity image of HCO$^+$ (1-0) is consistent with that reported by \citet{leroy_alma_2015} from earlier ALMA observations with a beam size of $1.9'' \times 1.3''$, equivalent to that of our data.
Emission of higher-$J$ transitions are more concentrated 
around the central star clusters because of the higher excitation conditions they require. 
Emission of H$^{13}$CO$^+$ is more compact than HCO$^+$ due to its lower abundance,
which means \httcop~ has a lower optical depth and less help for excitation from photon trapping.
For the moment 0 image of H$^{13}$CO$^+$ (1-0), there is a ``hole" in the emission
at the location of ``TH2" (near position M6), one of the brightest continuum positions reported by \citet{turner_1_1985} and close to the dynamical center of the galaxy.
As we discuss in Section \ref{sec:spec}, the spectra obtained towards this position
show that this is due to absorption 
against the strong continuum.
The absorption of H$^{13}$CO$^+$(1-0) is more visible than that of HCO$^+$(1-0) likely 
because its emission is more compact than HCO$^+$ 
within the beam. This difference in the extent of absorption becomes more apparent
in the spectra shown in Section \ref{sec:spec}.
We do not resolve the continuum source, 
and find emission surrounding TH2
and absorption against TH2 within the beam.
%For more compact emission like that of H$^{13}$CO$^+$, 
%the absorption becomes more apparent, while emission is still seen from components more extended than the %continuum source.
Nonetheless, the HCO$^+$(1-0) transition is also likely affected by absorption judging from the spectral shape (Section \ref{sec:spec}).
%Fig. \ref{fig:spectra}).
This may be the reason that there is no obvious clump 
in the TH2 position in the HCO$^+$(1-0) moment 0 image \citep[also noted in ][]{leroy_alma_2015}, although there is a molecular clump in 
higher $J$ transitions.

Emission of HOC$^+$(1-0) has a different distribution compared to HCO$^+$, H$^{13}$CO$^+$, and higher-$J$ transitions of HOC$^+$ near positions M2, M5, and M6.
There is a superbubble reported by \citet{krieger_molecular_2019} at position M5.
Unlike HCO$^+$ and H$^{13}$CO$^+$, which show relatively smooth distribution, \hocp~shows a hole-like structure 
clearly tracing this superbubble. This distribution implies that \hocp~is enhanced at locations with 
unusual physical conditions, instead of tracing dense gas in general.
Even in regions with hole-like suppressed intensities in HOC$^+$(1-0),
the intensities of higher $J$ transitions do not show any decrease in those regions, indicating higher excitation temperatures in these structures.
At position M2, there is enhancement of HOC$^+$(1-0) in the northern part. 
There is another superbubble reported by \citet{sakamoto_molecular_2006} near this position (the western-most superbubble shown in Figure \ref{fig:pos_mom0} middle), in addition to a supernova remnant towards the southwest direction (Figure \ref{fig:pos_mom0} bottom).
At TH2 near position M6, there is also suppression of \hocp(1-0) similar to \httcop(1-0).
%%%%%%%%%%%%%%%%%%%%%
\begin{deluxetable}{cccc} 
\tablecolumns{4} 
\tablewidth{0pc} 
\tabletypesize{\scriptsize} 
\tablecaption{Coordinates of analyzed positions.} \label{tab:pos_coord} 
\tablehead{\colhead{Position} &\colhead{RA(ICRS)} &\colhead{DEC(ICRS)} &\colhead{Remarks}\\
&\colhead{00h47m} &\colhead{-25d17m} &\colhead{}}
\startdata 
M2 &32.29s &19.10s &Near a superbubble and a SNR\\
 M3 &31.94 &29.10s &\\ 
  M4 &32.79s &21.10s &Near the base of an outflow (SW streamer), Super hot core\\ 
M5 &32.97s &19.50s &IR Core, In a superbubble, hard X-ray source, Super hot core\\ 
 M7 &33.32s &15.50s &Super hot core\\
 M8 &33.65s &13.10s &\\
 M9 &33.94s &11.10s &Near a superbubble\\
 M10 &34.15s &12.30s &Near a superbubble\\
 A1 &32.57s &27.07s &Near a superbubble\\ 
  A2 &32.85s &23.42s &On the outflow SW streamer\\ 
 A3 &32.96s &18.53s &Shell of a superbubble\\
  A4 &33.10s &22.27s &Shell of a superbubble\\ 
 A5 &33.09s &19.01s &Shell of a superbubble\\ 
 A6 &33.26s &18.05s &\\ 
 A7 &33.52s &14.51s &Near a superbubble\\ 
 A8 &33.69s &14.41s &Near a superbubble
\enddata 
\tablecomments{Superbubbles identified by \citet{sakamoto_molecular_2006} or \citet{krieger_molecular_2019},
supernova remnants (SNR) by \citet{ulvestad_vla_1997}, the southwest streamer by \citet{bolatto_suppression_2013}, super hot cores by \citet{2020MNRAS.491.4573R}, IR core by \citet{1996AJ....112..534W}, X-ray sources by \citet{2010ApJ...716.1166M}. } 
\end{deluxetable} 

%%%%%%%%%%%%%%%%%%%%%

\subsection{Ratio Maps}\label{sec:ratiomap}
\begin{figure*}[h]
\centering{
\includegraphics[width=1.\textwidth,trim = 100 0 100 0]{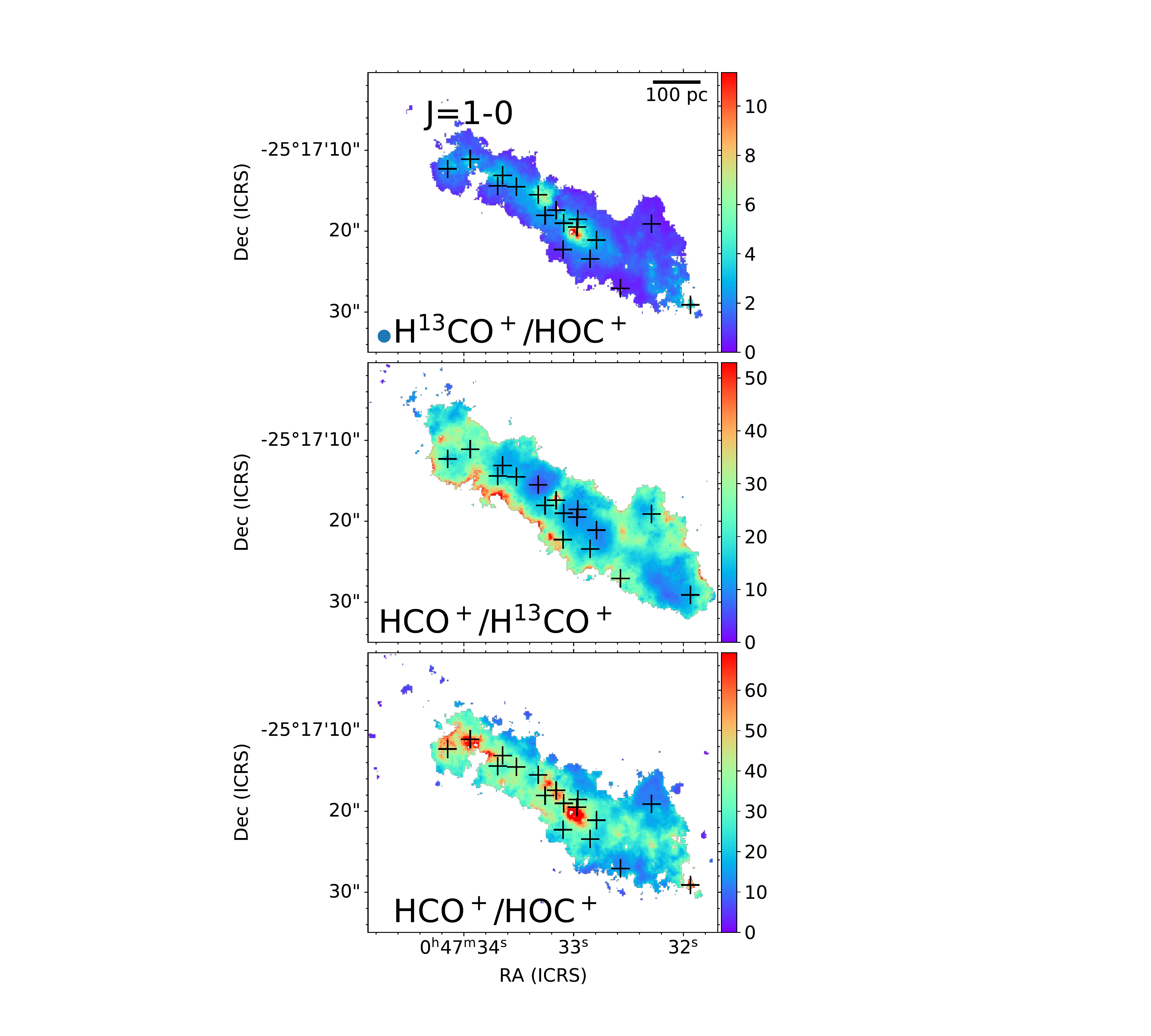}
}
\caption{Ratios of velocity-integrated intensity images for the $J=1-0$ transitions.
Grey crosses show the positions M2-10 and A1-8.\label{fig:ratio}}
\end{figure*}

\begin{figure*}[h]
\centering{
\includegraphics[width=1.\textwidth,trim = 100 0 100 0]{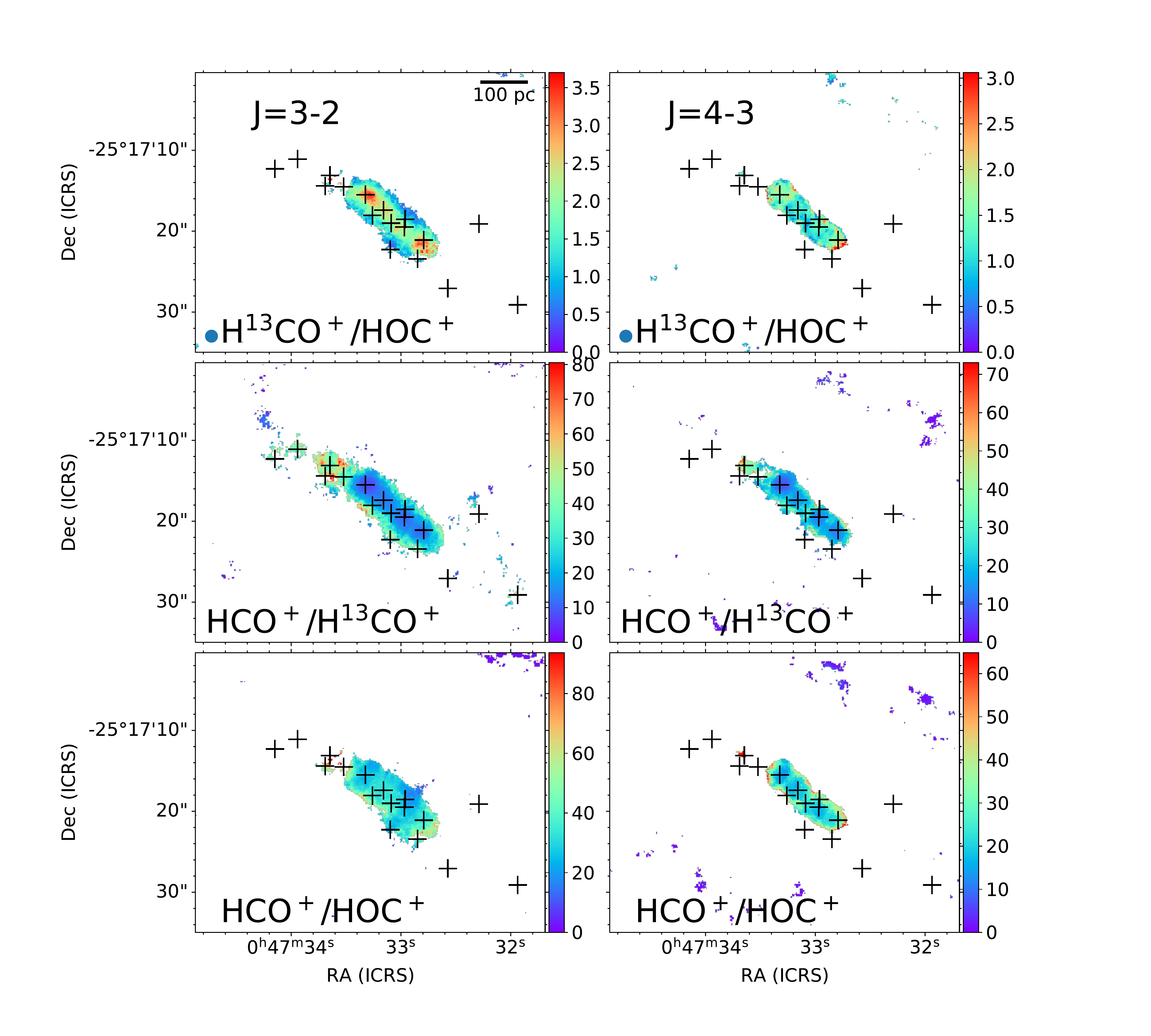}
}
\caption{Same as Figure \ref{fig:ratio}, but for $J=3-2$ and $J=4-3$ transitions.\label{fig:ratio_highJ}}
\end{figure*}

Figures \ref{fig:ratio} and \ref{fig:ratio_highJ} show velocity-integrated 
intensity ratio maps of HCO$^+$, HOC$^+$, and H$^{13}$CO$^+$
for the three transitions we observed.
In addition to the mask used to create these moment 0 images,
we eliminated any emission below $5\sigma$ from
each velocity-integrated intensity image before ratios are taken.
Note that $\sigma$ here is not $\sigma_{\rm typ}$, but the RMS
of the velocity-integrated intensity image estimated by considering the single-channel RMS
and velocity width, and the number of channels integrated as shown in Appendix \ref{sec:app_rms}.
The HCO$^+$/H$^{13}$CO$^+$(1-0) ratio map was created to assess
the optical depth of HCO$^+$(1-0). 
The range of the HCO$^+$/H$^{13}$CO$^+$(1-0) intensity ratio is $20\pm 8$.
If both lines are optically thin and the two species share the same excitation temperatures, their column density ratios can be obtained by dividing these intensity ratios by 1.1, the ratio between Einstein coefficients of these transitions.
On average, these ratios are consistent with values of $^{12}$C/$^{13}$C derived from $^{13}$C$^{18}$O by \citet{2019A&A...624A.125M} at $3''$ resolution and to that from HCO$^+$ at the $15''$ resolution ALCHEMI data extracted only from ALMA 7-m array measurements (Mart\'in et al. submitted).
%, although they are 
%much smaller than values obtained with single-dish observations such as $^{12}$C/$^{13}$C=40 \citep{2014A&A...565A...3H} and 81 in \citep{2010A&A...522A..62M}.
%Lower $^{12}$C/$^{13}$C ratios in higher-resolution data
%are reasonable because the $^{13}$C species should have a lower beam filling factor than the $^{12}$C species.
The low ratios of HCO$^+$/H$^{13}$CO$^+$(1-0) may also be affected by the optical depth of HCO$^+$,
especially for the central clumps (see discussion in Mart\'in et al. submitted).
The \hcop/\httcop~ratios in the two higher J transitions are similar to that in J=1--0 except for emission from the edge, which may be affected by low S/N ratios. 

If the regions are optically thin, which may not be the case for the densest clumps,
and if \hcop~and \hocp~share similar excitation temperatures,
the HCO$^+$/HOC$^+$(1-0) intensity ratios multiplied by 0.5 (the ratio between the Einstein A coefficients of these transitions)
can be used as the HCO$^+$/HOC$^+$ abundance ratios.
The H$^{13}$CO$^+$/HOC$^+$ ratios are less affected by
the optical depth, so they are potentially better at constraining the \hcop/\hocp~ratio. However, these ratios are not proportional to 
HCO$^+$/HOC$^+$ if the $^{12}$C/ $^{13}$C ratio is not uniform throughout the image as in \citet{2019A&A...624A.125M}.
Overall, the H$^{13}$CO$^+$/HOC$^+$ and HCO$^+$/HOC$^+$ ratios show
similar distributions, indicating that the saturation of \hcop~is not significant for most regions.
The lowest ratios are measured near the southwest part of the CMZ, position M2.
High opacity is not expected at this position, which suggests that the \hcop/\hocp~abundance ratio is actually low.
The highest ratio is seen near the position M5.
Although this position is associated with a superbubble,
the position inside the bubble does not have an enhanced HOC$^+$, which is counter-intuitive.
However, this region is also very crowded with the IR core, hard X-ray source, and super hot cores,
and thus the interpretation of the high ratio of \hcop/\hocp~is not straightforward.
The \hcop/\hocp~ratios tend to become lower with higher galactic latitude,
which likely has a lower density. 
%This may be caused by the lower densities 
%there, compared with molecular clumps.
%In fact, as we discuss later in Section \ref{sec:model}, the HCO$^+$/HOC$^+$ ratio decreases with lower density.
The density dependence of the HCO$^+$/HOC$^+$ ratios is discussed in Section \ref{sec:model}.

The \hcop/\hocp~and \httcop/\hocp~ratios of two higher-$J$ transitions 
show different variations compared to $J=1-0$ transition.
In $J=1-0$ transition, the above ratios are slightly higher in clumps 
(high column density regions such as positions M4-M8)
compared with more extended components. This trend is not seen in $J=4-3$.
The \hcop/\hocp(4-3) ratio is rather lower in clumps
compared with extended regions. This is likely due to blending with 
SO$_2$ as we discuss in Section \ref{sec:spec}.

%%%%%%%%%%%%%%%%%%%%%
%%%%%%%%%%%%%%%%%%%%%
%%%%%%%%%%%%%%%%%%%%%
%%%%%%%%%%%%%%%%%%%%%
\section{Spectra at selected positions}\label{sec:spec}
\begin{figure*}[h]
\centering{
\includegraphics[width=1.\textwidth,trim=0 100 0 150]{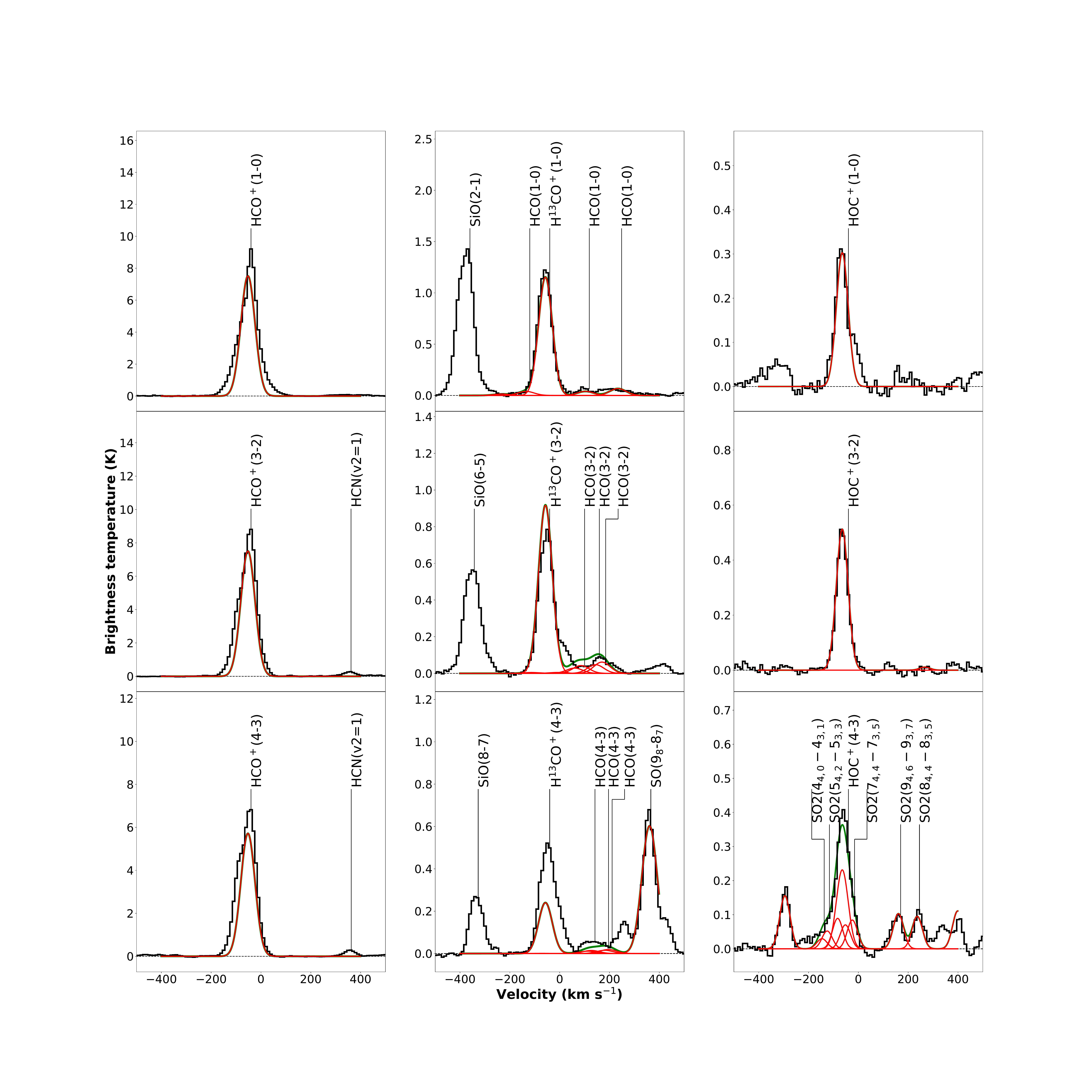}
}
\caption{Spectra of lines at the M7 position.
Spectra are centered at the systemic velocity $v_{\rm sys, LSRK}= 243 $ km \ps.
The fit to each molecular transition is shown as a red solid line, while the sums of line intensities at each frequency are shown as a green line.
\label{fig:spectra}}
\end{figure*}

\begin{figure*}
\centering{
\includegraphics[width=1.\textwidth,trim = 0 100 0 150]{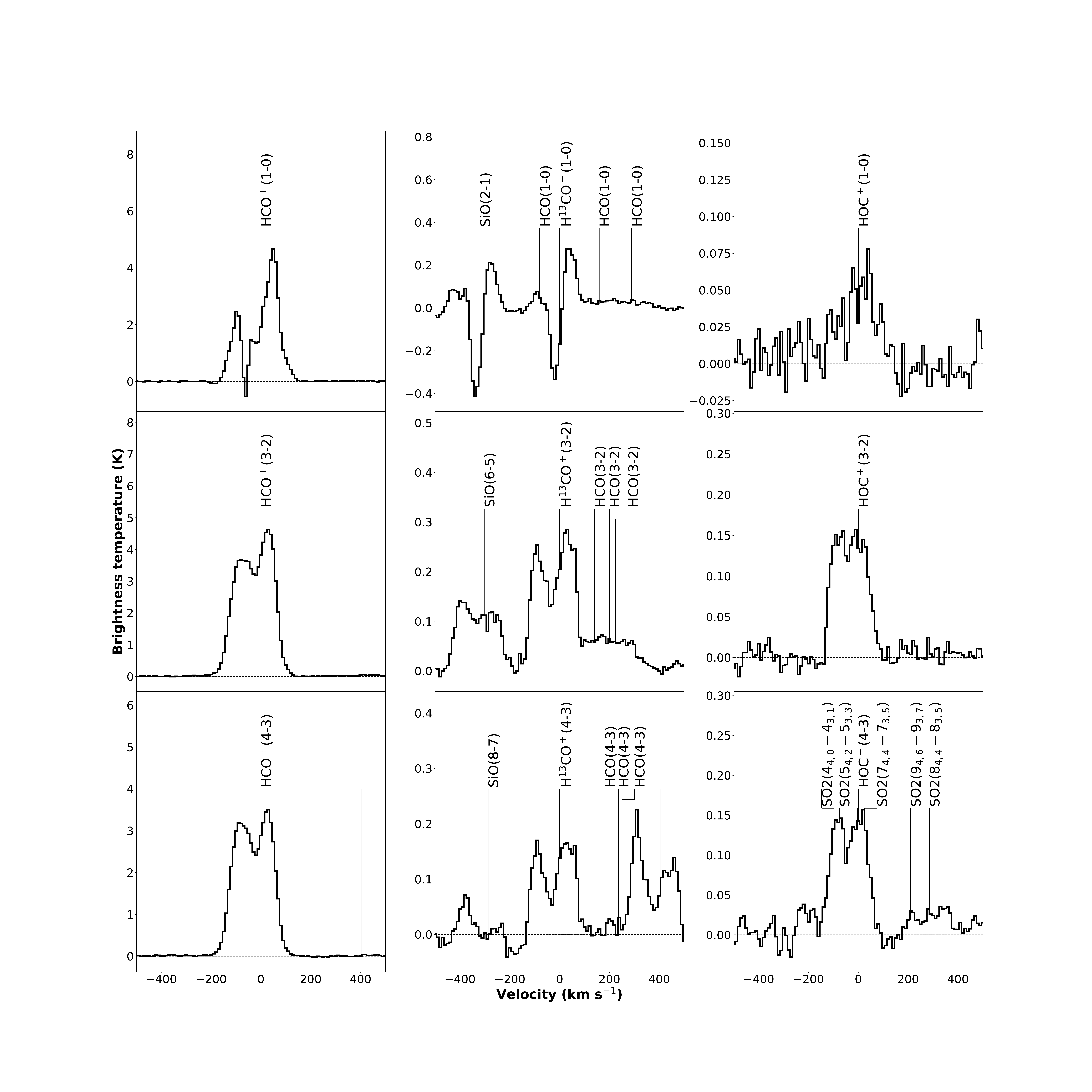}
}
\caption{Same as Figure \ref{fig:spectra}, but for the position M6. Because position M6 is not analyzed due to the absorption feature, the fit is not shown. \label{fig:spectraM6}}
\end{figure*}

Figures \ref{fig:spectra} and \ref{fig:spectraM6} show the spectra of lines of interest at positions M7 and M6 (next to TH2).
Their intensities are taken from a pixel corresponding to the coordinates of these positions, and are converted to brightness temperature (e.g., Equation 3.31 in the ALMA Cycle 8 Technical Handbook). 
Position M7 is the chemically richest among all the positions, while position M6 is known to possess absorption in H$^{13}$CO$^+$(1-0), and HCO$^+$(1-0) against the continuum emission from TH2. 
We observe similar decrements in spectral shapes of \hocp(3-2) and (4-3) but that in \hocp(1-0) cannot be confirmed
at the current sensitivity. The spectra for other positions are shown in Appendix \ref{sec:app_spectra}.

There are some neighboring and partially blended transitions
adjacent to our transitions of interest labeled in Figure \ref{fig:spectra}. 
However, many of the neighboring transitions and some blended transitions are unlikely to be causing problems, 
because we decompose the blended or neighboring lines through a multi-line spectral fitting for our analysis.
Based on our spectral fitting analysis in Section \ref{sec:col},
we show that contamination does not affect our analysis except for
SO$_2$ transitions blended with HOC$^+$(4-3).
For positions with strong SO$_2$ emission (positions M4, M5, and M7),
we fit the \hocp~transitions together with SO$_2$ transitions. 
This is possible because there are multiple neighboring SO$_2$ transitions 
in addition to the ones blended into the transition of \hocp(4-3).
The decrement in the \hocp(4-3) line shape may be partly (or mostly) due to the blended lines or multiple velocity components of \hocp.

We note that some of the spectral features of \httcop~(1-0) could appear like the P-Cygni profile (red-shifted emission plus blue-shifted absorption), but this spectral feature is not likely due to an outflow. 
The P-Cygni profile is reported as the evidence of outflow features in other locations in 28 mas resolution observations by \citet{2021ApJ...912....4L}.
 In our case, the spectra are more affected by the strong contamination by emission surrounding the continuum in our coarser 1\farcs6 resolution images.

%%%%%%%%%%%%%%%%%%%%%

%%%%%%%%%%%%%%%%%%%%%

\section{Column Densities and Ratios}\label{sec:col}
We derived column densities towards the positions shown in Figure \ref{fig:pos_mom0}, whose coordinates are listed in Table~\ref{tab:pos_coord}.
The derived column densities and excitation temperatures of HOC$^+$, HCO$^+$, H$^{13}$CO$^+$ and C$^{18}$O are listed in Table~\ref{tab:columns}.
These column densities were derived using MADCUBA \citep{2019A&A...631A.159M}.
MADCUBA determines the column densities and excitation temperatures through spectral fitting under the assumption of local thermodynamic equilibrium
with consideration of optical depth in the modeled spectra.
In MADCUBA, the continuum-subtracted brightness temperature of the transition $T_L$
is modeled as 
\begin{equation}
T_L = [J(T_{\rm ex}) - J(T_c) - J(T_{bg})](1 - e^{-\tau_{\nu}})    
\end{equation}
where $J$ is the intensity in the temperature unit, $T_{ex}$ is the excitation temperature, 
$T_c$ is the continuum brightness temperature, $T_{bg}$ is the background temperature,
and $\tau_{\nu}$ is the optical depth at the frequency $\nu$ \citep[Equation (6) of ][]{2019A&A...631A.159M}.
It is apparent that the intensity is insensitive to an extremely high optical depth.
We assumed a large source size (10$''$) to simulate emission filling the beam 
(i.e., no beam dilution) so the line brightness temperature is similar to 
the synthesized beam temperature.
If only one transition was detected, we assume an excitation temperature of 5.0 K for the derivation
of column densities of H$^{13}$CO$^+$ or HOC$^+$.
This excitation temperature is similar to that of 
HCO$^+$ in positions
where high-$J$ transitions of 
\hocp~or \httcop~are not detected.
Note that all 3 transitions of \hcop~are detected in all the positions analyzed here.
The line widths are in the range of 50-100 km\,\ps in full-width at half maximum (FWHM),
except for a component in the position A5 with $\sim 150\,$km\,\ps.

\subsection{HCO$^+$/H$^{13}$CO$^+$}
Table~\ref{tab:column_ratio} shows the column density ratios of $N$(HCO$^+$)/$N$(H$^{13}$CO$^+$), 
$N$(HCO$^+$)/$N$(HOC$^+$), and $N$(H$^{13}$CO$^+$)/$N$(HOC$^+$). 
Before presenting HCO$^+$/HOC$^+$ abundance
ratios, we discuss the H$^{12}$CO$^+$/H$^{13}$CO$^+$ ratio
to see if the $^{12}$C/$^{13}$C ratio in HCO$^+$ is 
reasonable compared with the literature values of 
the carbon isotopic ratio in NGC\,253 in order to check 
if the HCO$^+$ transitions are optically thick.
As mentioned earlier, this optical depth effect is taken into account in the derivation
of column densities in MADCUBA, but it should still be checked in the case of very high 
optical depths.
Values of $N$(HCO$^+$)/$N$(H$^{13}$CO$^+$) are 
relatively low $\sim$ 10-30.
These values are similar to the intensity ratios of 
HCO$^+$/H$^{13}$CO$^+$ discussed in Section \ref{sec:mom0}.
MADCUBA fitting results show that 
HCO$^+$ is optically thick ($\tau > 4$) in few positions (M4, M5, and M8)
of the CMZ, $\tau\sim 2-4$ in A6, A8, M7, M9, and M10, and optically thin in other positions
when a transition with the highest optical depth is considered (See Table \ref{tab:columns}).
The $^{12}$C/$^{13}$C ratios obtained in NGC\,253 in the literature are $21\pm6$ \citep{2019A&A...624A.125M} from C$^{18}$O/$^{13}$C$^{18}$O, which is close to our obtained values.
However, note that the $^{12}$C/$^{13}$C ratio can also depend on
the species because of fractionation \citep{2020A&A...640A..51C,2020MNRAS.497.4333V}.

%%%%%%%%%%%%%%%%%%%%%

\begin{longrotatetable}
\begin{deluxetable}{cccccccccc} 
\tablecolumns{10} 
\tablewidth{0pc} 
\tabletypesize{\scriptsize} 
\tablecaption{Observed column densities and excitation temperatures} \label{tab:columns} 
\tablehead{\colhead{Region} &\colhead{$N$(HCO$^+$)}  &\colhead{$T_{\rm ex}$(HCO$^+$)}    &\colhead{$N$(HOC$^+$)} 
&\colhead{$T_{\rm ex}$(HOC$^+$)} &\colhead{$N$(H$^{13}$CO$^+$)} &\colhead{$T_{\rm ex}$(H$^{13}$CO$^+$)} 
&\colhead{$N$(C$^{18}O$)} &\colhead{$T_{\rm ex}$(C$^{18}$O)} &\colhead{$\tau(HCO^+,max)$}\\ 
\colhead{} &\colhead{($10^{14}$ cm$^{-2}$)}  &\colhead{(K)} &\colhead{($10^{12}$ cm$^{-2}$)}  &\colhead{(K)} &\colhead{($10^{12}$ cm$^{-2}$)}  &\colhead{(K)} &\colhead{($10^{16}$ cm$^{-2}$)}  &\colhead{(K)} &\colhead{} }
\startdata 
M2 & $1.6 \pm 0.2$  & $6.4 \pm 0.2$   & $14.1 \pm 1.9$  & $4.2 \pm 0.3$   & $11.3 \pm 1.2$  & $4.4 \pm 0.3$   & $5.1 \pm 0.2$  & $8.0 \pm 0.2$  &0.6\\ 
 & $0.7 \pm 0.1$  & $6.5 \pm 0.4$   & $5.1 \pm 0.8$  & $5.0 \pm ...$   & $3.6 \pm 0.4$  & $5.0 \pm ...$   & $2.1 \pm 0.2$  & $7.4 \pm 0.5$  \\ 
M3 & $3.3 \pm 0.1$  & $7.3 \pm 0.2$   & $6.7 \pm 2.1$  & $4.2 \pm 0.8$   & $19.0 \pm 1.3$  & $4.8 \pm 0.2$   & $5.0 \pm 0.3$  & $8.6 \pm 0.4$  &0.7\\ 
M4 & $23.9 \pm 3.7$  & $10.5 \pm 0.2$   & $25.3 \pm 0.7$  & $7.5 \pm 0.1$   & $79.2 \pm 3.4$  & $7.0 \pm 0.2$   & $17.3 \pm 0.6$  & $17.5 \pm 0.7$  &9.7\\ 
M5 & $11.0 \pm 0.8$  & $13.0 \pm 0.3$   & $13.4 \pm 0.5$  & $17.8 \pm 0.7$   & $71.1 \pm 3.9$  & $9.1 \pm 0.3$   & $13.9 \pm 0.2$  & $25.0 \pm 0.9$  &4.4\\ 
 & $8.0 \pm 3.8$  & $5.2 \pm 0.3$   & $5.1 \pm 0.9$  & $6.1 \pm 0.9$   & $15.4 \pm 2.5$  & $5.0 \pm ...$   & $5.4 \pm 0.4$  & $10.2 \pm 0.6$  \\ 
M7 & $12.8 \pm 0.8$  & $13.5 \pm 0.3$   & $21.2 \pm 0.7$  & $12.3 \pm 0.3$   & $105.0 \pm 5.1$  & $8.1 \pm 0.2$   & $20.6 \pm 0.4$  & $23.6 \pm 0.9$  &3.6\\ 
M8 & $13.2 \pm 2.3$  & $9.9 \pm 0.2$   & $20.7 \pm 1.5$  & $4.8 \pm 0.3$   & $60.5 \pm 2.0$  & $5.3 \pm 0.1$   & $12.1 \pm 0.4$  & $13.4 \pm 0.4$  &4.3\\ 
M9 & $6.5 \pm 0.5$  & $6.6 \pm 0.2$   & $6.1 \pm 0.6$  & $4.8 \pm 0.4$   & $20.6 \pm 1.2$  & $4.8 \pm 0.2$   & $5.2 \pm 0.3$  & $10.2 \pm 0.4$  &2.4\\ 
M10 & $6.3 \pm 0.6$  & $6.0 \pm 0.1$   & $6.8 \pm 0.4$  & $5.0 \pm ...$   & $21.7 \pm 1.4$  & $4.6 \pm 0.2$   & $5.9 \pm 0.2$  & $8.7 \pm 0.2$  &2.4\\ 
A1 & $1.0 \pm 0.1$  & $5.4 \pm 0.2$   & $6.0 \pm 0.5$  & $5.0 \pm ...$   & $3.7 \pm 0.3$  & $5.0 \pm ...$   & $3.3 \pm 0.2$  & $6.6 \pm 0.2$  &0.4\\ 
A2 & $4.6 \pm 0.2$  & $6.9 \pm 0.2$   & $10.7 \pm 0.8$  & $5.4 \pm 0.3$   & $18.8 \pm 1.0$  & $4.9 \pm 0.2$   & $4.0 \pm 0.2$  & $12.1 \pm 0.5$ & 1.2\\ 
A3 & $6.9 \pm 0.6$  & $9.8 \pm 0.3$   & $13.0 \pm 0.8$  & $12.1 \pm 0.5$   & $39.0 \pm 2.0$  & $7.7 \pm 0.2$   & $9.2 \pm 0.2$  & $19.6 \pm 0.6$  &1.7\\ 
 & ... &... & ... &... & ... &... & $2.7 \pm 0.2)$  & $9.3 \pm 0.6$  \\ 
A4 & $4.1 \pm 0.2$  & $6.1 \pm 0.1$   & $9.3 \pm 0.5$  & $6.2 \pm 0.2$   & $17.6 \pm 1.3$  & $4.5 \pm 0.2$   & $3.4 \pm 0.1$  & $11.4 \pm 0.4$  &1.2\\ 
A5 & $3.0 \pm 1.3$  & $6.4 \pm 1.0$   & $8.0 \pm 2.8$  & $4.2 \pm 0.9$   & $12.4 \pm 3.2$  & $4.5 \pm 0.9$   & $2.9 \pm 0.4$  & $8.2 \pm 0.9$  &1.9\\ 
 & $4.3 \pm 0.2$  & $11.7 \pm 0.4$   & $6.9 \pm 0.4$  & $20.2 \pm 1.3$   & $34.5 \pm 2.6$  & $8.1 \pm 0.3$   & $8.8 \pm 0.4$  & $18.9 \pm 1.1$  \\ 
A6 & $4.6 \pm 0.8$  & $8.1 \pm 0.3$   & $7.8 \pm 0.5$  & $6.6 \pm 0.2$   & $16.7 \pm 0.9$  & $6.4 \pm 0.2$   & $3.1 \pm 0.1$  & $10.0 \pm 0.3$  &2.3\\ 
 & $3.0 \pm 0.9$  & $5.9 \pm 0.5$   & $3.6 \pm 0.5$  & $5.0 \pm ...$   & $9.8 \pm 0.9$  & $5.6 \pm 0.3$   & $3.8 \pm 0.1$  & $16.6 \pm 0.6$  \\ 
 & $2.3 \pm 0.1$  & $9.9 \pm 0.3$   & $4.6 \pm 0.4$  & $14.8 \pm 0.9$   & $8.8 \pm 0.7$  & $8.6 \pm 0.4$   & $2.8 \pm 0.1$  & $23.8 \pm 1.9$  \\ 
A7 & $4.2 \pm 0.3$  & $8.0 \pm 0.2$   & $6.2 \pm 0.7$  & $5.1 \pm 0.4$   & $24.7 \pm 1.2$  & $5.3 \pm 0.2$   & $3.1 \pm 0.2$  & $11.4 \pm 0.5$  &1.5\\ 
 & $2.4 \pm 1.1$  & $6.4 \pm 0.5$   & $8.1 \pm 0.8$  & $5.3 \pm 0.3$   & $6.5 \pm 0.8$  & $5.7 \pm 0.3$   & $5.7 \pm 0.2$  & $14.1 \pm 0.5$  \\ 
A8 & $7.0 \pm 0.6$  & $7.1 \pm 0.2$   & $10.2 \pm 1.2$  & $5.0 \pm 0.4$   & $20.9 \pm 1.6$  & $4.8 \pm 0.2$   & $5.7 \pm 0.2$  & $11.6 \pm 0.3$  &2.2\\ 
 & $0.8 \pm 0.1$  & $7.0 \pm 0.5$   & $4.4 \pm 0.5$  & $5.0 \pm ...$   & $11.8 \pm 2.7$  & $4.2 \pm 0.5$   & $1.7 \pm 0.2$  & $13.1 \pm 1.2$  \\ 
\enddata 
\tablecomments{As mentioned in Section \ref{sec:col}, the excitation temperature was fixed when the species is only detected in one transition. For these cases, the errors of excitation temperature are not shown. Errors shown are only from spectral fitting, and do not contain observational errors. These column densities are derived with the consideration of optical depth. The maximum value of optical depth of \hcop~is shown in the last column. Only the component with the highest value is shown for each position.} 
\end{deluxetable} 
\end{longrotatetable}
%%%%%%%%%%%%%%%%%%%%%
%%%%%%%%%%%%%%%%%%%%%
\begin{deluxetable}{cccc} 
\tablecolumns{4} 
\tablewidth{0pc} 
\tabletypesize{\scriptsize} 
\tablecaption{Column density ratios} \label{tab:column_ratio} 
\tablehead{\colhead{Region} &\colhead{$N$(HCO$^+$)/$N$(H$^{13}$CO$^+$)}  &\colhead{$N$(HCO$^+$)/$N$(HOC$^+$)} 
&\colhead{$N$(H$^{13}$CO$^+$)/$N$(HOC$^+$)}} 
\startdata 
M2 & $15.2 \pm 2.5$   & $11.9 \pm 2.3$   & $0.8 \pm 0.1$   \\ 
M3 & $17.2 \pm 1.3$   & $48.5 \pm 15.4$   & $2.8 \pm 0.9$   \\ 
M4 & $30.1 \pm 4.8$   & $94.5 \pm 14.7$   & $3.1 \pm 0.2$   \\ 
M5 & $22.0 \pm 5.5$   & $102.8 \pm 25.8$   & $4.7 \pm 0.5$   \\ 
M7 & $12.2 \pm 1.0$   & $60.3 \pm 4.3$   & $5.0 \pm 0.3$   \\ 
M8 & $21.9 \pm 3.8$   & $63.8 \pm 11.8$   & $2.9 \pm 0.2$   \\ 
M9 & $31.6 \pm 3.1$   & $106.0 \pm 13.1$   & $3.4 \pm 0.4$   \\ 
M10 & $29.2 \pm 3.5$   & $93.8 \pm 11.4$   & $3.2 \pm 0.3$   \\ 
A1 & $26.8 \pm 2.8$   & $16.8 \pm 1.6$   & $0.6 \pm 0.1$   \\ 
A2 & $24.6 \pm 1.9$   & $43.4 \pm 3.9$   & $1.8 \pm 0.2$   \\ 
A3 & $17.7 \pm 1.7$   & $52.9 \pm 5.5$   & $3.0 \pm 0.2$   \\ 
A4 & $23.2 \pm 1.9$   & $44.1 \pm 2.9$   & $1.9 \pm 0.2$   \\ 
A5 & $15.6 \pm 3.8$   & $49.1 \pm 14.7$   & $3.2 \pm 0.8$   \\ 
A6 & $28.1 \pm 5.7$   & $61.9 \pm 12.9$   & $2.2 \pm 0.2$   \\ 
A7 & $21.4 \pm 4.7$   & $46.6 \pm 10.8$   & $2.2 \pm 0.3$   \\ 
A8 & $23.8 \pm 3.8$   & $53.5 \pm 7.8$   & $2.2 \pm 0.4$   \\ 
\enddata 
\tablecomments{Errors shown are only from spectral fitting, and do not contain observational errors.} 
\end{deluxetable} 

%%%%%%%%%%%%%%%%%%%%%

\subsection{HCO$^+$/HOC$^+$} \label{sec:col_hcop_hocp}
\begin{figure*}[h]
\centering{
\includegraphics[width=0.99\textwidth]{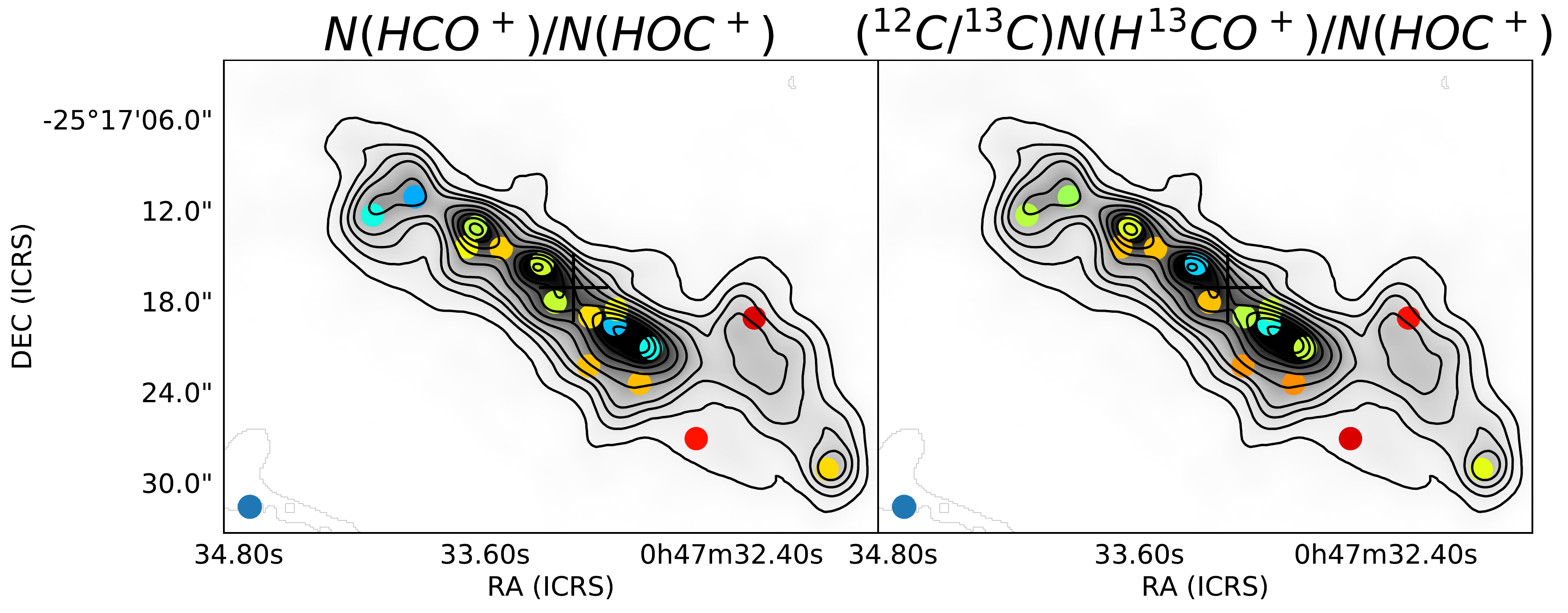}
}
\centering{
\includegraphics[width=0.88\textwidth]{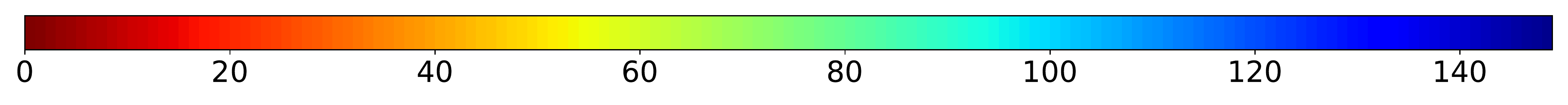}
}
\centering{
\includegraphics[width=1.\textwidth]{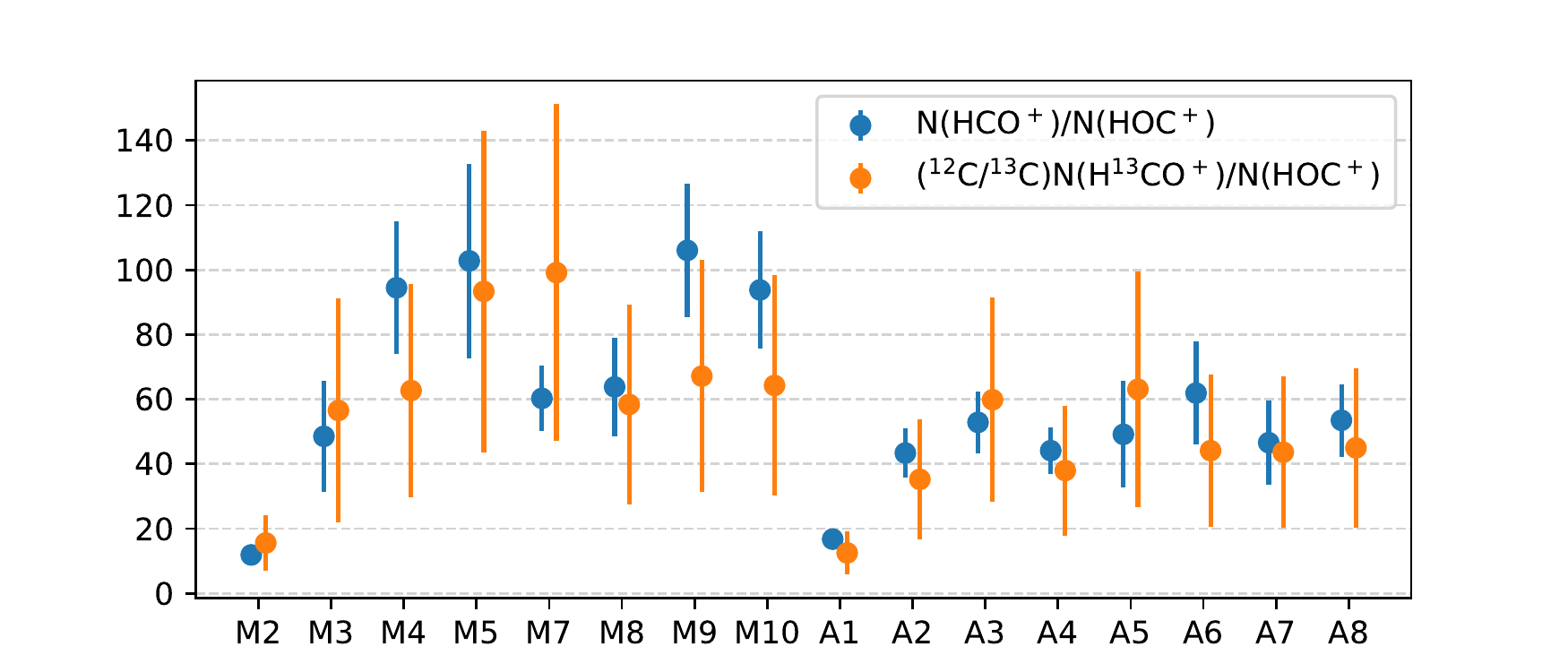}}
\caption{Abundance ratios of HCO$^+$/HOC$^+$ plotted as colored dots over HCO$^+$ moment 0 images (greyscale + contours) using $N$(\hcop)/$N$(\hocp) (Method 1, top left) and  ($^{12}$C/$^{13}$C)$N$(\httcop)/$N$(\hocp) (Method 2, top right). The \hcop/\hocp~abundance ratios estimated from two methods in top panels are shown for individual positions together with error bars (bottom). \label{fig:ratio_column}}
\end{figure*}

We derive the abundance ratio of HCO$^+$/HOC$^+$
using two methods to account for possible effects of line saturation. 
The first method is to directly use the column density ratios $N$(HCO$^+$)/$N$(HOC$^+$) (Method 1). This method should be a good indicator of HCO$^+$/HOC$^+$ in optically thin regions, which likely holds for most of the observed regions.
The other method is to take the column density ratios
of H$^{13}$CO$^+$/HOC$^+$ and multiply them by an assumed a constant $^{12}$C/$^{13}$C across the entire region ($\frac{^{12}C}{^{13}C}N$(H$^{13}$CO$^+$)/$N$(HOC$^+$), Method 2).
This method may shed light on the errors from the optically thick regions, but this method is also 
dependent on the isotopic ratios of $^{12}$C/$^{13}$C.
Here we assume the carbon isotopic ratio to be $^{12}$C/$^{13}$C =$20\pm 10$.
This value is suggested by our \hcop/\httcop(1-0) intensity ratio (see Section \ref{sec:ratiomap}), 
and is similar to the value obtained by \citet{2019A&A...624A.125M}.
The column density ratios of HCO$^+$/HOC$^+$ and H$^{13}$CO$^+$/HOC$^+$ are also included in Table \ref{tab:column_ratio}.
Values of HCO$^+$/HOC$^+$ abundance ratios estimated from the above-mentioned methods are also displayed in Figure \ref{fig:ratio_column}.
As already shown in the ratio map of \hcop/\hocp (1-0), the abundance ratio of 
\hcop/\hocp is low in the position M2. In addition, the position A1 also has a comparably
low ratio.

%%%%%%%%%%%%%%%%%%%%%
%%%%%%%%%%%%%%%%%%%%%

\subsection{Fractional Abundances}
In addition to column density ratios, we derived fractional abundances defined as column density of species over total hydrogen column density (here denoted as $X({\rm species})$)
To obtain fractional abundances, we estimated the molecular hydrogen column densities from
the dust continuum (see continuum images in Appendix \ref{sec:app_cont}) and 
C$^{18}$O column densities (see  Appendix \ref{sec:app_coltot}), and assumed N(H$_{total}$)=2N($H_2$) to obtain total hydrogen column densities. It is common for observational studies to show fractional abundances over molecular hydrogen column densities, not total hydrogen column densities. 
On the other hand, chemical models often use fractional abundances over total hydrogen column densities. We show fractional abundances of N(\hocp)/N(H$_{total}$) in Table \ref{tab:obs_frac_ratio}.
 
 Errors associated with quantities shown in Table \ref{tab:obs_frac_ratio} are as follows.
For column densities of \hcop, \httcop, and \hocp, errors from spectral fitting were added
to an observational error of 15\% (Mart\'in et al., submitted). 
We used the value of isotopic ratio with a large error $^{12}$C/$^{13}$C $=20\pm 10$ to 
account for the discrepancies between the observed ratios \citep{2019A&A...629A...6T}, 
although we favor values around 20 from \hcop/\httcop~in optically thin regions.
For the error of log(N(HOC$^+$)/N(H$_{total}$)),  we used a conservative error log(2.)=0.30
because the total column densities from dust can easily change with different values of the dust emissivity power law $\beta$ or the dust temperature $T_{d}$,
and because there are some discrepancies amongst total column densities derived from dust and C$^{18}$O 
as discussed in Appendix \ref{sec:app_coltot}. %(see also Table \ref{tab:tot_columns}).

%%%%%%%%%%%%%%%%%%%%%
\begin{deluxetable}{cccc} 
\tablecolumns{4} 
\tablewidth{0pc} 
\tabletypesize{\scriptsize} 
\tablecaption{Observed \hocp~fractional abundances and \hcop/\hocp~ratios} \label{tab:obs_frac_ratio} 
\tablehead{\colhead{Region}  &\colhead{log(N(HOC$^+$)/N$_{tot}$)} &\colhead{log(N(HCO$^+$)/N(HOC$^+$))} &\colhead{log(($^{12}$C/$^{13}$C)N(H$^{13}$CO$^+$)/N(HOC$^+$))} \\
\colhead{} &\colhead{} &\colhead{Method 1} &\colhead{Method 2}}
\startdata 
M2  & $-9.76\pm 0.30$  &$1.07\pm 0.11$  &$1.19 \pm 0.24$\\ 
M3  & $-10.20\pm 0.30$  &$1.69\pm 0.15$  &$1.75 \pm 0.27$\\ 
M4  & $-10.53\pm 0.30$  &$1.98\pm 0.09$  &$1.80 \pm 0.23$\\ 
M5  & $-10.84\pm 0.30$  &$2.01\pm 0.13$  &$1.97 \pm 0.23$\\ 
M7  & $-10.75\pm 0.30$  &$1.78\pm 0.07$  &$2.00 \pm 0.23$\\ 
M8  & $-10.27\pm 0.30$  &$1.80\pm 0.10$  &$1.77 \pm 0.23$\\ 
M9  & $-10.31\pm 0.30$  &$2.03\pm 0.08$  &$1.83 \pm 0.23$\\ 
M10  & $-10.31\pm 0.30$  &$1.97\pm 0.08$  &$1.81 \pm 0.23$\\ 
A1  & $-10.22\pm 0.30$  &$1.22\pm 0.08$  &$1.10 \pm 0.23$\\ 
A2  & $-10.02\pm 0.30$  &$1.64\pm 0.08$  &$1.55 \pm 0.23$\\ 
A3  & $-10.69\pm 0.30$  &$1.72\pm 0.08$  &$1.78 \pm 0.23$\\ 
A4  & $-10.04\pm 0.30$  &$1.64\pm 0.07$  &$1.58 \pm 0.23$\\ 
A5  & $-10.68\pm 0.30$  &$1.69\pm 0.15$  &$1.80 \pm 0.25$\\ 
A6  & $-10.53\pm 0.30$  &$1.79\pm 0.11$  &$1.64 \pm 0.23$\\ 
A7  & $-10.25\pm 0.30$  &$1.67\pm 0.12$  &$1.64 \pm 0.23$\\ 
A8  & $-10.09\pm 0.30$  &$1.73\pm 0.09$  &$1.65 \pm 0.24$\\ 
\enddata 
\tablecomments{Both \hcop/\hocp,  \hocp~fractional abundances and their errors are shown on a logarithmic scale. Observational errors of 15 \% is taken into account. } 
\end{deluxetable} 

%%%%%%%%%%%%%%%%%%%%%

\section{Chemical Modeling}\label{sec:model}

We conducted modeling of chemical abundances in order to
interpret our observational results. In particular, we
examine the effect of the cosmic-ray ionization rate and the UV photons on molecular abundances.
A modified version of the time-dependent, gas-grain code Nautilus \citep{2009A&A...493L..49H} was used to model the fractional abundances of the species of interest.
This chemical network includes about 500 gas-phase species, and about 200 of them are also considered on the grain surface. The total number of gas-phase, accretion to or desorption from grains, and grain-surface reactions are about 8800.
The basic network was taken from \citet{2015MNRAS.447.4004R}, but excluded the complexes formed through Eley-Rideal reactions on grain surfaces
and other species that are newly added in their work because those species are not relevant 
to our work and because our focus in this paper is not complex molecules formed through Eley-Rideal reactions. 
We ran the chemical model in a grid of physical conditions with densities of $n_H=10^3 - 10^6\,$cm$^{-3}$, where $n$ is the total hydrogen volume density, not $H_2$ density. 
For the gas and dust temperatures, we use values obtained from the Meudon PDR code\footnote{http://ism.obspm.fr} \citep{2006ApJS..164..506L},
which has a detailed calculation of thermal structure including various heating and cooling mechanisms and radiative transfer. Although the Meudon code also derives chemical abundances including HCO$^+$ and HOC$^+$, 
we used Nautilus for the calculation of chemical abundances
in order to examine the effect of grain-related reactions as well as 
time dependence. 
While this is not a truly consistent treatment because the chemistry 
is coupled with heating and cooling of the gas, the
approximate behavior of chemical compositions in PDRs and cosmic-ray dominated regions (CRDR) should be reproduced.
Overall, comparisons between results from the different chemical models
show that the main conclusion of this paper
does not change with different models.
Further details of the input parameters used for the chemical modeling are included in Appendix \ref{sec:app_chem_model}.

\subsection{Chemical Reactions Leading to \hcop~and \hocp}\label{sec:chem_route}
\begin{figure}[h]
\centering{
\includegraphics[width=0.9\textwidth]{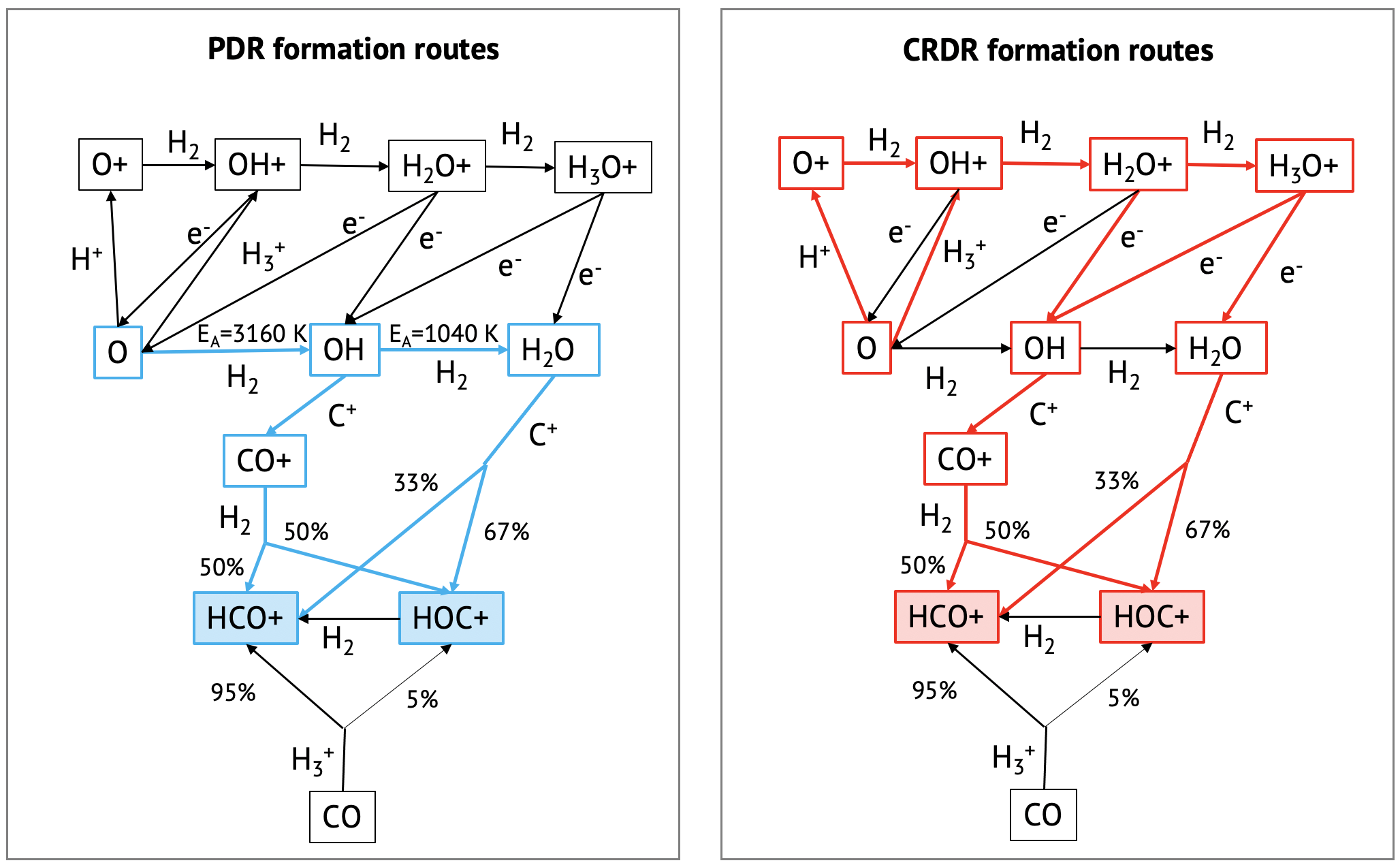}
\caption{Graphical presentation of formation routes of HCO$^+$ and HOC$^+$ in the cases of PDRs and CRDRs.
The blue boxes and arrows are species and reactions involved in the formation of HCO$^+$ and HOC$^+$ in PDRs, while red boxes and arrows are the ones in CRDRs. The activation barriers in the gas-phase formation of OH and water are 
indicated as $E_{\rm A}$. \label{fig:routes}}}
\end{figure}

In this section, we describe the chemical reactions related to the production and destruction
of HCO$^+$ and HOC$^+$. 
Reactions discussed in this section are also shown schematically in Figure \ref{fig:routes}.
Two main gas-phase chemical reactions leading to formation of HOC$^+$ discussed in 
\citet{2004AA...428..117L} are
\begin{eqnarray}\label{hocp_prod1}
{\rm C^+ + H_2O \longrightarrow HOC^+ + H} \\
{\rm \longrightarrow HCO^+ + H,} \nonumber
\end{eqnarray}

and 
\begin{eqnarray}\label{hocp_prod2}
{\rm CO^+ + H_2 \longrightarrow HOC^+ + H}\\
{\rm \longrightarrow HCO^+ + H}\nonumber.
\end{eqnarray}

As shown above, both of these reactions also produce HCO$^+$. The branching ratios assumed in our model to produce both HCO$^+$ and HOC$^+$ are HCO$^+$:HOC$^+$ = 33\%:67\% for the former reaction \citep{1986ApJ...303..392J}
 and HCO$^+$:HOC$^+$ = 50\%:50\%  for the latter reaction. Therefore, there is a relatively high abundance of HOC$^+$ when the region is influenced by UV-photons or cosmic rays to enhance fractional abundances of C$^+$ or CO$^+$.

We find that the following channels lead to the production of \hocp.
The high gas temperature ($T\gtrsim 300$ K) helps both of the reactions above because of the following reactions with barriers \begin{equation}\label{eq:hydro_o}
{\rm O + H_2 \longrightarrow OH + H}
\end{equation}
 with an activation barrier of 3160 K, and 
\begin{equation}\label{eq:hydro_oh}
{\rm OH + H_2 \longrightarrow H_2O + H}
\end{equation}
with an activation barrier of 1040 K, which produces OH and water efficiently. Thus, high water abundance makes  Reaction \ref{hocp_prod1} faster, while high OH abundance makes Reaction \ref{hocp_prod2} more efficient because CO$^+$ is dominantly produced via reaction 
\begin{equation}
{\rm C^+ + OH \longrightarrow CO^+ + H.} 
\end{equation}

On the other hand, the dominant production routes in CRDRs are as follows.
At lower temperatures ($T_{\rm gas} \lesssim 300\,$K), there are alternative formation routes for OH and H$_2$O. 
When the cosmic-ray ionization rates are high, an oxygen atom can be ionized by its reactions with H$^+$. 
The ionized oxygen can successively become hydrogenated from reactions with molecular hydrogen, 
\begin{equation} 
{\rm OH_n^+ + H_2 \longrightarrow OH_{n+1}^+ + H }
\end{equation}
for $n=0 - 2$, as described in \citet{2010A&A...521L..10N}. 
There is another formation route of OH$^+$ such as 
\begin{equation}
{\rm O + H_3^+ \longrightarrow OH^+ + H_2,}
\end{equation}
which also becomes efficient at the high cosmic-ray ionization rate due to an enhancement of H$_3^+$ \citep[e.g., ][]{2003Natur.422..500M}. 
The ions produced in reactions above, H$_2$O$^+$ and H$_3$O$^+$, 
can recombine with electrons to form water or OH.

Even without strong effects of UV-photons or cosmic rays (e.g., $A_V > 10$ or $\zeta < 10^{-17}$ s$^{-1}$), HOC$^+$ can still be produced by the reaction 
${\rm CO~+~H_3^+}$. However, this reaction produces HCO$^+$ with a much higher branching ratio than HOC$^+$, with only 5\% going to HOC$^+$ \citep{2009ApJS..185..273W}. In addition, high ratios of HCO$^+$/HOC$^+$ in quiescent regions (i.e., without strong UV radiation or cosmic-ray flux) are promoted by a reaction that directly converts HOC$^+$ to HCO$^+$:

\begin{equation}\label{eq:hocp_molh}
{\rm HOC^+ + H_2 \longrightarrow HCO^+ + H_2.}
\end{equation}
This reaction itself can occur in PDRs or CRDRs, but the \hocp~production rates in PDRs/CRDRs are
comparable to that of Reaction \ref{eq:hocp_molh}, which can maintain low \hcop/\hocp.

We employed the rate coefficient of $4\times 10^{-10}$ cm$^3$ s$^{-1}$ for this reaction in our model following the experiments by \citet{2002ApJ...578L..87S}. This rate is different from the one listed in KIDA 2014 \citep{2015ApJS..217...20W}, which is the rate estimated by \citet{1996ApJ...463L.113H,1996ApJ...471L..73H}
because we favor experimental rates over theoretical estimation.
For most other reaction rates, we use rates taken from KIDA 2014. 

From the formation path via Reaction \ref{hocp_prod2}, CO$^+$ is expected in both PDRs and CRDRs. In our observations, we do detect CO$^+$, but we do not discuss this species because there is severe blending from neighboring transitions, including unidentified lines.

%%%%%%%%%%%%%%%%%%%%%
%%%%%%%%%%%%%%%%%%%%%

\subsection{The effects of UV-photons}\label{sec:effecUV}
\begin{figure*}[h]
\centering{
\includegraphics[width=0.45\textwidth]{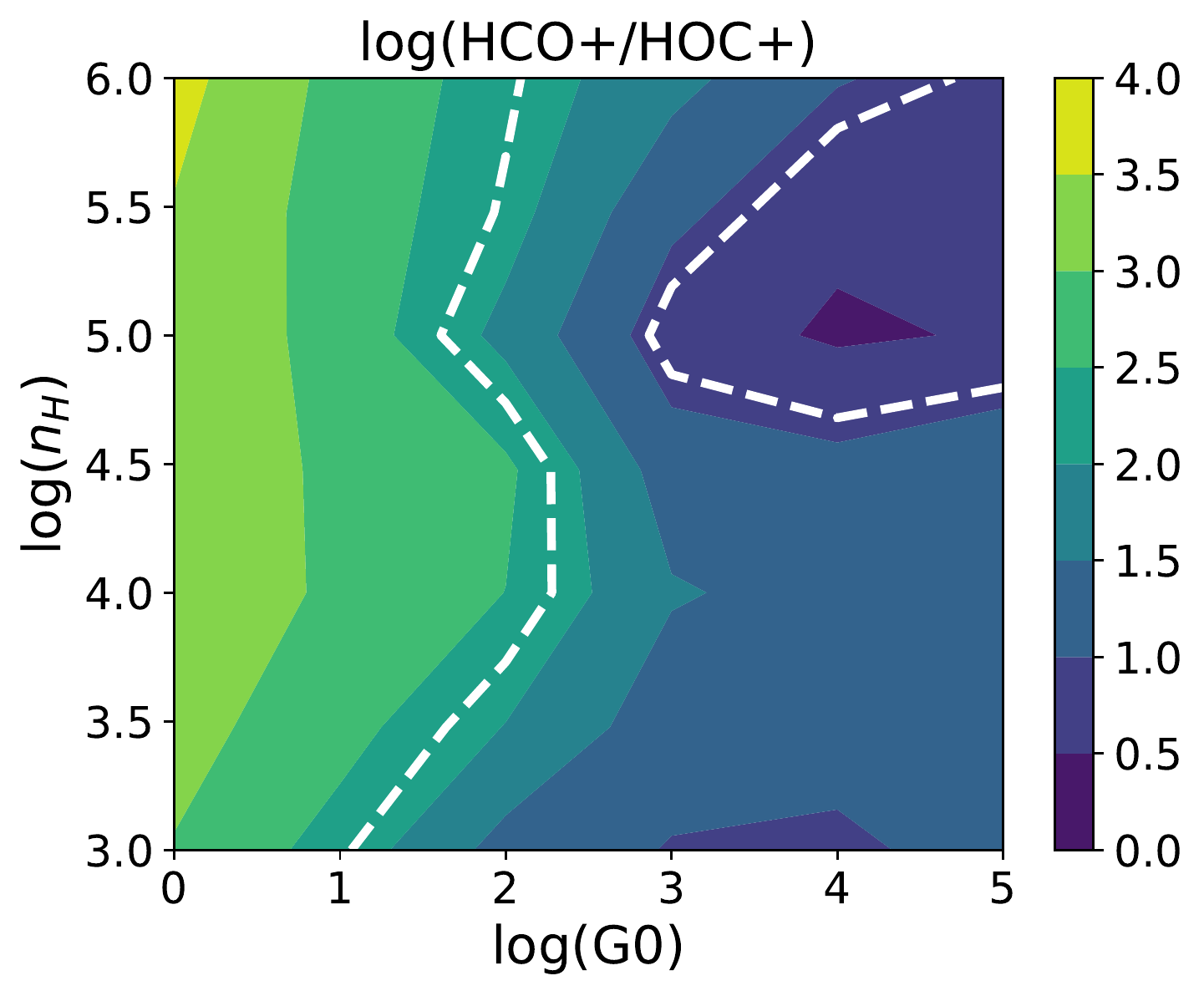}
\includegraphics[width=0.45\textwidth]{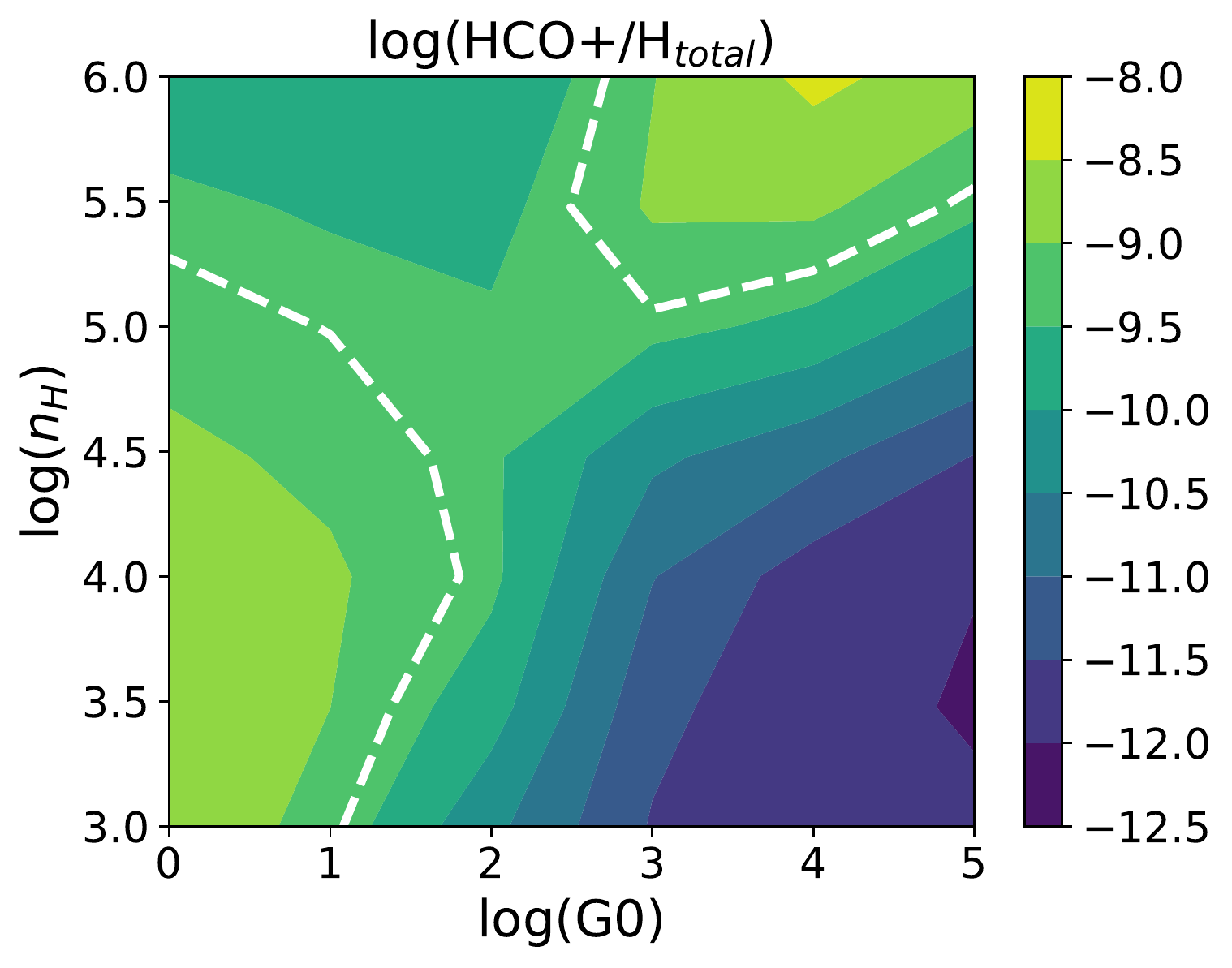}
\includegraphics[width=0.45\textwidth]{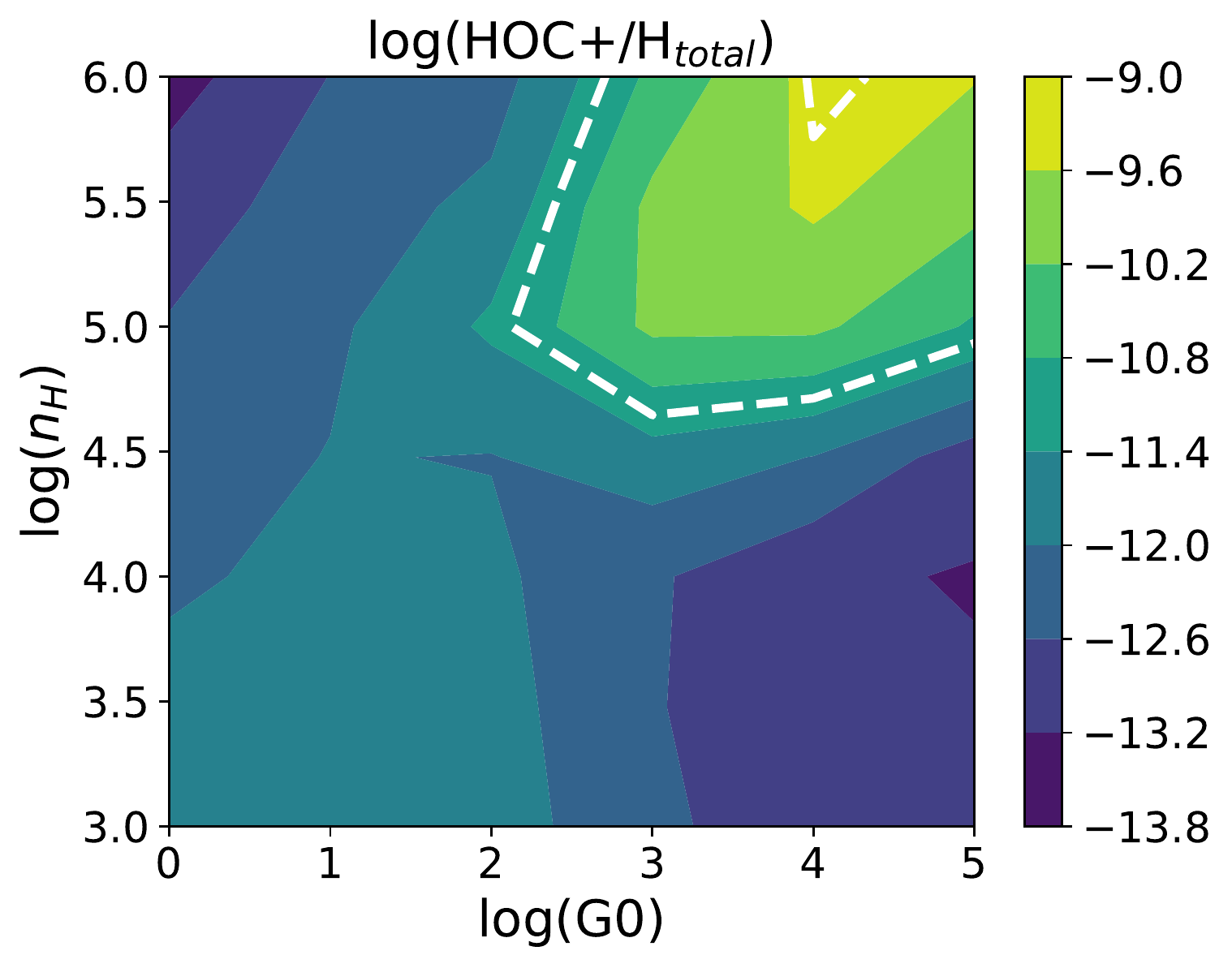}
}
\caption{The abundance ratios of HCO$^+$/HOC$^+$, fractional abundances of HCO$^+$ and HOC$^+$ are shown as functions of density $n_H$ and interstellar radiation field $G_0$. All of them are shown in the logarithmic scale. The abundances are averaged over $A_{\rm V}=$0--5 mag. For higher values of $A_{\rm V,max}$, \hcop/\hocp~will be higher and \hocp/H$_{total}$ will be lower. Observed ranges of log(\hcop/\hocp),  log(\hcop/H$_{total}$), and log(\hocp/H$_{total}$)  among all the analyzed positions are shown with white dashed lines. The observed \hcop/\hocp~range shown here is a union of ranges obtained from both Methods 1 and 2 (See Section \ref{sec:col_hcop_hocp} for the description of two methods).
\label{fig:pdr_model}}
\end{figure*}

We examine the effects of UV-photons by varying the interstellar radiation field $G_0=1-10^5$ 
in the Habing field (Interstellar radiation field, \enum{1.6}{-3} erg\,\psc\,\ps). 
First, we evaluate the average line-of-sight abundances by integrating fractional abundances over visual extinction $A_{\rm V}$ in the range of 0-5 mag.
This maximum visual extinction is the value suggested from \citet{2008AA...492..675F} in the case 
of the galactic center of the starburst galaxy M82,
which has been suggested to act as a giant PDR.
The actual maximum $A_{\rm V}$ may be different from this value, but an increase in the $A_{\rm V}$
will always lead to higher HCO$^+$/HOC$^+$ ratios because an increase in HOC$^+$ abundance only occurs at low $A_{\rm V} < 5$,
where the effects of UV-photons are more effective.
A fiducial value of the cosmic-ray ionization rate $\zeta = 10^{-17}$\,s$^{-1}$ was used for models presented in this section.
Abundance ratios of HCO$^+$/HOC$^+$ are shown as a function of $G_0$ and $n$ in Figure \ref{fig:pdr_model} (top left). 
These ratios decrease with increasing $G_0$, and are particularly low at the high density ($n_H\gtrsim 10^5$ cm$^{-3}$), in high $G_0$ regions. 
The fractional abundance of HCO$^+$ remains relatively high ($\gtrsim 10^{-10}$) for
most values of the density and interstellar radiation field, except for extremely high $G_0$ and low $n_H$ environments.
On the other hand, HOC$^+$ has 
fractional abundances higher than $\sim 10^{-10}$ only at high density ($n_H \gtrsim 10^{5}$ cm$^{-3}$) and 
high interstellar radiation field ($G_0\gtrsim 10^3$).

%%%%%%%%%%%%%%%%%%%%%
%%%%%%%%%%%%%%%%%%%%%

\begin{figure*}[h]
\centering{
\includegraphics[width=0.45\textwidth]{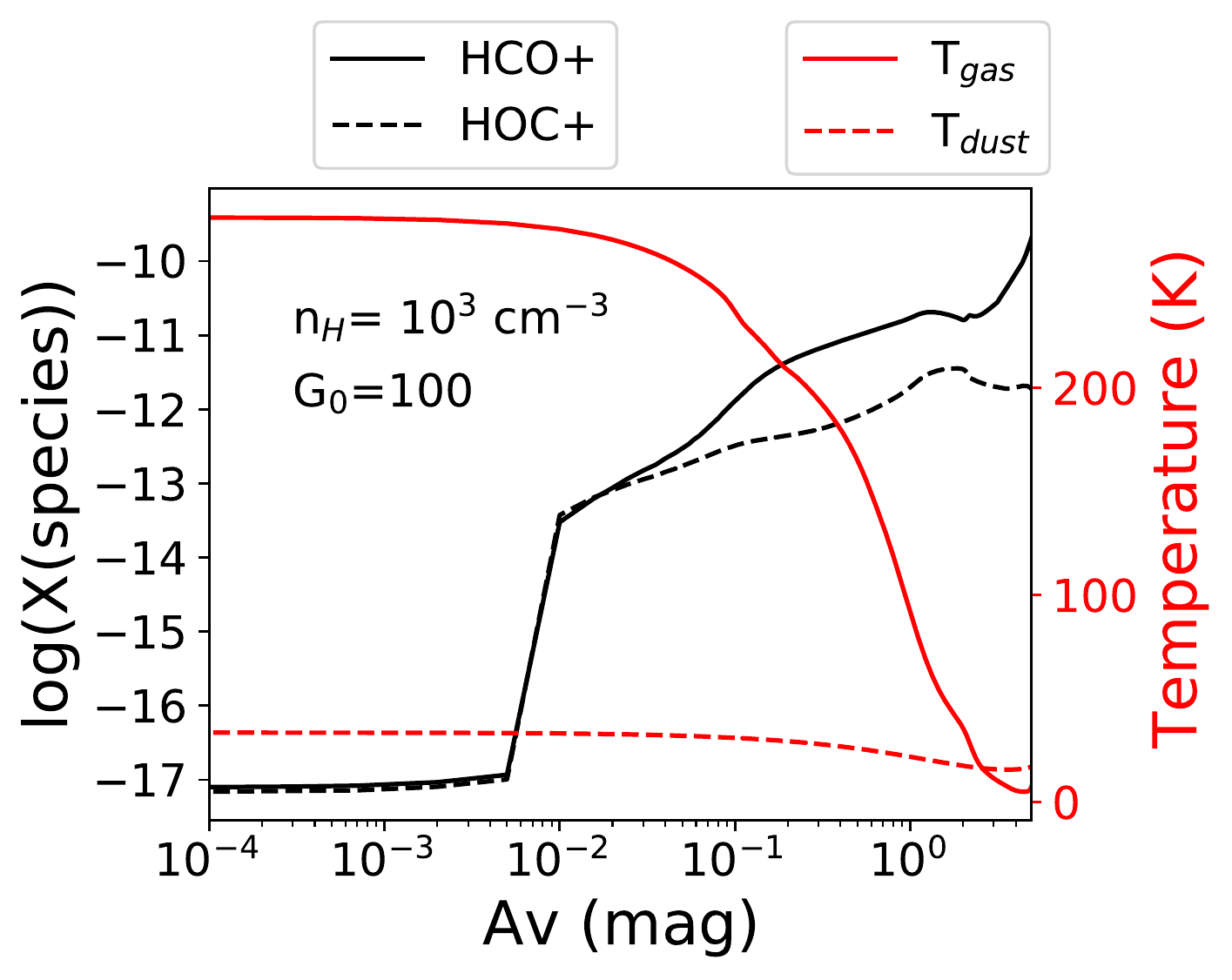}
\includegraphics[width=0.45\textwidth]{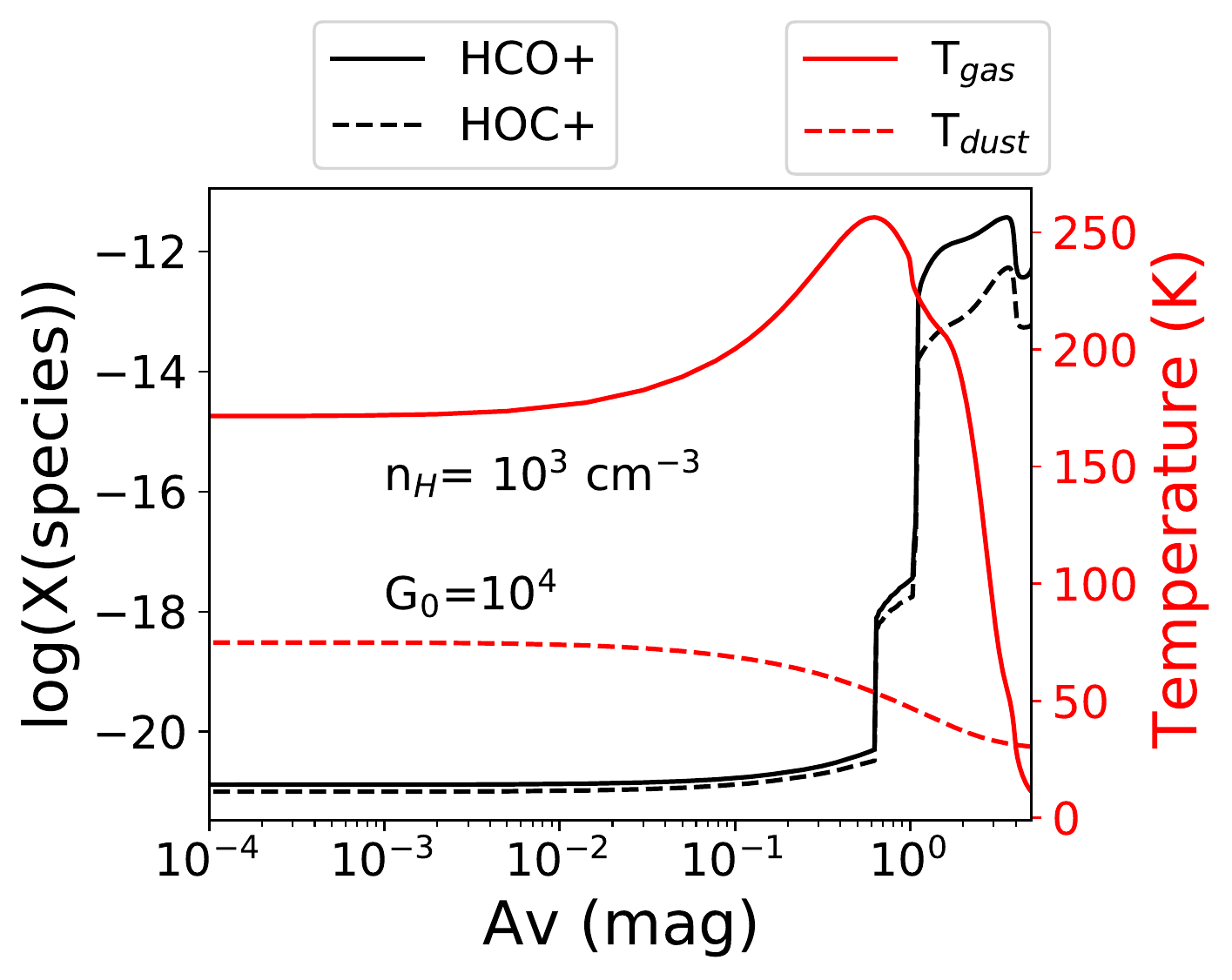}
}
\centering{
\includegraphics[width=0.45\textwidth]{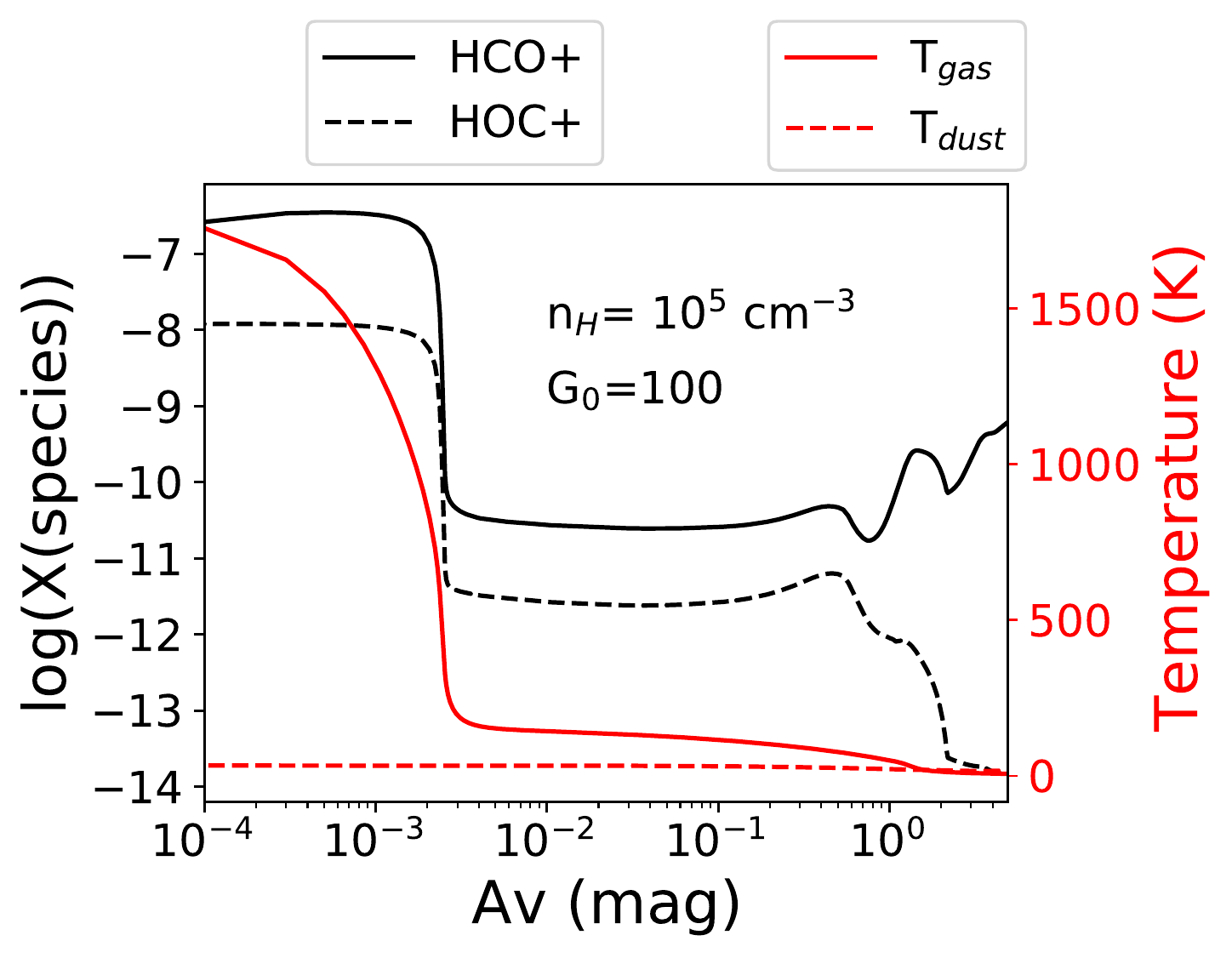}
\includegraphics[width=0.45\textwidth]{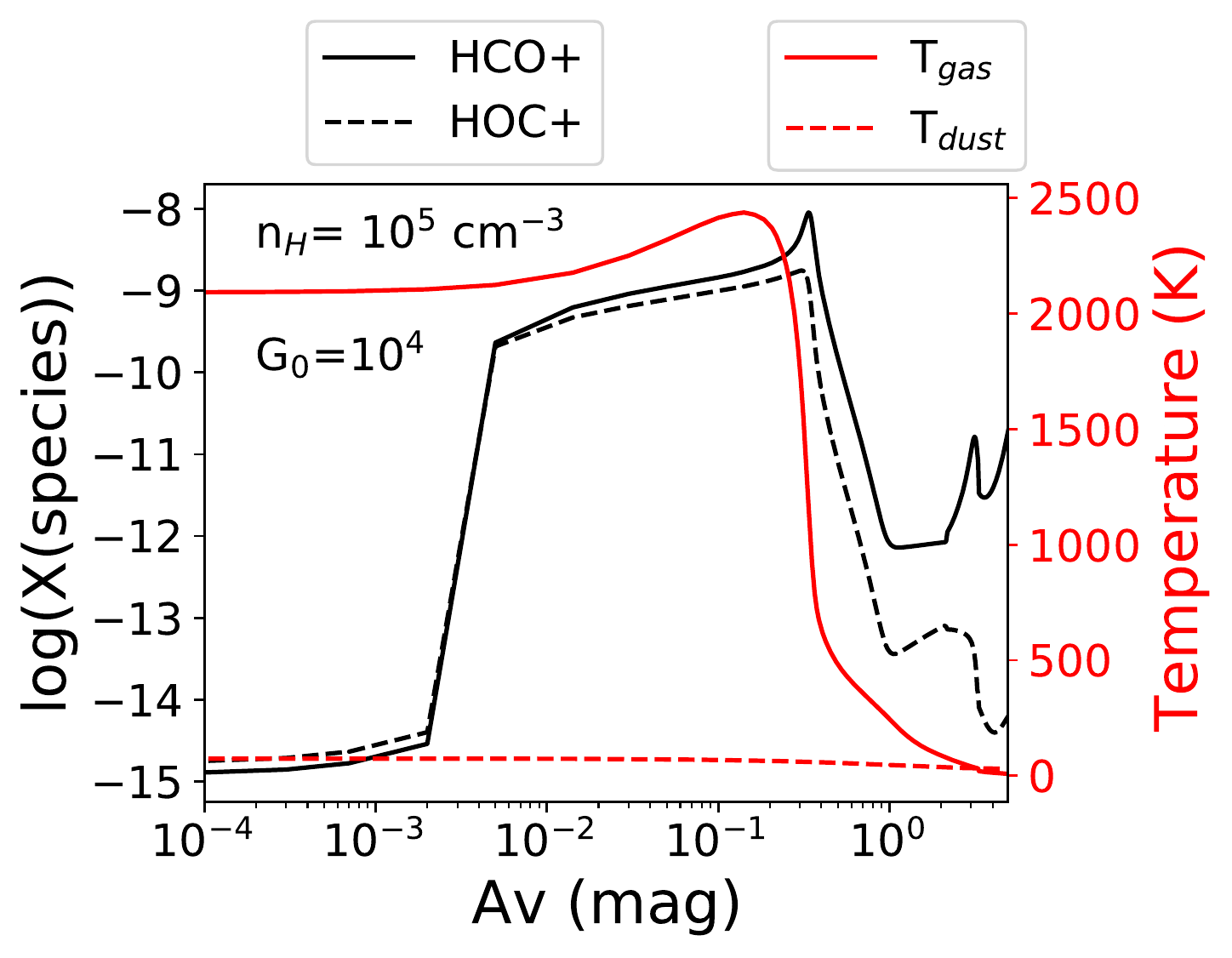}
}
\caption{The fractional abundances of HCO$^+$ (black solid) and HOC$^+$ (black dotted) and gas temperature (red solid) and dust temperature (red dotted) are shown. The density and interstellar radiation field used for the model is $n_H=10^3$ cm$^{-3}$ $G_0=10^2$ (top left), $n_H=10^3$ cm$^{-3}$ $G_0=10^4$ (top right), $n_H=10^5$ cm$^{-3}$ $G_0=10^2$ (bottom left), and $n_H=10^5$ cm$^{-3}$ $G_0=10^4$ (bottom right).\label{fig:pdr_model_Av}}
\end{figure*}

The $A_{\rm V}$ dependence of the gas temperature, dust temperature, and fractional abundances of 
HCO$^+$ and HOC$^+$ are shown in Figure \ref{fig:pdr_model_Av}. 
In the low-density case shown in the top panels, the fractional abundance of HOC$^+$ is less than
10$^{-11}$, failing to produce the observed values.  
On the other hand, higher density models (bottom panels) have high peak fractional abundances
at certain $A_{\rm V}$. For the case of $G_0=10^2$, the range of $A_{\rm V}$ where HOC$^+$ is enhanced is narrow,
and the integrated fractional abundance ratio of HCO$^+$/HOC$^+$ is rather high as shown in Figure \ref{fig:pdr_model}. Meanwhile, the case of higher $G_0 = 10^4$ shows \hocp~enhancement at a wide enough range of $A_{\rm V}$, \enum{3}{-3} to 0.3 
to have a significant effect on the HCO$^+$/HOC$^+$ ratio even after integrating to $A_V = 5$ mag.
Note that the temperature drops from $\sim 2000$ K to $< 1000$ K at $A_{\rm V} \sim 0.03$,
where the \hocp~enhancement drops.
The high fractional abundance of HOC$^+$ in high-density cases is attributed to the endothermic reactions 
that produce water and OH, precursors of HOC$^+$, efficiently at high gas temperatures as discussed in Section \ref{sec:chem_route}.

%%%%%%%%%%%%%%%%%%%%%
%%%%%%%%%%%%%%%%%%%%%

Our result shows that it is possible to produce relatively high abundances (10$^{-10}$) of HOC$^+$ with
a high enough density and interstellar radiation field. This result is consistent with the work by \citet{2008AA...492..675F}, which claimed that the source of enhancement of HOC$^+$ is by UV photons in dense gas. 
Meanwhile, this appears somewhat inconsistent with the result by \citet{2007ApJ...664L..23S}, who
concluded that PDRs cannot produce observable column densities of CO$^+$ and HOC$^+$. 
\citet{2007A&A...461..793M}, upon which the result of \citet{2007ApJ...664L..23S} is based, ran the calculation of
a grid of densities and interstellar radiation field values. In all of their PDR models, the HCO$^+$/HOC$^+$ 
ratios are $> 10^4$, failing to produce a high enough enhancement of HOC$^+$ in comparison with
the HCO$^+$. However, these ratios are taken from column densities integrated to a certain
maximum visual extinction ($A_V =160$), defined by their ``standard'' cloud described in \citet{2007A&A...461..793M}.
 In fact, in these models by \citet{2007A&A...461..793M}, there is a certain range of $A_{\rm V}$ where the HCO$^+$/HOC$^+$ ratio is low ($\sim 10$) in particular for high-density,
 and high-$G_0$ cases.

\begin{figure*}[h]
\centering{
\includegraphics[width=0.45\textwidth]{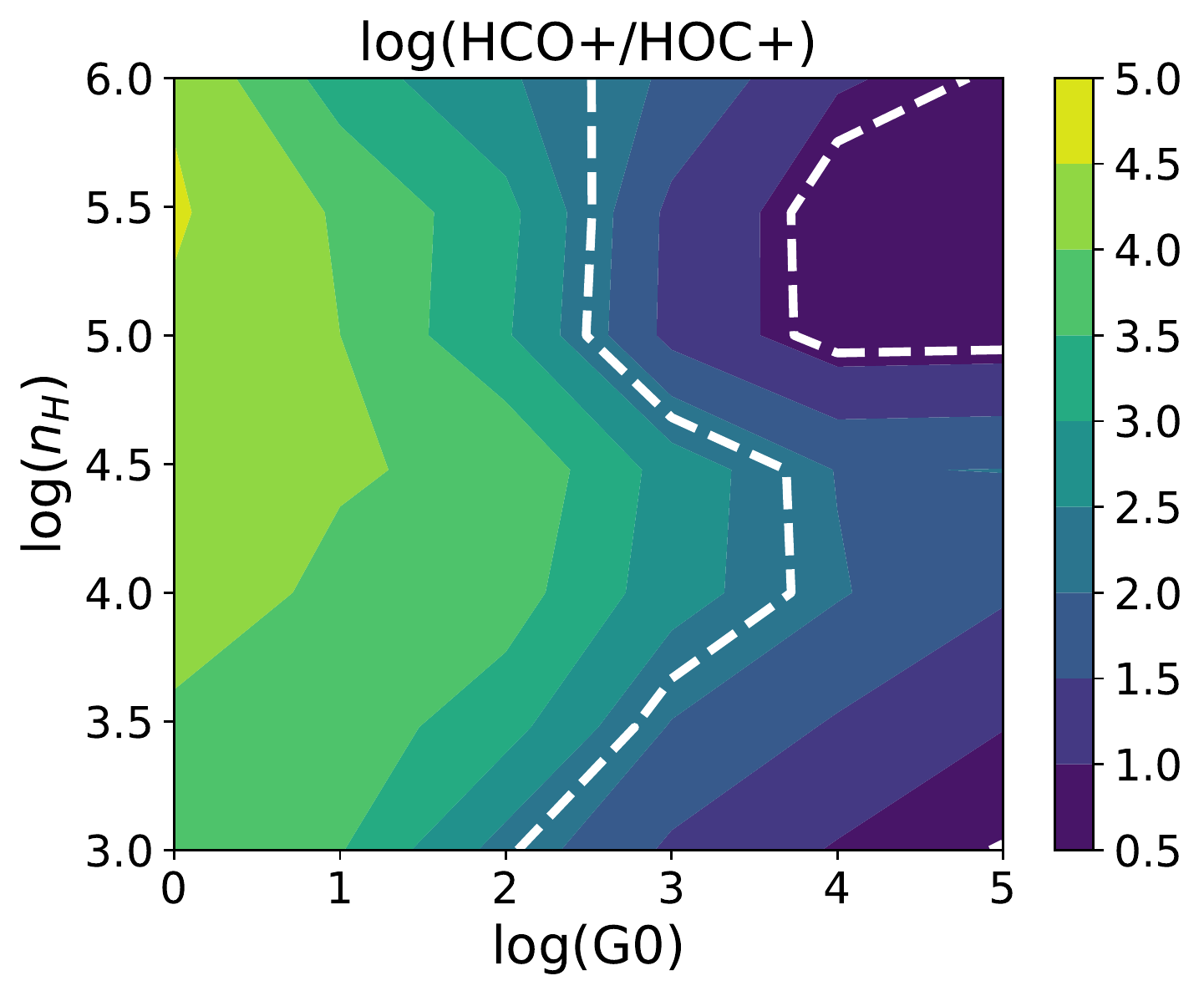}
\includegraphics[width=0.45\textwidth]{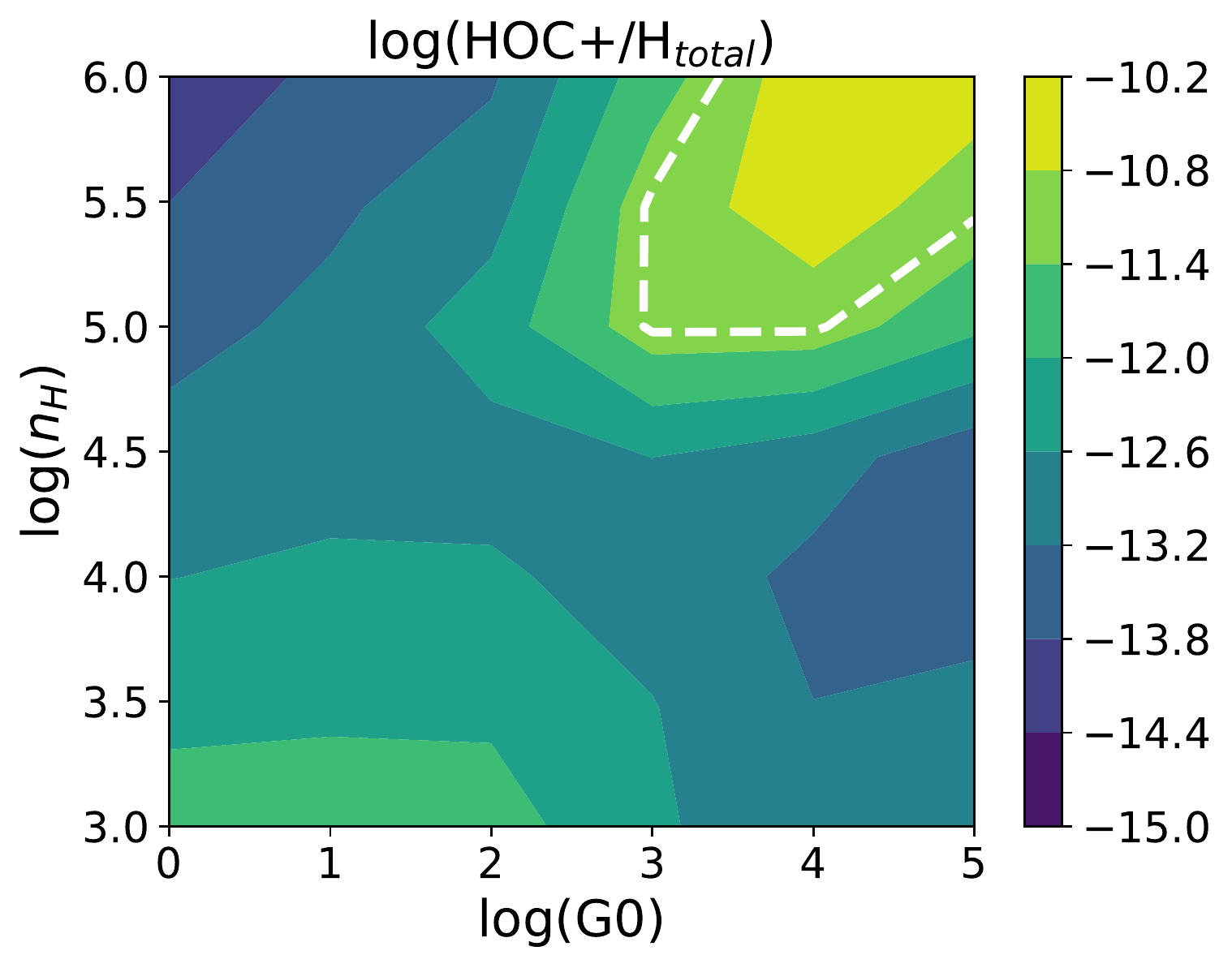}
}
\centering{
\includegraphics[width=0.45\textwidth]{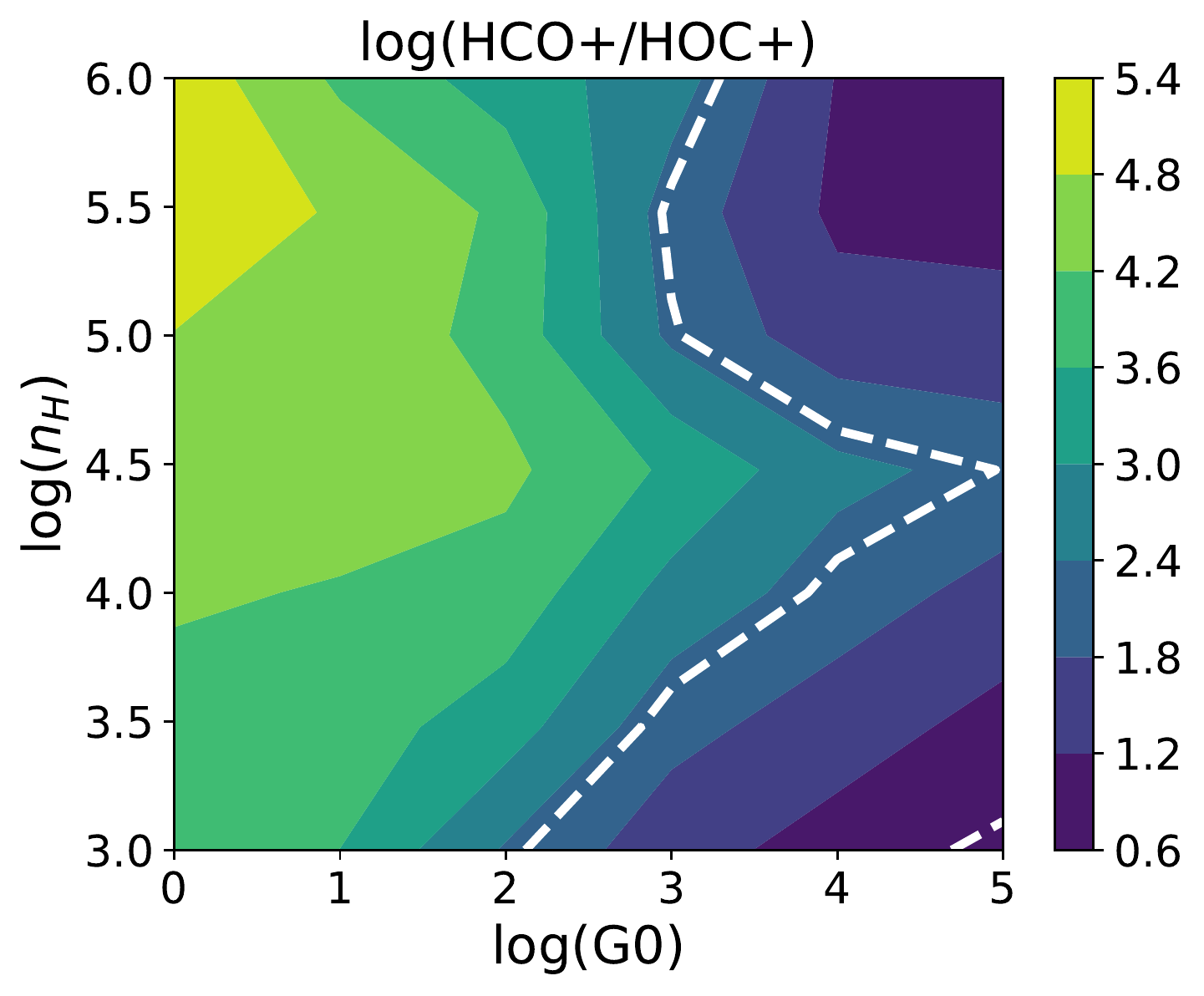}
\includegraphics[width=0.45\textwidth]{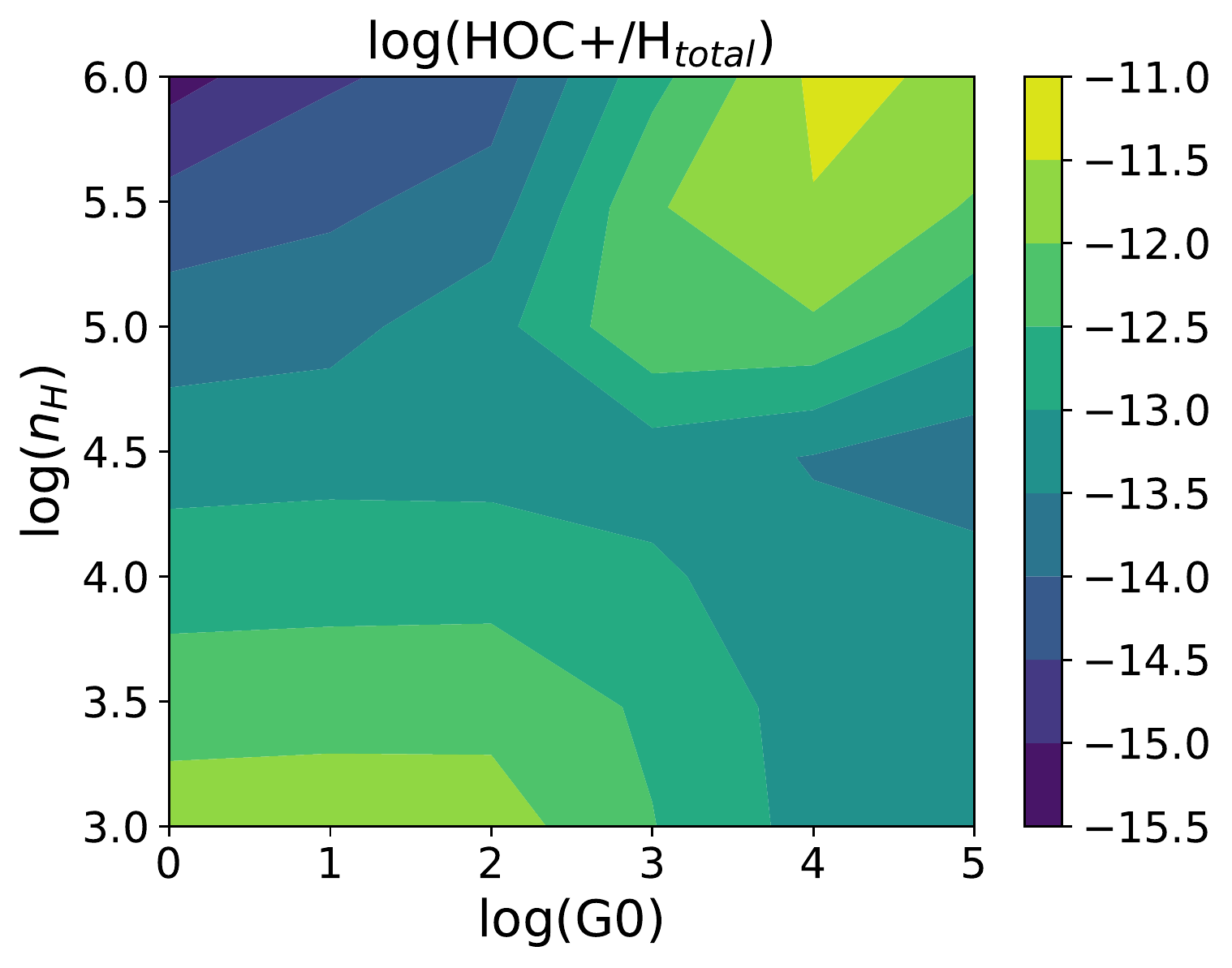}
}
\caption{Same as \ref{fig:pdr_model}, but for $A_{\rm V,max} =50$ (top) and $500$ (bottom). In the right bottom figure, white dashed lines are not shown because the observed value is out of range. \label{fig:pdr_model_Avmax}}
\end{figure*}

The fractional abundance of \hocp and \hcop/\hocp~ratio depend on the maximum visual extinction $A_{\rm V,max}$.
In chemical models shown above, we only integrate $A_{\rm V}=0-5$. 
Multi-band infrared observations with Very Large Telescope (VLT) by \citet{2009MNRAS.392L..16F} show that the values of visual extinction are $\lesssim 10$,
which is similar to the maximum $A_{\rm V}$ used in our models. On the other hand, \citet{leroy_forming_2018} and \citet{mangum_fire_2019} have revealed that some clumps have $N({\rm H_2}) \sim 10^{24}$\,\psc~($A_{\rm V} \sim 1000$,  assuming $N(H_{total})/A_V = 1.8 \times 10^{21}\,$ cm$^{-2}$), which is also consistent with total column densities we obtained for these locations (see Appendix \ref{sec:app_coltot}). This visual extinction suggests that the maximum values of $A_{\rm V}$ should be $\sim 500$ assuming that the star clusters at the centers of GMCs.
Meanwhile, this value of visual extinction may not be used to measure the effect of UV photons.
The visual extinction that affects the chemistry is the effective visual extinction, which considers the effect of UV-photons averaged in all directions \citep[e.g., ][]{2010MNRAS.404....2G}.
While the effective visual extinction may be lower than the actual visual extinction in a clumpy medium, it is likely that visual extinction in some regions of the CMZ in NGC\,253 may be much higher than 5. 
Therefore, the \hcop/\hocp~ratios for each $n$ and $G_0$ we obtain here with $A_{\rm V}=5$ are likely lower than actual values
if $A_{\rm V}> 5$.
In order to give a rough estimate of the effect of higher maximum $A_{\rm V}$, we calculated the average fractional abundance over $A_{\rm V} = 0-A_{\rm V,max}$, $\bar{X}$($A_{\rm V,max}$) via 
\begin{equation}\label{eq:av_approx}
\bar{X}(A_{\rm V,max}) \sim \left(\int_0^{5}X(A_{\rm V})d(A_{\rm V}) + X(A_{\rm V}=5)\times (A_{\rm V,max} - 5) \right)/A_{\rm V,max}   
\end{equation}
where X($A_{\rm V}$) is a fractional abundance as a function of $A_{\rm V}$. This estimate is valid under the assumption that effects of UV photons are negligible for $A_{\rm V} > 5$.  
When $A_{\rm V,max}=500$, the observed fractional abundance of \hocp~cannot be reproduced even with $G_0 =10^5$, and \hocp~enhancement due to PDRs can be excluded unless $G_0 > 5$ (Figure \ref{fig:pdr_model_Avmax}). 
For a lower $A_{\rm V,max}=50$, observations can be reproduced for a higher radiation field of $G_0= 10^3 - 10^4$, thus yielding higher values than for $A_{\rm V,max}=5$.

\begin{figure*}[h]
\centering{
\includegraphics[width=0.45\textwidth]{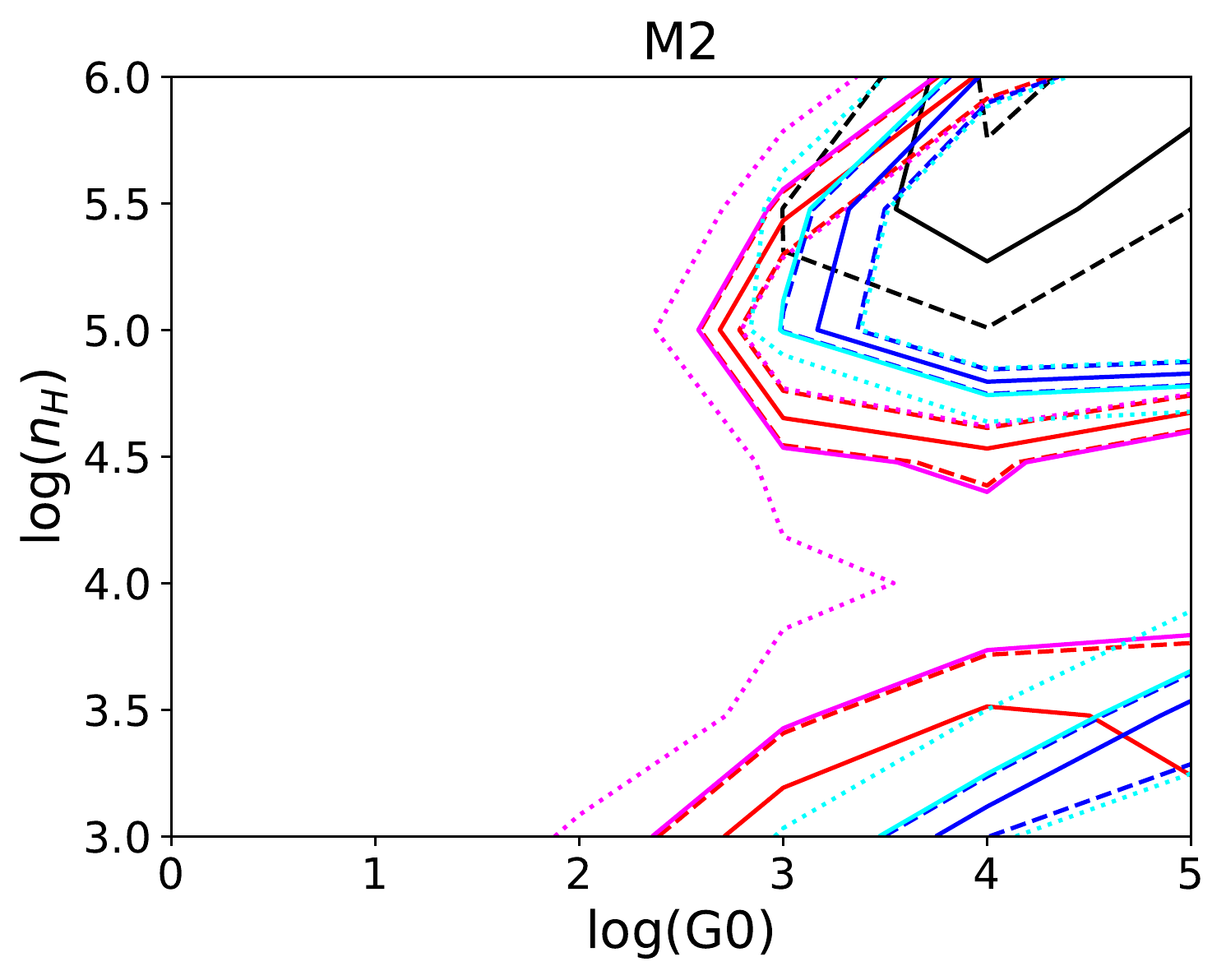}
\includegraphics[width=0.45\textwidth]{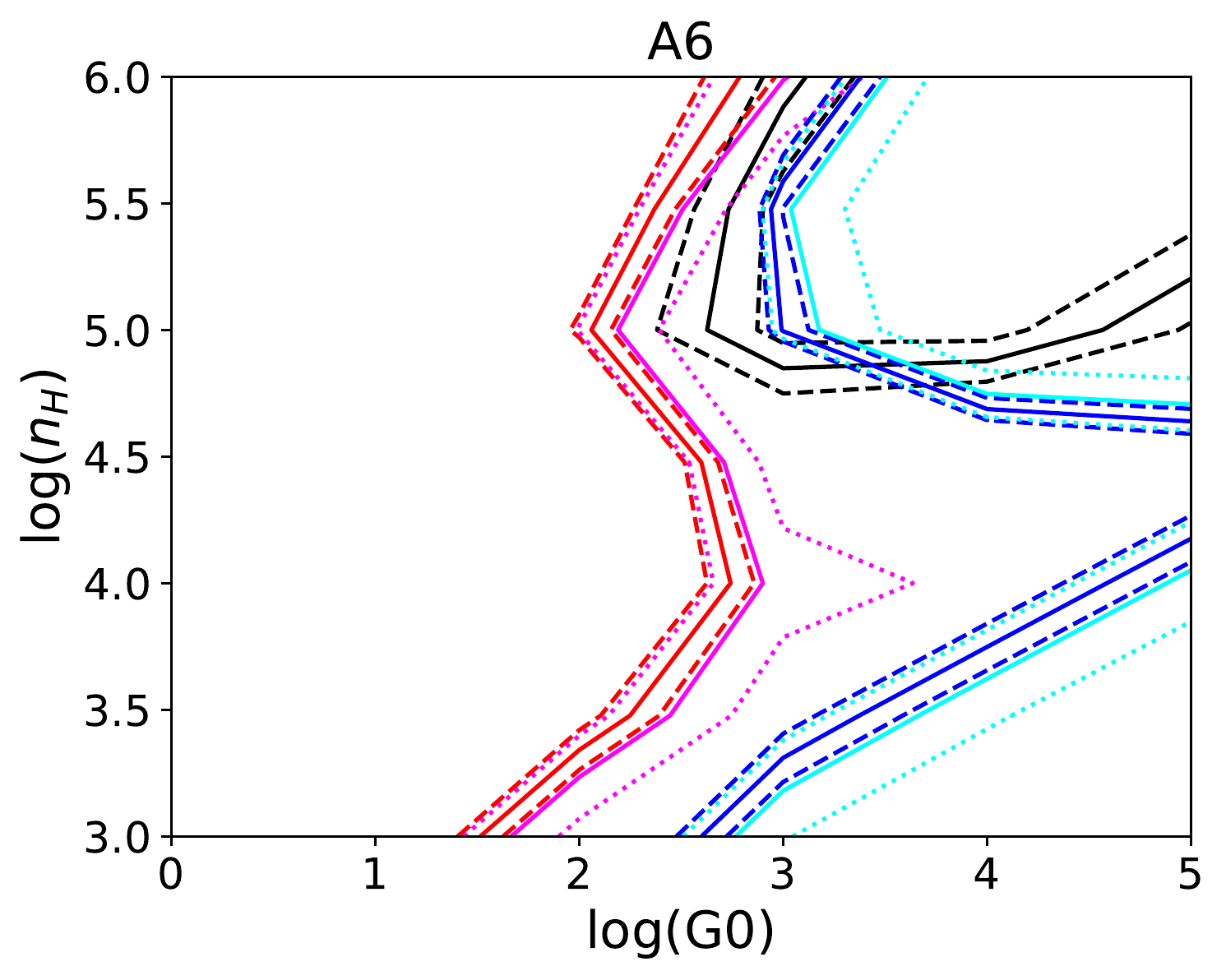}}
\centering{
\includegraphics[width=0.45\textwidth]{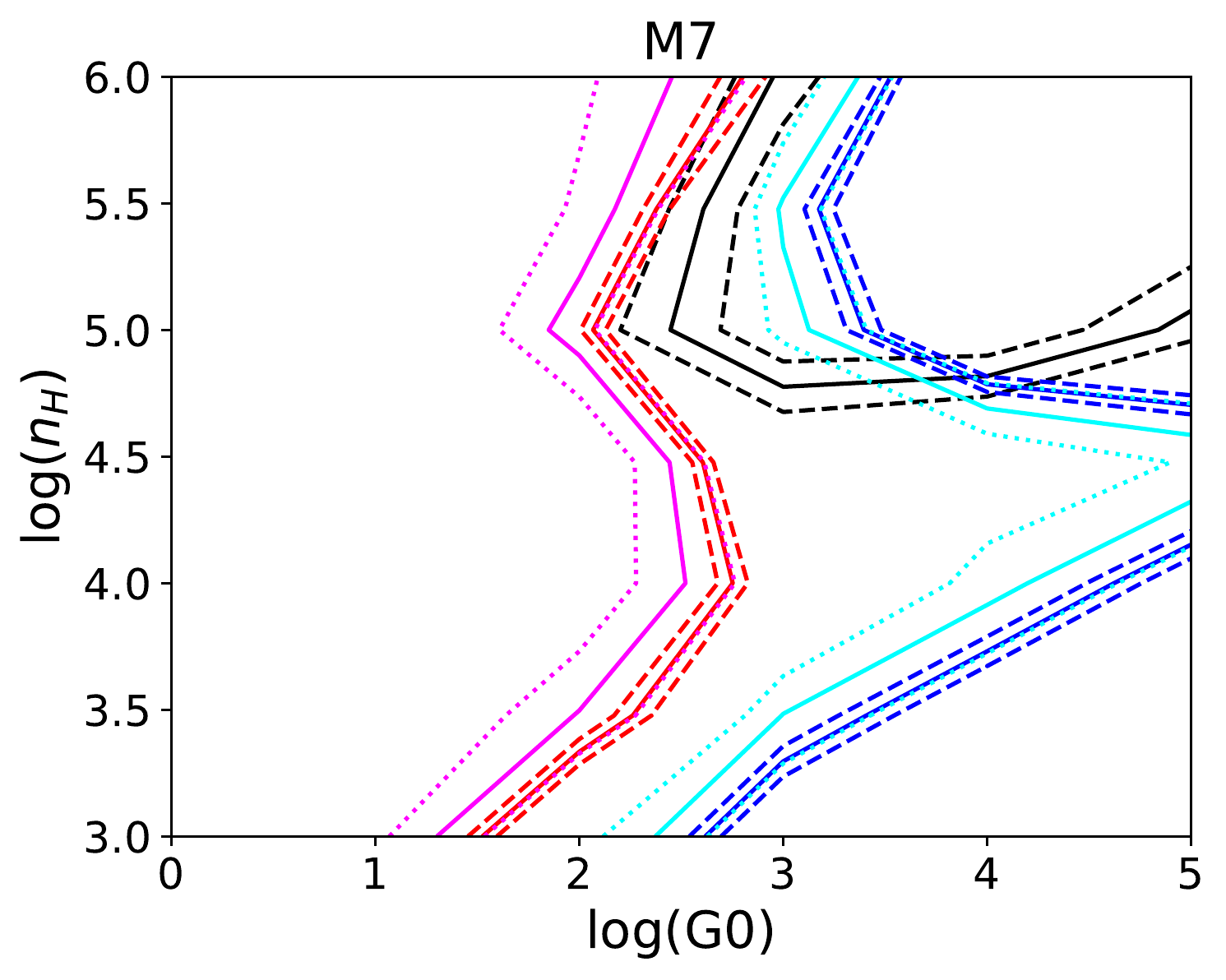}
\includegraphics[width=0.45\textwidth]{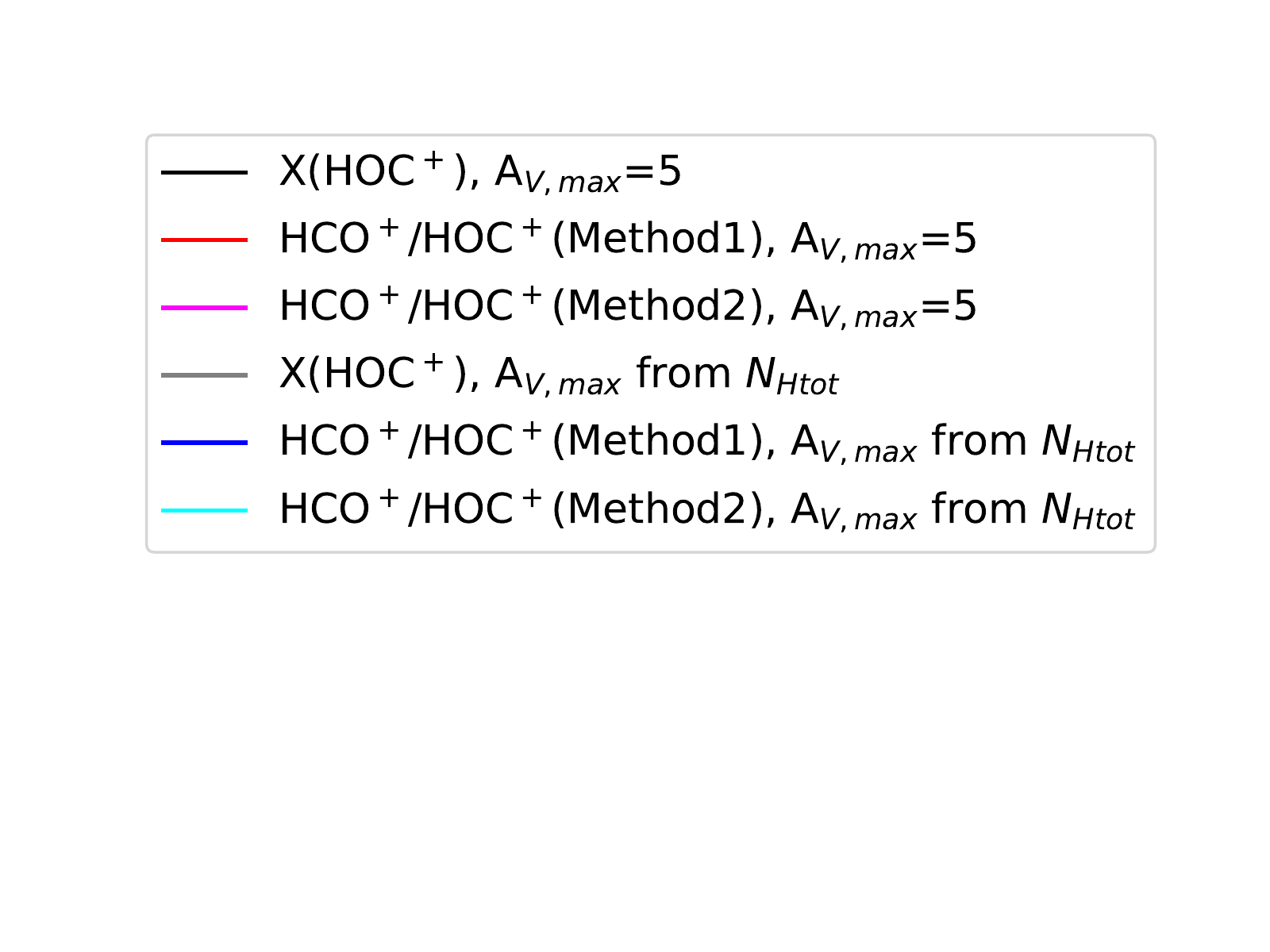}
}
\caption{X(HOC$^+$) (black) and HCO$^+$/HOC$^+$ with method 1(red) and method 2 (magenta) show parameter regions ($G_0$ and $n$) where the observed values towards M2 (top left), A6 (top right), and M7 (bottom left) match values in the PDR models for $A_{V,max}=5$.  Solid lines show the observed values while dashed lines (for Method 1) and dotted lines (for Method 2) show the range within errors.
Results at positions A3, M4, M5, M6, M9, and M10 resemble that of M7.
The same quantities are shown for X(HOC$^+$) (grey) and HCO$^+$/HOC$^+$ with method 1(blue) and method 2 (cyan) when $A_{V,max}$ is taken from the total column density obtained in Appendix \ref{sec:app_coltot}. Note that none of these positions matches the fractional abundances of \hocp~with this assumption of visual extinction, 
and grey lines do not appear in plots.
%The modeled values are for $A_{\rm V,max}=5$, and results at higher maximum visual extinction have better agreement 
%with observations with higher $G_0$.
\label{fig:pdr_fit_Av}}
\end{figure*}

In order to compare the results from chemical modeling and 
observations, we plotted observed ranges of 
log$_{10}(N({\rm HCO}^+)/N(\rm{HOC}^+))$ and log$_{10}(N(\rm{HOC}^+)/N(\rm{H}_{total}))$
against the modeled values as a function of $G_0$ and $n$ (Fig. \ref{fig:pdr_fit_Av}).
Because we use two methods to derive HCO$^+$/HOC$^+$ as explained in Section \ref{sec:col_hcop_hocp},
we compare values derived by both methods with the models. We also examine two different assumptions for $A_{\rm V,max}$. One is simply $A_{\rm V,max}=5$. The other assumption of $A_{\rm V,max}$ is obtained by the half of total column densities as described in Appendix \ref{sec:app_coltot}, and models are approximated with Equation \ref{eq:av_approx}. This factor of 2 comes assuming the GMCs are uniform and star clusters are located at the center. For the positions shown in Figure \ref{fig:pdr_fit_Av}, values of $A_{\rm V, max}$ through this assumption are 30, 150, and 330, for positions M2, A6, and M7 respectively. The assumption of $A_{\rm V,max}=5$ is likely a lower limit of the actual value, while the $A_{\rm V,max}$ from the total column density is an upper limit due to the clumpiness and non-uniform distribution of star clusters within our beam.

In Figure \ref{fig:pdr_fit_Av}, the parameter space where agreement between the observation
and the models is achieved can be found from the ranges of $G_0$ and $n_H$ that reproduce observations
of $X$(HOC$^+$) and HCO$^+$/HOC$^+$ overlap.
First, we discuss the case where $A_{\rm V,max}=5$.
A good agreement is achieved for $G_0\sim 10^{3.5}$, and $n_H \sim 10^5 - 10^6$ cm$^{-3}$
for the position M2 (Figure \ref{fig:pdr_fit_Av} top left).  The position A1 leads to a similar result. 
The position A6 has the best agreement at a slightly lower radiation field of $G_0\sim 10^{3}$,
which is similar to other positions A2, A4, A5, A7, A8, M3, and M8.
Although the observed ranges do not overlap for the position M7, 
if we allow larger errors considering that the chemical models also have uncertainty in rates, 
the best agreement should be met around $G_0\sim 10^{2.5}$ and $n_H\sim 10^5 - 10^6$ cm$^{-3}$ . 
Positions A3, M4, M5, M6, M9, and M10 have similar results as M7.
If $A_{V,max}$ from total column densities is used, the observed \hcop/\hocp~ratio can be reproduced with higher $G_0$, but there is difficulty reproducing sufficiently high \hocp~fractional abundances.
Therefore, if the effective visual extinction is equivalent to that obtained from total
column densities, PDRs can be ruled out.
Many molecular clumps with names ``M" \citep[from ][]{2015ApJ...801...63M} have the best-fit models with low $G_0$. The low $G_0$ implied from chemical models in these clumps may be explained by a large visual extinction. 
%%%%%%%%%%%%%%%%%%%%%
%%%%%%%%%%%%%%%%%%%%%

%%%%%%%%%%%%%%%%%%%%%
%%%%%%%%%%%%%%%%%%%%%

\subsection{The Effects of Cosmic Rays}\label{sec:model_cr}
In addition to UV photons, cosmic rays can enhance the abundance of HOC$^+$,
as already reported by \citet{2011MNRAS.414.1583B} and \citet{2018ApJ...868...40A}.
This enhancement is because cosmic rays increase the abundances of C$^+$ and CO$^+$, which leads to faster production of HOC$^+$ 
through Reactions \ref{hocp_prod1} and \ref{hocp_prod2}. The chemistry of CRDRs is known to be similar to that in XDRs although the heating in XDRs
is more efficient than in the CRDRs.
The enhancement of HOC$^+$ in XDRs is also shown in the modeling by \citet{2007ApJ...664L..23S}.
Here we examine the effect of varying the cosmic-ray ionization rate. 
The cosmic-ray ionization rate $\zeta$ is expressed in terms of the rate per H$_{2}$.
This is the total ionization rate including the ionization caused by protons and secondary electrons.
In the following, we use $G_0=1$, and take results from $A_{\rm V}=4$, where the effects of PDRs are negligible. 
Here we do not integrate over a range of $A_{\rm V}$, but we consider the values for this particular $A_{\rm V}$.
The model also provides an idea of the chemistry in XDRs. We also note that the cosmic-ray ionization rate is dependent on the column densities because the lower-energy cosmic rays have higher ionization cross section, and they become attenuated at lower column densities \citep[e.g., ][]{2009ApJ...694..257I,2009A&A...501..619P,2017ApJ...845..163N}. However, how exactly the cosmic-ray ionization rate depends on column densities is uncertain in our case where the energy spectrum and the density distribution are uncertain. Therefore, we only show modeled abundances and their ratios as a function of $\zeta$ instead of integrating over certain column densities with varying $\zeta$.

\begin{figure*}[h]
\centering{
\includegraphics[width=0.45\textwidth]{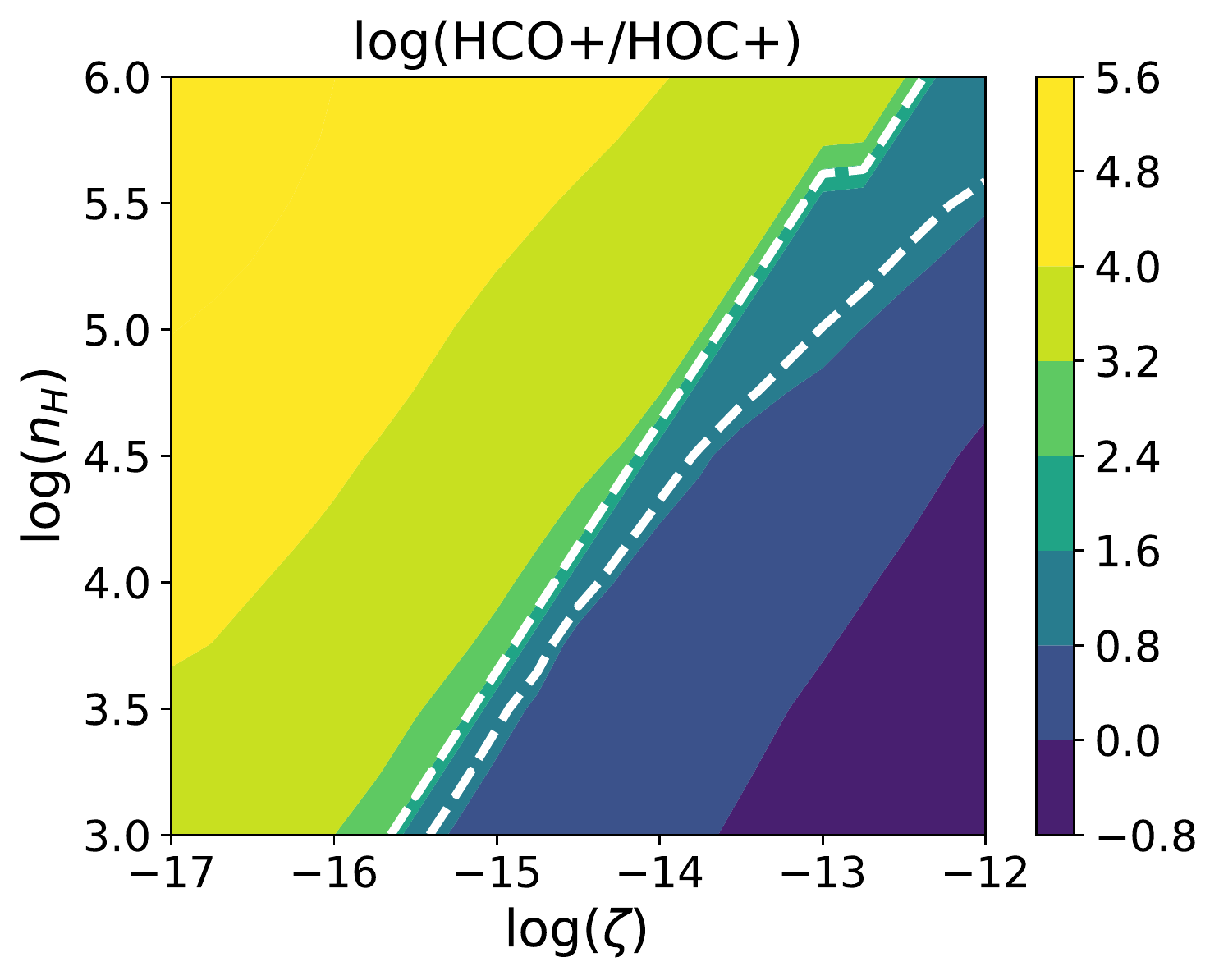}
\includegraphics[width=0.45\textwidth]{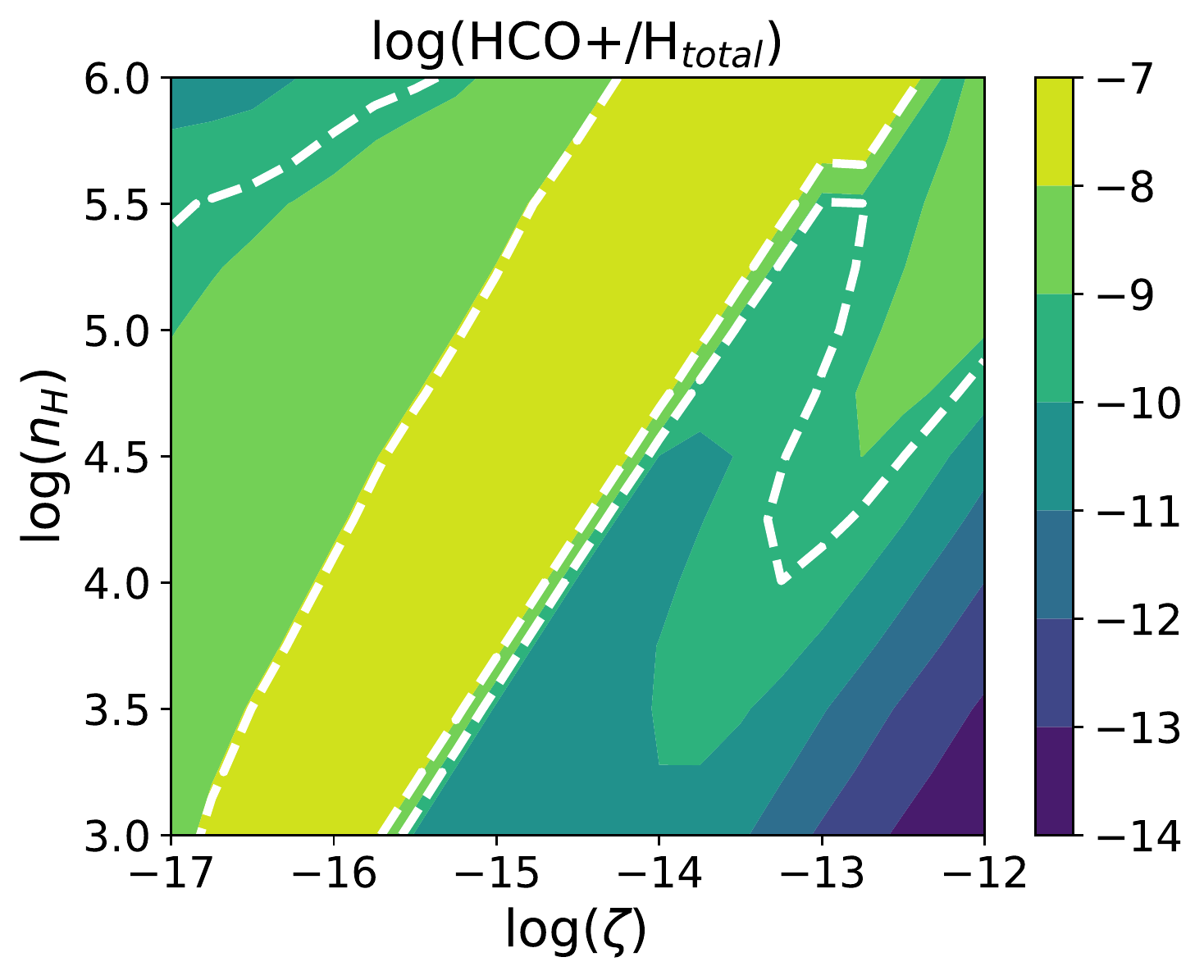}
}
\centering{
\includegraphics[width=0.45\textwidth]{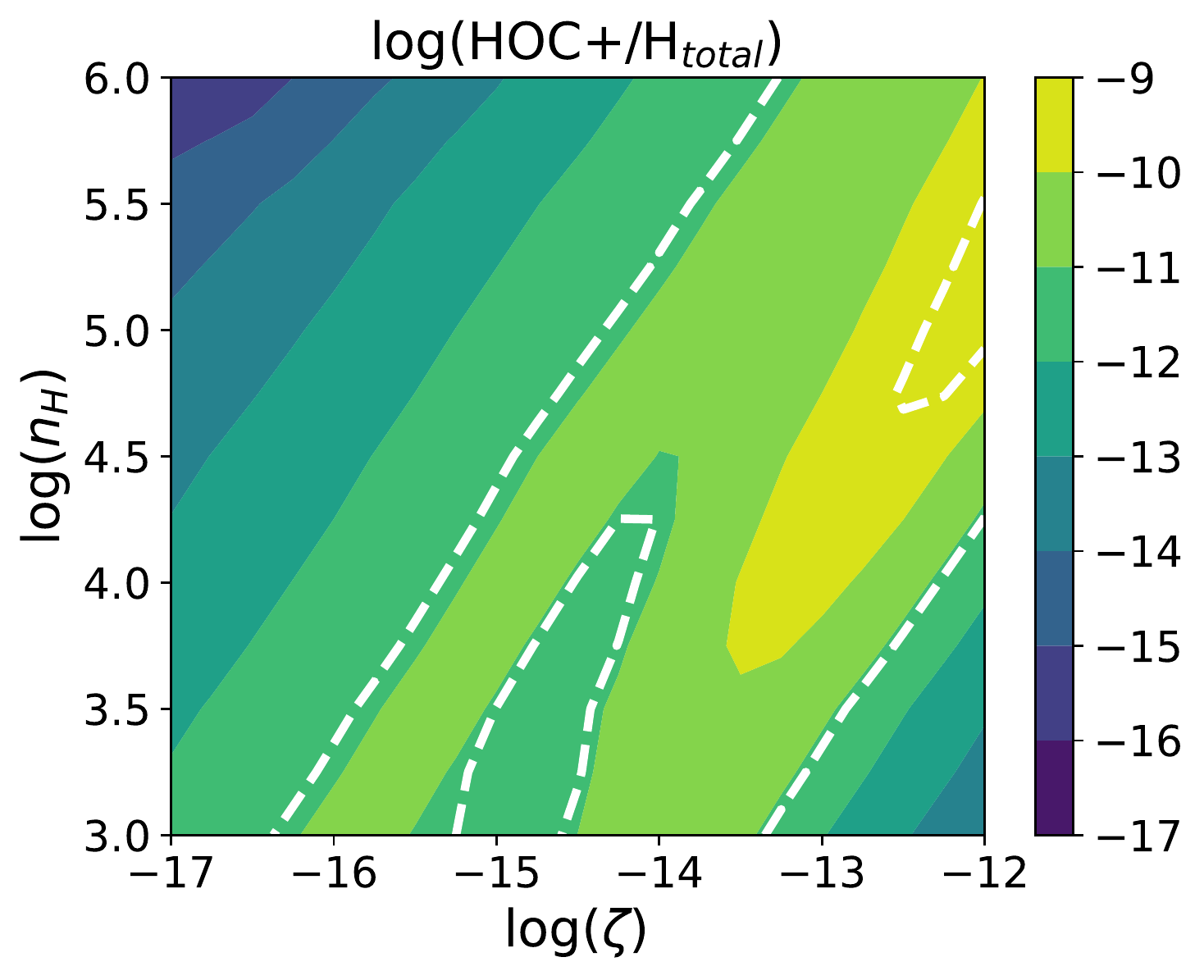}
\includegraphics[width=0.45\textwidth]{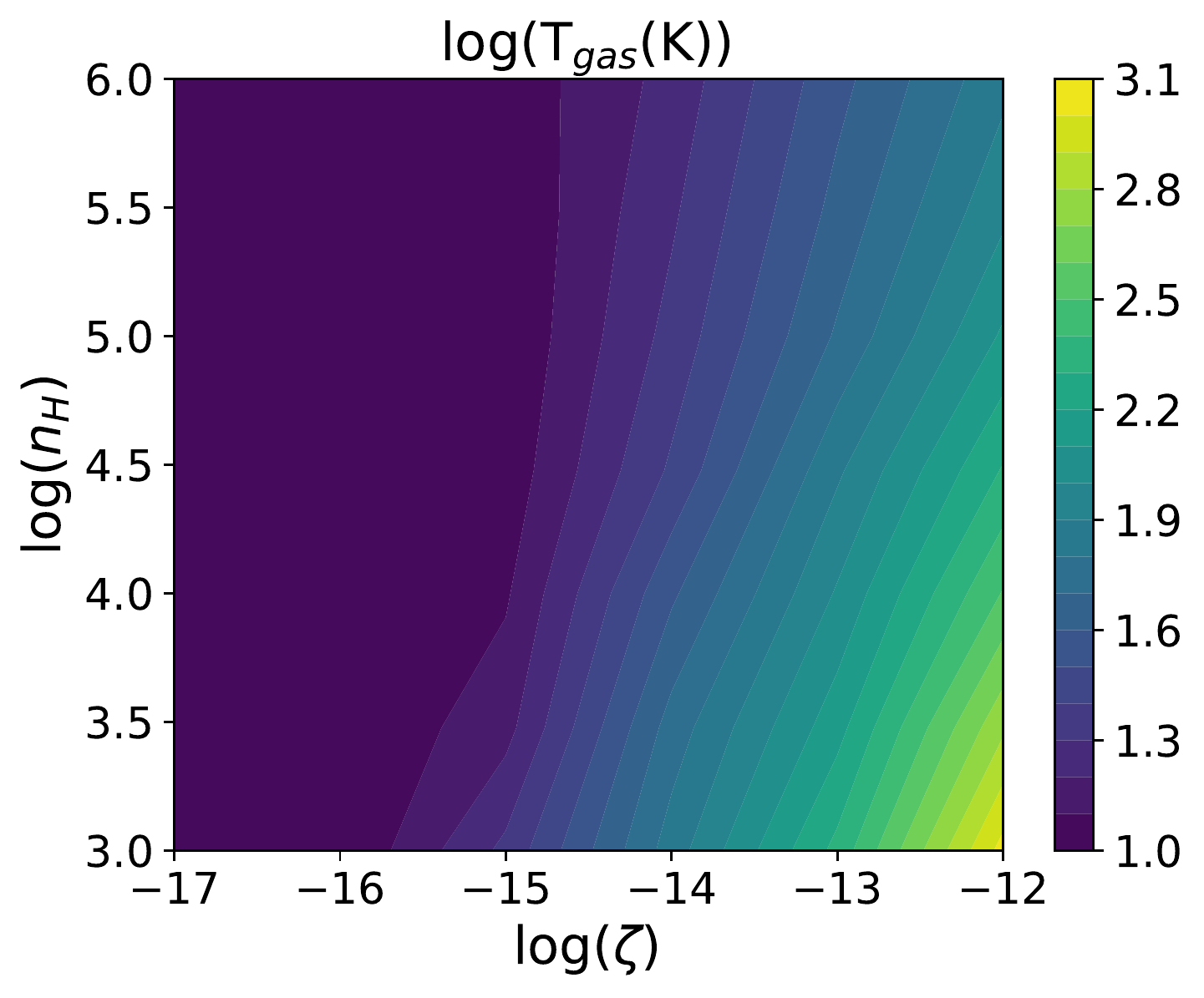}
}
\caption{The abundance ratio of HCO$^+$/HOC$^+$ (top left), the fractional abundance of HCO$^+$ (top right), the fractional abundance of HOC$^+$ (bottom left), and the gas kinetic temperature (bottom right). 
These quantities are shown on a logarithmic scale. 
The abscissa is the logarithm of cosmic-ray ionization rate $\zeta$ s$^{-1}$, while the ordinate is the logarithm of the density. The ranges of observed values in all analyzed positions are shown as white dashed lines for the \hcop/\hocp~ratio and fractional abundances of \hcop~and \hocp. \label{fig:cr_model}}
\end{figure*}

Figure \ref{fig:cr_model} (top left) shows the abundance ratios of HCO$^+$/HOC$^+$ with varying cosmic-ray ionization rate $\zeta$ and density $n_H$.  
It is apparent that the abundance ratios vary roughly proportional to the quantity $\zeta/n_H$, and that
the ratio decreases with higher $\zeta/n_H$. The fractional abundance of HCO$^+$ (Fig \ref{fig:cr_model}: top right) peaks when $\zeta/n_H \sim 10^{-20}-10^{-19}$ s$^{-1}$ cm$^{3}$, although \hcop~still keeps a moderate fractional abundance $\gtrsim 10^{-9}$ even at lower $\zeta/n$. At high values of $\zeta/n_H \sim 10^{-16}$ s$^{-1}$ cm$^{3}$, the medium starts to become atomic, not molecular, and the molecular abundances of almost all the species tend to decrease. The fractional abundance of HOC$^+$ peaks when $\zeta/n_H \sim 10^{-17}$ s$^{-1}$ cm$^{3}$, at higher ionization rates compared to the peak values of HCO$^+$ (Figure \ref{fig:cr_model}: bottom left). The gas temperature can be as high as $>1000$\,K in low-density, high-$\zeta$ environments,
but is quite low ($< 150$\,K) for most of the parameter space that we explored. The dust temperature is not shown as the modeled value was 14.6\,K with little variation for all densities and cosmic-ray ionization rates. Unlike PDR regions, the enhancement of HOC$^+$ in CRDRs does not need high gas temperature. 
This is because the formation route of HOC$^+$ is different from that in PDRs.
As discussed in Section \ref{sec:chem_route}, the high cosmic-ray ionization rate can produce OH and water through the electron recombination of H$_3$O$^+$ and H$_2$O$^+$. This will lead to HCO$^+$ and HOC$^+$ production via
Reactions \ref{hocp_prod1} and \ref{hocp_prod2}.

 %%%%%%%%%%%%%%%%%%%%%
%%%%%%%%%%%%%%%%%%%%%

\begin{figure*}[h]
\centering{
\includegraphics[width=0.45\textwidth]{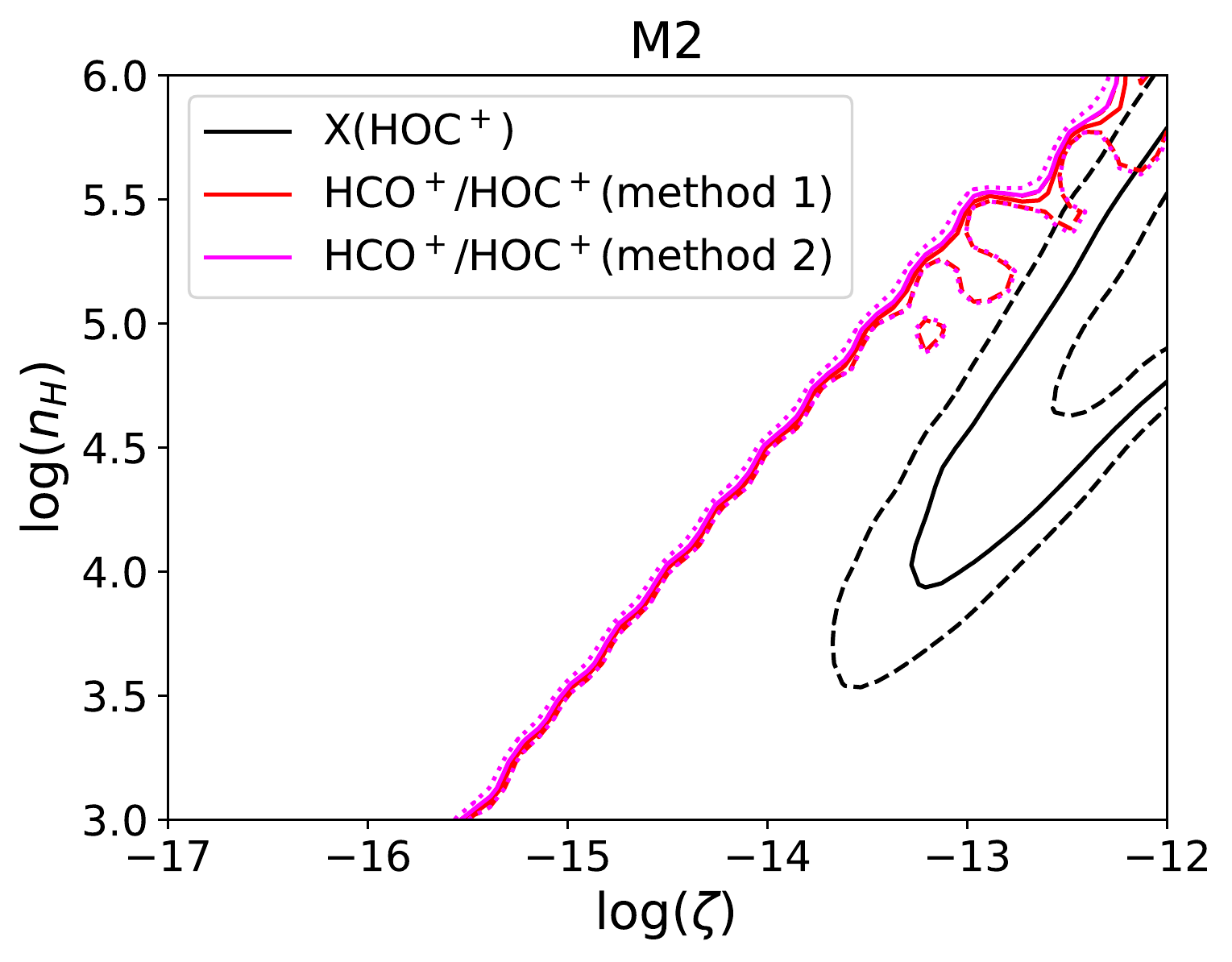}
\includegraphics[width=0.45\textwidth]{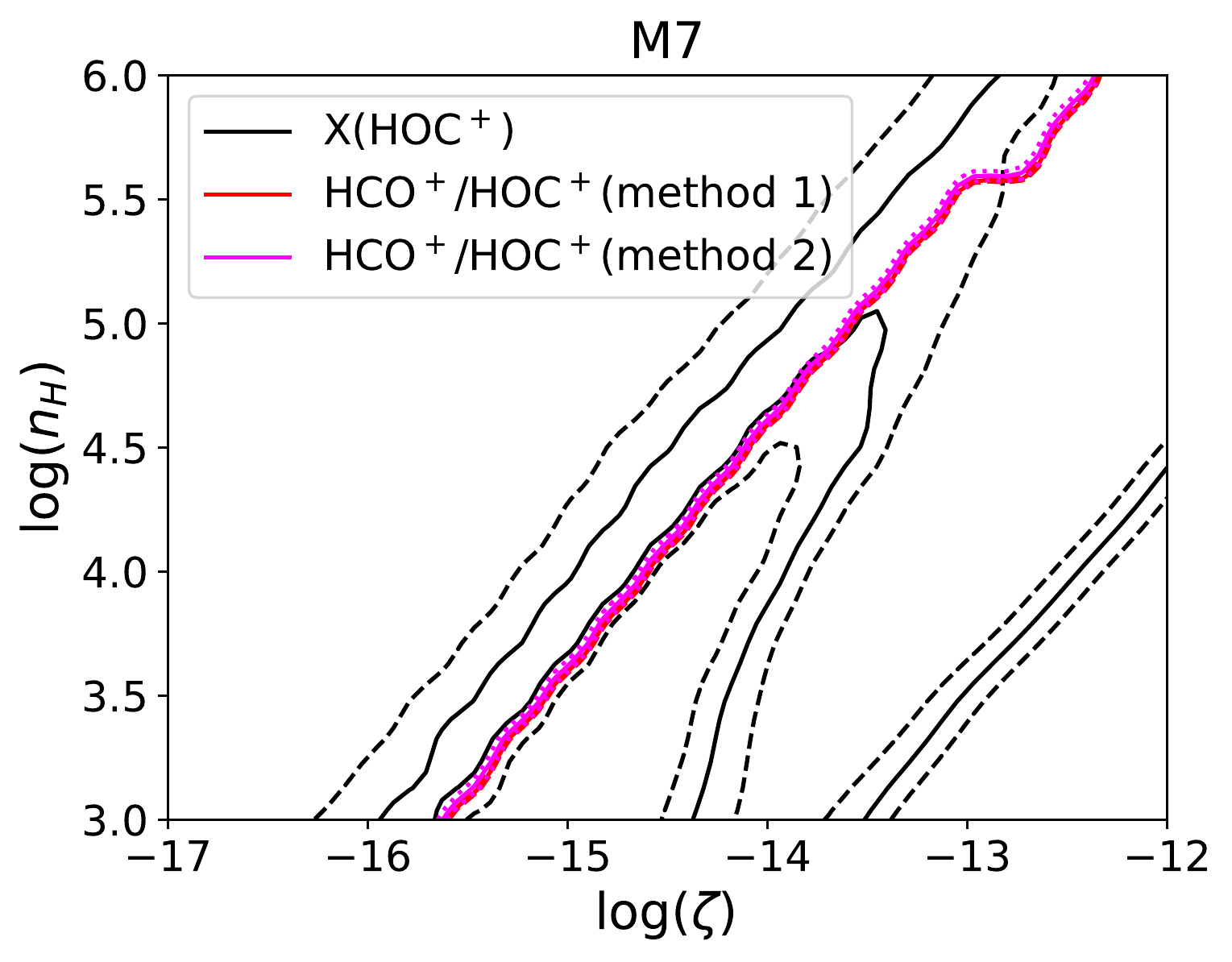}
}
\caption{X(HOC$^+$) (black) and HCO$^+$/HOC$^+$ with method 1 (red) and method 2 (magenta) that correspond to the observed values in the CRDR models are
shown as a function of $\zeta$ and $n_H$, for positions M2 (left) and M7 (right).  Solid lines show the observed values while dashed lines (for Method 1) and dotted lines (for Method 2) show the range within errors.
\label{fig:cr_fit}}
\end{figure*}

Figure \ref{fig:cr_fit} shows the same information as Figure \ref{fig:pdr_fit_Av},
but for CRDR models. In these figures, the observed range of HCO$^+$/HOC$^+$ 
appears very narrow, as this ratio changes drastically near $\zeta/n_H \sim 10^{-19}-10^{-18}$ s$^{-1}$ cm$^{3}$
in the chemical model while the observational error is only about 20\%.
In the position M2, observations are reproduced well around $n_H\gtrsim 10^{5.5}$ cm$^{-3}$, $\zeta \gtrsim 10^{-13}$ s$^{-1}$.
Many positions (A1, A2, A4, A7, A8, M3, M8, M9, M10) have good agreement with
observations in the same parameter range. 
Other positions have good agreement with observations in similar parameter space as 
the position M7 (Fig \ref{fig:cr_fit} right), when $\zeta/n_H \sim 10^{-18.5}$ s$^{-1}$ cm$^{3}$.
This includes lower density, lower $\zeta$ regions, but does not uniquely constrain the density and cosmic-ray ionization rate.
A large velocity gradient (LVG) analysis from our team (Tanaka et al., in preparation) suggests that 
the density of major molecular clumps (M4-M8) are $\sim 10^{5}$ cm$^{-3}$, which suggests $\zeta \sim 10^{-14}$ s$^{-1}$.

As already discussed, for M2 it appears that the parameters needed to reproduce observations 
are uniquely constrained to high density, high cosmic-ray ionization rate.
However, the comparison between the observed X(HOC$^+$), HCO$^+$/HOC$^+$ 
and the results of the chemical model should be taken 
with caution. 
It should be noted that the chemical model has uncertainties related to reaction rates, which are currently not considered. If larger errors were considered because of these uncertainties, the best-fit parameter could be at a lower density and lower cosmic-ray ionization rate
because the \hcop/\hocp~ratio varies with $\zeta/n$.
Because of the density obtained from the LVG analysis (Tanaka et al. in preparation), it is likely that the cosmic-ray 
ionization rate is still $\zeta \gtrsim 10^{-14}$ s$^{-1}$.
At the same time, $\zeta$
should not be orders of higher than 10$^{-14}$\,\ps,
due to reasons discussed in Section \ref{sec:ci_co}.

%%%%%%%%%%%%%%%%%%%%%
%%%%%%%%%%%%%%%%%%%%%

\section{Discussion}\label{sec:disc}
\subsection{Comparison of Abundance Ratios with Previous Work and Other Sources}

Our derived values of \hcop/\hocp $= 10-150$ are overall in agreement with the previous single-dish studies in NGC\,253.
\citet{2009ApJ...706.1323M} derived the HCO$^+$/HOC$^+$ ratios of $80\pm30$ and $63\pm17$ for two different velocity components,
while \citet{2015AA...579A.101A} obtained
the \hcop/\hocp~ratio of 30.
The results of ALCHEMI survey from ACA data alone (convolved to $15''$; Mart\'in et al. submitted) shows \hcop/\hocp $=50$,
a result between the two values provided from single-dish observations.
Our higher-angular-resolution study highlights the variation within the region
covered by the beam of the single-dish telescope.

As stated in the Introduction, observed values of the
\hcop/\hocp~ratios are $>1000$ in quiescent dense clouds,
far greater than in NGC 253.
Our work suggests similar values to Galactic PDR values ($\sim 100$), either in diffuse clouds and dense PDRs. The starburst galaxy M82 also has a \hcop/\hocp ratio similar to that 
in NGC 253, showing similarity in the properties of the ISM in these two local starburst galaxies.
Despite being another Galactic center and having a high cosmic-ray ionization rate ($\zeta \sim 10^{-15}$ \ps), the range of \hcop/\hocp~ratios in the Sgr~B2 clouds near the Galactic Center is 300-1500 \citep{2020ApJ...895...57A},
at least a factor of a few higher than values obtained in NGC\,253.
Among the ratios obtained in the CMZ of NGC 253, lower ratios are close to that 
suggested by
\citet{2015AA...574A..85A} and \citet{2018AA...617A..20A} from the intensity ratio of HCO$^+$/HOC$^+$(J=3-2)$\sim 10$ towards the quasars Mrk~231 and Mrk~273.
If these lines are optically thin and these species have the same excitation temperatures,
the \hcop/\hocp abundance ratios would be 5.
Contributing factors to these ratios are discussed in the following subsections.

\subsection{Derived physical parameters}
In Section \ref{sec:model}, we have constrained physical parameters 
such as $G_0$ and $\zeta$ in PDR and CRDR models.
We discuss those parameters in comparison with other sources here.

Relatively high values of interstellar radiation field $G_0 \sim 10^3$ are suggested from models
in Section \ref{sec:effecUV} for the case of $A_{V,max} =5$. While this value is higher
than the Galactic interstellar radiation field ($G_0=1$), it is not high compared with 
a strong PDR such as the Orion Bar \citep[$G_0 =3\times 10^4$;][]{1998A&A...330..696M}.
It should be noted that our beam size of 27 pc is much larger than the size
of the Orion Bar, and extreme PDR may have a smaller beam filling factor in our beam.
On the other hand, as discussed in Section \ref{sec:effecUV}, $A_{V,max}$ may be
much larger than 5, and the actual $G_0$ may be $10^4 - 10^5$.

The high value of cosmic-ray ionization rate we derived in Section \ref{sec:model_cr} is consistent with 
another work from our team that concluded that $\zeta > 10^{-14}$ \ps~ 
because C$_2$H is somewhat abundant even in extremely high $A_V$ regions \citep{2021arXiv210704580H}.
The derived value of cosmic-ray ionization rate is higher than those usually obtained in our Galaxy.
The cosmic-ray ionization rate in the Galactic spiral arm dense clouds was found to be around
$(1-5)\times 10^{-17}$ s$^{-1}$ \citep{2000A&A...358L..79V}, while this value can be an order of magnitude higher for diffuse clouds \citep{2012ApJ...745...91I}.
However, in the Galactic center, orders of magnitude higher  $\zeta$ has been inferred. Observational results from diffuse clouds in the Galactic center using H$_3^+$ tend to be higher than one from dense clouds, with $\zeta \gtrsim 10^{-14}$ s$^{-1}$  \citep{2014ApJ...786...96G,2019ApJ...883...54O}. 
The values in dense clouds are
$\zeta = (3-10)\times 10^{-16}$ s$^{-1}$ \citep{2016A&A...593A..43V,2019A&A...628A..27B},
although a higher value of $\zeta \gtrsim 10^{-15}$ s$^{-1}$ is possible within several parsecs of the Galactic center even in dense clouds \citep{2015A&A...584A.102H}.
The high cosmic-ray ionization rate of $\zeta \sim 10^{-15}$\,\ps~is also found 
in a molecular cloud in the vicinity of the supernova remnant W51C \citep{2011ApJ...740L...4C}.
High values of the cosmic-ray ionization rate of $\zeta=$\enum{2}{-14}-\enum{3}{-15} \ps
were found in two lines of sight in the $z=0.89$ molecular absorber towards
the quasar PKS1830-211 \citep{2016A&A...595A.128M}.
Although it is not possible to compare directly between CMZs of the Milky Way and NGC\,253,
the higher $\zeta$ in NGC\,253 is reasonable considering the difference in the star formation rate 
by a factor of 20 \citep[0.1~M$_{\odot}$\,yr$^{-1}$ in the Milky Way CMZ;][]{2013MNRAS.429..987L}.

In both PDR and CRDR models, our results suggest high densities of $n \sim 10^5 - 10^6\,$cm$^{-3}$. This is consistent with densities obtained for the central molecular clumps with dust observations by \citet{leroy_forming_2018} ($n_H \sim 4\times 10^5$\,cm$^{-3}$). 
Although our beam sizes of 1\farcs 6 are larger than their 0$\farcs$2 beam, \hcop~and \hocp~emission are likely 
originating from these dense clumps within our beam.

\subsection{Association of HCO$^+$/HOC$^+$ with Superbubbles }
%Show the bubble locations on HCO+ and HOC+ mom0 maps
The moment 0 image of HOC$^+$(1-0) show differences from HCO$^+$(1-0)
at locations related to some of the superbubbles (Figures \ref{fig:mom0} and \ref{fig:pos_mom0}).
It is not known whether these superbubbles are produced from
expanding HII regions or supernovae for most of the sources. The spectral indices of radio observations are often used to distinguish between supernovae emitting Synchrotron radiation and HII regions causing free-free emission with the spectral indices of $\sim -0.7$ and $\sim -0.1$, respectively. 
From the spectral index, it is suggested that a radio source at the superbubble southwest of M2 is produced from a supernova
remnant (Figure \ref{fig:pos_mom0}).
Although this is a useful measure for young and small superbubbles, the temperature of the bubbles cools down as they age and expand, making it difficult to observe. 
Therefore, some superbubbles do not have corresponding radio sources
(e.g., the bubbles southeast of M2 and one near A1).
Another cause of difficulty in separating HII and supernova remnants
is that many different sources are gathered in the CMZ of NGC 253.
In crowded regions, there are multiple radio sources close to each other, making it 
difficult to know which particular radio source is causing the superbubble.
This applies to the case of the superbubble right besides M5, where the neighboring
radio sources are identified to be HII regions or ``unknown". 
The dominant driving force of chemistry 
depends on the cause of the superbubbles.

Position M2 is the location with one of the lowest HCO$^+$/HOC$^+$ ratios ($\sim 14$).
It has a superbubble nearby
as shown in Figure \ref{fig:pos_mom0}.
In a position slightly southwest from M2, there is a supernova remnant (Figure \ref{fig:pos_mom0} bottom).
A position with similarly low HCO$^+$/HOC$^+ \sim 15$,
A1, also has a shell-like structure likely associated with a superbubble
according to higher-resolution observations by \citet{krieger_molecular_2019}.
We are not able to detect this bubble because this bubble is too small, and its shell is too thin
to be detected by our beam. 

Morphologically, HOC$^+$(1-0) also appears to have an association with the superbubble 
near the position M5 (Figure \ref{fig:pos_mom0}). 
At this position, the \hcop/\hocp~ratio is around 100.
There, \hocp~also shows the hole-like structure that is seen
in the channel map of CO data around $v = 293$ km s$^{-1}$(+50 km s$^{-1}$ from $v_{\rm sys}$).
The position M5 coincides with the hard X-ray source X-2 in \citet{2010ApJ...716.1166M} with the luminosity of $L_{2-10keV} \sim 10^{38}$ erg \ps. 
This area also has emission in soft X-ray, stretched towards the center 
of the superbubble. M5 is located in the northwestern part of the superbubble shell.
This is not the point where the HCO$^+$/HOC$^+$ ratio is low in comparison with 
other positions in NGC\,253 CMZ. 
This may be because multiple unresolved components are included in this position, and the enhancement 
of HOC$^+$ around the superbubble does not necessarily show up in the ratio. 
On the other hand, it is also possible that HOC$^+$ is not enhanced in this position. 
The southeast part of this bubble, A4, has a lower ratio of HCO$^+$/HOC$^+$ than M5,
around 40.

While an enhancement of \hocp~is suggested near the superbubble from M2 positions,
the observed \hcop/\hocp~ratio in M5 near the superbubble makes it less conclusive.
However, the position M5 may have multiple components in the line of sight,
and \hocp~may be enhanced near the superbubble.

\subsection{PDR, CRDR, or XDR?}

Our modeling results show that HOC$^+$ can be enhanced both with PDRs and CRDRs,
and that neither scenario can be ruled out simply from the observed values of HCO$^+$/HOC$^+$ or
the fractional abundances of HOC$^+$.
The chemistry of XDRs is very similar to that of CRDR,
so it is expected that XDR can also reproduce the observed ratios and abundances. 
In fact, the \hcop~fractional abundance in our CRDR model peaks at $\zeta_{\rm CR}/n= 10^{-19}$ \ps cm$^{3}$,
similar to that in the XDR model by \citet{1996A&A...306L..21L}.

We note here that PDRs and CRDRs are products of star formation, and both regions are expected to exist in a starburst galaxy like NGC\,253. \citet{2017ApJ...843...38G} proposed a following approximation for the relationship between the UV radiation field and cosmic-ray ionization rate; $\zeta = $\enum{1.8}{-16}$\sqrt{G_0}$ s$^{-1}$. According to this relationship, $\zeta \sim 2\times 10^{-14}$ s$^{-1}$ if $G_0=10^4$. These values of $G_0$ and $\zeta$ are similar to values needed to reproduce our observational results. Therefore, it is likely that the CMZ in NGC 253 is affected by the combination of these levels of the interstellar radiation field and cosmic-ray ionization rates. At the same time, we also expect that there is a deviation from the above relationship by \citet{2017ApJ...843...38G} in our observed regions, because this relationship is derived as a general behavior of the ISM. Depending on the distance from star clusters and supernova remnants, a region may be more affected by UV-photons or cosmic rays.

In an attempt to distinguish which scenario might be suitable
for the case in the CMZ of NGC\,253, we could use constraints from the physical parameters.
For example, models by \citet{2007A&A...461..793M} did not predict very low ratios of HCO$^+$/HOC$^+$ in their PDR models,
partly because they made some assumptions on standard clouds. For dense clouds, 
they chose the cloud size to be
1 pc, which corresponds to $N_H = 3\times 10^{23}$ cm$^{-2}$ ($A_V = 160$ mag) for the density of $n_H=10^5$ cm$^{-3}$.
After integrating to this $A_{\rm V}$, the HCO$^+$/HOC$^+$ ratio becomes very high, as already discussed in Section \ref{sec:effecUV}. 
Therefore, whether PDRs are the major sources of observed  \hocp~or not highly depends on the effective visual extinction. The value of effective visual extinction is most affected by the lowest visual extinction in all the directions, and can get significantly lower than the line-of-sight visual extinction when the medium is clumpy or filamentary. We consider the presence of such a clumpy medium is possible, and do not exclude PDRs even in the clumps with high column densities ($N_{\rm H_2} \sim 10^{24}$ \psc).   

X-rays can also produce \hocp.
\citet{2010ApJ...716.1166M} show three hard X-ray sources in the CMZ of NGC\,253 (Figure \ref{fig:pos_mom0} middle).
Out of them, X-1 has the highest luminosity ($L_{2-10keV}=10^{40}$ erg \ps),
but this is the position close to TH2, which is excluded from our analysis due to the absorption.
Other sources have the luminosities $L_{2-10keV}\sim 10^{38}$\,erg~\ps,
and this level of luminosity can contribute to the X-ray ionization rate $\zeta_X \sim 10^{-15}$ \ps, 
if we use the approximation formula \citep{1996ApJ...466..561M} as follows
\begin{equation}
\zeta_X = 1.4\times 10^{-15} \frac{L_X}{\rm 10^{38}\,erg\,s^{-1}} \left(\frac{r}{\rm 10\,pc}\right)^{-2} \left(\frac{N}{\rm 10^{22} \,cm^{-2}}\right)^{-1}.
\end{equation}
where $L_X$ is the X-ray luminosity, $r$ is the distance from the X-ray source,
and $N$ is the column density from the X-ray source. 
Note that the observed X-ray luminosity was derived assuming 
an obscuring column density of $N=10^{22}$ cm$^{-2}$.
Since our beam size is 27 pc, this level of luminosity can exceed 
the ionization by cosmic rays locally, but not on a large scale. Therefore, we suggest that X-rays do not significantly contribute to the overall abundance of \hocp.

One possible way to differentiate PDRs and CRDRs is from excitation. In PDRs, high temperatures are achieved in high-density regions.
This high temperature is necessary for the main route to form HOC$^+$
because the efficient water production plays a key role (Equations \ref{eq:hydro_o} and \ref{eq:hydro_oh}).
In such high-temperature and high-density regions, we expect excitation temperatures
of molecules to be very high. 
High kinetic temperatures of $\sim 300\,$K have been derived by \citet{mangum_fire_2019} from the excitation of multiple H$_2$CO transitions in all the major molecular clumps, namely, positions M4-8. 
The density of molecular clumps are also high ($n \gtrsim 10^5$\,\pcc),
while the density of the extended part of the CMZ is lower ($n_H \sim 10^{4.5}$\,\pcc) from the LVG analysis by our team (K. Tanaka, in preparation).
Meanwhile, the formation route of HOC$^+$ in CRDRs does not require high temperature nor high density,
which means that it can have low excitation. 
The high excitation does not exclude CRDRs, but it seems necessary that a low HCO$^+$/HOC$^+$ ratio and 
a high fractional abundance of HOC$^+$ in PDRs implies high excitation. 
In our moment 0 images of HOC$^+$, there are regions with the detection of only the $J=1-0$ transition,
while there are other regions with detection of all three transitions.
This difference in excitation may have resulted from different chemical scenarios.
In fact, the regions where higher-$J$ transitions of HOC$^+$ are detected correspond well to 
the 3-mm continuum peaks in the image shown in Appendix \ref{sec:app_cont}.
Unfortunately, there are no collisional coefficients available for HOC$^+$(F. van der Tak, private communication),
and we are unable to conduct the quantitative examination of this proposed method of
distinguishing scenarios. Once such data are published, radiative transfer calculations from each scenario
would be helpful in deciding whether we can indeed use excitation for differentiating scenarios.
We note that the difference in excitation is already seen with different excitation temperatures between molecular clumps such as M5 and M7 (15-20K) and other positions $T_{\rm ex} < 10\,$K (Table \ref{tab:columns}). 

\subsection{High Cosmic-Ray Ionization Rate and CI/CO Ratio}\label{sec:ci_co}
\begin{figure*}[h]
\centering{
\includegraphics[width=0.45\textwidth]{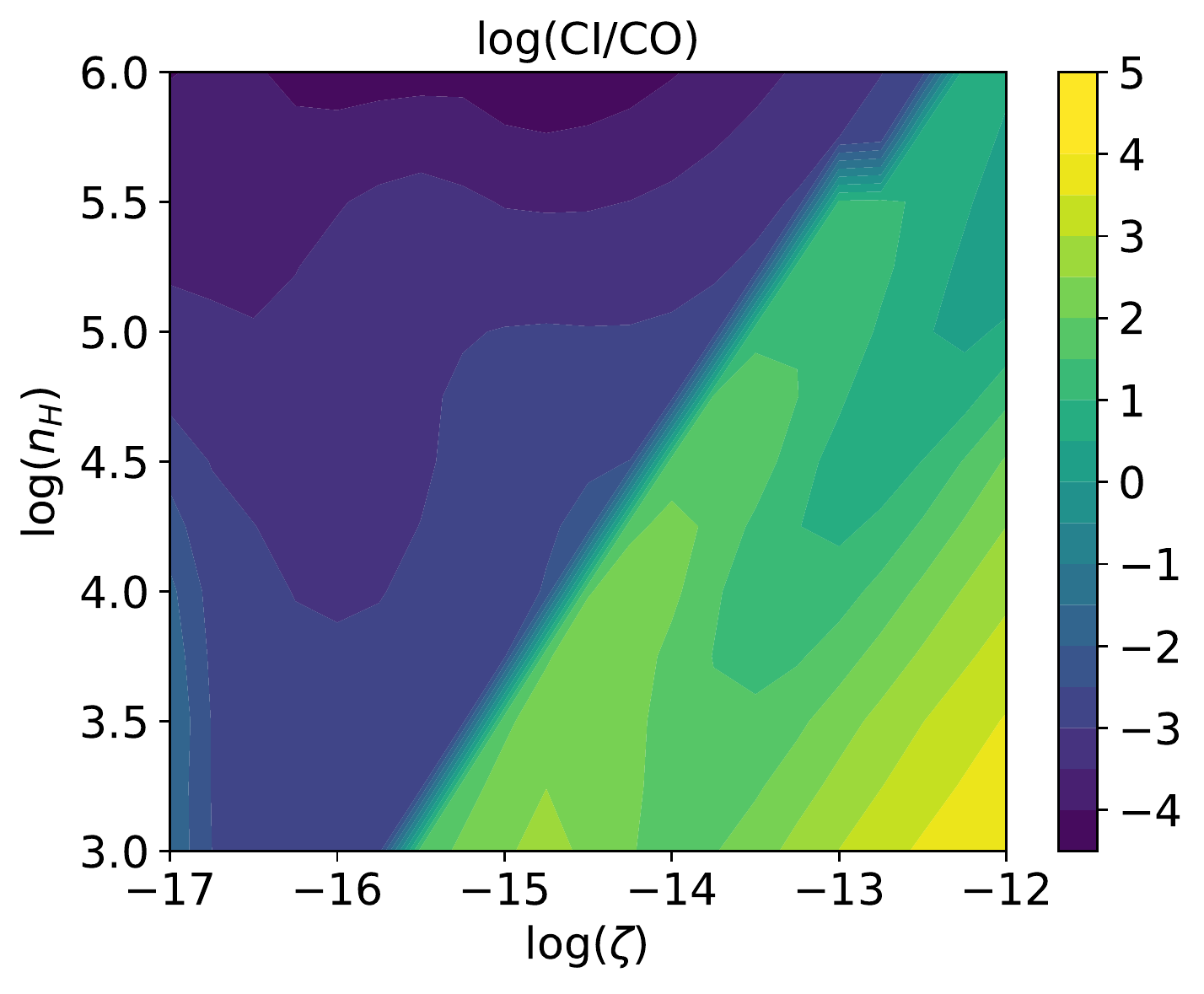}
}
\caption{Ratio of CI/CO as a function of $\zeta$ and $n$.
\label{fig:cr_ci}}
\end{figure*}

The extremely high cosmic-ray ionization rates of $\zeta \gtrsim 10^{-14}\,$\ps~we derived can pose challenges in retaining enough CO.
It has been proposed that at high cosmic-ray ionization rates, most of the carbon in the molecular
clouds will be in the form of CI instead of CO \citep{2017ApJ...839...90B}. This is not the case for NGC\,253,
as there is still a large amount of CO.
The observations of CI in NGC\,253 \citep{2016A&A...592L...3K} show the abundance ratio
[CI]/[CO] to be 0.5 - 1, which means that the CO abundance is equivalent to or larger than
that of CI. 
Here we shall check our derived value of cosmic-ray ionization rates is consistent with CI and CO observations.
Figure \ref{fig:cr_ci} shows the ratios of CI/CO with varying $\zeta$ and $n$ in our chemical model. 
The value of $\zeta \sim 10^{-13}$ s$^{-1}$ and $n_H\sim 10^5$ cm$^{-3}$ is in agreement with 
the observed value of [CI]/[CO] near unity. However, if this high value of $\zeta$ is
widespread, most of the molecular mass will be in atomic carbon instead of CO
in the lower density regions.
Because CI and CO can both be excited relatively easily, it is expected that most of the molecular
mass traced by CI and CO has a low density ($n_H \lesssim 10^4$ \pcc).
Therefore, if the enhancement of HOC$^+$ is caused by the high value of $\zeta$,
we claim that the enhancement of $\zeta$ should be relatively localized.
Although this is somewhat against the common conception that cosmic rays should penetrate 
into high column density gas, lower-energy cosmic rays can increase $\zeta$
in a relatively localized way due to the higher ionization cross section at lower energy, only affecting low-column-density regions \citep[e.g., ][]{2009A&A...501..619P}.
In addition, on average, the value of the cosmic-ray ionization rate should not exceed $10^{-14}$\,\ps~by orders of magnitude to retain enough CO.

\subsection{Outflows: An Alternative Formation Mechanism for HOC$^+$}
We have explored the effects of UV-photons and cosmic rays on HOC$^+$, 
but an alternative scenario is also possible. For example, outflows can provide a 
high abundance of H$_2$O by heating the gas, and ionization that leads to C$^+$ by creating
an outflow cavity that allows the ionization source (cosmic rays/ UV-photons) to travel further.
This may be the case for Mrk~231, where there seems to be an association between a radio jet and off-nuclear HOC$^+$.  It is unlikely that all the HOC$^+$ emission 
in our observations is affected by the outflow because \hocp~is detected in molecular clumps as well. However, it is also possible that there is some contribution from outflows as the \hcop/\hocp ratio becomes lower with higher galactic latitude. 
%%%%%%%%%%%%%%%%%%%%%
%%%%%%%%%%%%%%%%%%%%%

\section{Summary}\label{sec:sum}
We have used the high-sensitivity imaging spectral scan from the ALCHEMI Large Program to investigate the abundance ratios between HCO$^+$ and its metastable isomer HOC$^+$.  These measurements have allowed us to study the abundances of HCO$^+$ and HOC$^+$ within the CMZ of NGC\,253 and their relationship with photodissociation regions or cosmic rays in the galactic center of this starburst galaxy.  Our main findings are as follows.
\begin{itemize}
\item The \hocp(1-0) emission shows significantly different distribution from \hcop~or \httcop~J=1-0 emission towards a few superbubbles identified in CO data. This association of \hocp~emission with superbubbles has never been observed previously.
 %interesting morphology of HOC$^+$ (1-0)
\item The observed abundance ratios of HCO$^+$/HOC$^+$ range from $\sim 10$ to 150.  This is $1-3$ orders of magnitude lower than the ratios seen towards quiescent dense clouds 
or even some of the PDRs in the Galaxy, indicating physical and chemical processes by energetic photons or cosmic-ray particles enhancing \hocp. The ratio is low near the position of the western superbubble (position M2) and another superbubble in the southwest part of the CMZ (A1). On the other hand, the ratios are higher at the centers of molecular clumps (e.g., position M5), which may be due to the higher density or high column density. %abundance ratios, compared with other observations
\item We derived the fractional abundances of \hocp~to be [\hocp]/[H$_{total}$] = $(0.7 - 30)\times 10^{-11}$ ([\hocp]/[H$_2$] = $(1.5 - 60)\times 10^{-11}$). This is equivalent to or higher than Galactic PDR values. 
\item We ran models of photodissociation regions (PDR) to see what mechanism is likely producing the observed abundance ratios [HCO$^+$]/[HOC$^+$] and the fractional abundances. The PDR models produce high fractional abundances of HOC$^+$ and low ratio of HCO$^+$/HOC$^+$ when the density is high ($n_H\gtrsim 10^5$ cm$^{-3}$) and the interstellar radiation field $G_0 \sim 10^{2.5} - 10^{3.5}$ if we assume $A_{V,max}=5$. For higher $A_{V,max}$, higher values of $G_0$
reproduce the \hcop/\hocp ratios, but cannot reproduce high enough \hocp~fractional abundances.%comparison with the chemical model
\item In addition to PDRs, we also ran models of cosmic-ray dominated regions (CRDR). If HOC$^+$ is enhanced due to cosmic rays, our models suggest the cosmic-ray ionization rate $\zeta \gtrsim 10^{-14}$ s$^{-1}$. It is about 3-4 orders of magnitude higher than the Galactic spiral-arm dense clouds, and 2 orders of magnitude higher than the Galactic center dense cloud values. 
%However, it is also possible that the observed region is a mixture of regions with a lower $\zeta \sim 10^{-14}$ s$^{-1}$ and with higher $\zeta$.
\item In our PDR models, the formation of HOC$^+$ with high abundance requires a high temperature, high-density environment. On the other hand, such conditions are not required in models for cosmic-ray dominated regions. Therefore, we suggest that the region traced with a high luminosity of HOC$^+$ (1-0) is caused by cosmic rays, while regions traced with higher-$J$ transitions of HOC$^+$ can be either caused by PDRs or CRDRs. 
%This hypothesis needs to be tested once the collisional coefficients of HOC$^+$ become available. 
\end{itemize}

These analyses of \hocp~have shown that this molecule can be used to study feedback in the starburst ISM.

\acknowledgments

\sloppypar{We thank the anonymous referee for the constructive comments.
We are grateful to Francesco Costagliola, the original PI of ALCHEMI, for his initial leadership of this program that got us started on this journey. 
We thank the ALMA staff for observations, quality assessments, and help at the local regional center. 
N.H. thanks Eric Herbst for his comments on the reaction rate of  \hocp + H$_{2}$. N.H. also thanks Franck Le Petit, Jacques Le Bourlot, and Evelyne Roueff for their help in Meudon code, especially the insight on the bistability in certain parameter spaces. 
This paper makes use of the following ALMA data: ADS/JAO.ALMA\#2017.1.00161.L. 
ALMA is a partnership of ESO (representing its member states), NSF (USA) and NINS (Japan), together with NRC (Canada), 
MOST and ASIAA (Taiwan), and KASI (Republic of Korea), in cooperation with the Republic of Chile. 
The Joint ALMA Observatory is operated by ESO, AUI/NRAO and NAOJ. Data analysis was in part carried out on the Multi-wavelength Data Analysis System operated by the Astronomy Data Center (ADC), National Astronomical Observatory of Japan.
This research made use of APLpy, an open-source plotting package for Python
    \citep{aplpy2012,aplpy2019}.
N.H. acknowledges support from JSPS KAKENHI Grant Number JP21K03634. 
K.S. has been supported by grants 
MOST 108-2112-M-001-015 and 109-2112-M-001-020 from the Ministry of Science and Technology, Taiwan.
Y.N. is supported by the NAOJ ALMA Scientific Research Grant Number 2017-06B.
%({\bf Should  I use KAKENHI, EA ALMA, NAOJ publication fund, or DoS research budget?}). 
V.M.R. and L.C. are funded by the Comunidad de Madrid through the Atracci\'on de Talento Investigador (Doctores con experiencia) Grant (COOL: Cosmic Origins Of Life; 2019-T1/TIC-15379).}


\vspace{5mm}
\facilities{ALMA}


\software{CASA \citep{2007ASPC..376..127M}, APLpy \citep{aplpy2012,aplpy2019}, MADCUBA\citep{2019A&A...631A.159M}, Meudon PDR code \citep{2006ApJS..164..506L}, Nautilus \citep{2009A&A...493L..49H}}
          
      % \bibliography{alchemi}



\appendix

\section{Estimation of Noise in Velocity-Integrated Intensity Images}\label{sec:app_rms}
\begin{figure}
    \centering
    \includegraphics[width=0.45\textwidth]{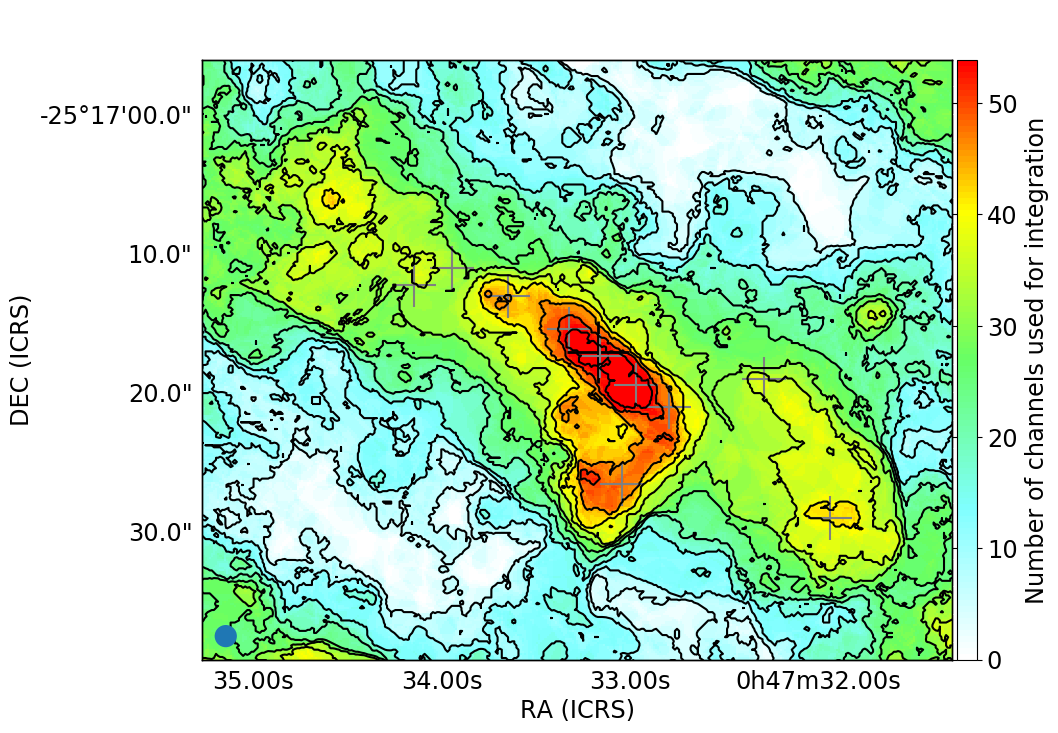}
    \caption{Number of channels used for integration.}
    \label{fig:nchan}
\end{figure}

The RMS noise level of our velocity-integrated intensity images (moment 0 images) varies over the image because we masked out the regions in velocity and position space where no significant emission is expected.
These noise levels can be estimated by 
\begin{equation}
    \sigma = \sigma_{1ch}\sqrt{N}\Delta v,
\end{equation}
where $\sigma$ is the RMS noise of the velocity-integrated intensity image at a certain pixel, $\sigma_{1ch}$
is the RMS of a single channel, $N$ is the number of 
channels integrated in the moment 0 image,
and $\Delta v$ is the velocity resolution of a single channel.
The resolution of the data in this work is $\Delta v$ of 10 km s$^{-1}$. The number of channels used to integrate
for each position is shown in Figure \ref{fig:nchan}.
From this image, we used $N=40$ to obtain a typical error of the image to draw contours in Figure \ref{fig:mom0}.
Although there are regions with $N \gtrsim 50$, those regions have enough $S/N$ ratio, and there is no doubt in detection there.
The RMS of a single channel for each image is shown in Section \ref{sec:obs}.

\section{Spectra at Selected Positions}\label{sec:app_spectra}

\begin{figure*}
\centering{
\includegraphics[width=0.99\textwidth,trim= 0 0 0 0]{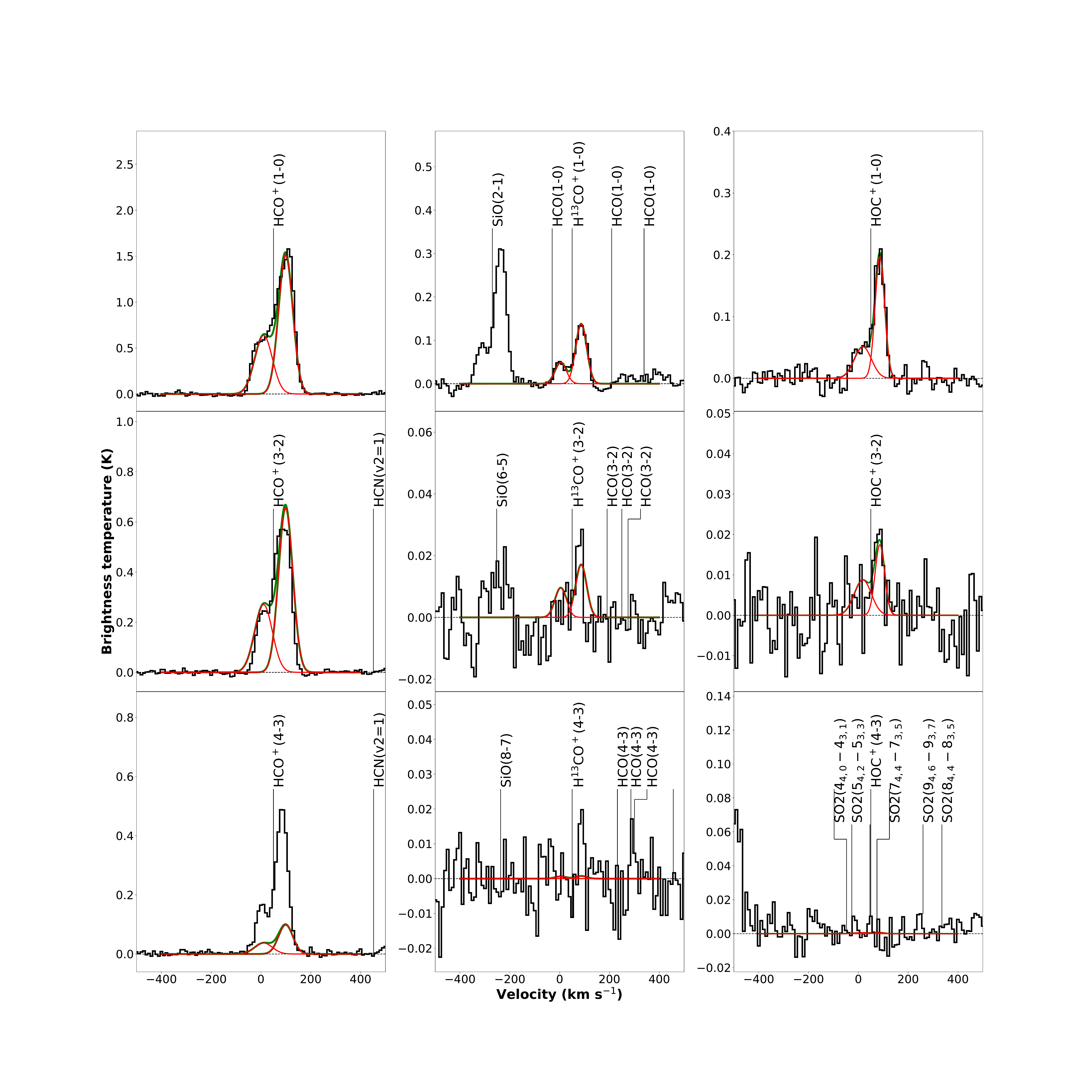}
}
\caption{Spectra at position M2. \label{fig:spec_m2}}
\end{figure*}

\begin{figure*}
\centering{
\includegraphics[width=0.99\textwidth,trim= 0 0 0 0]{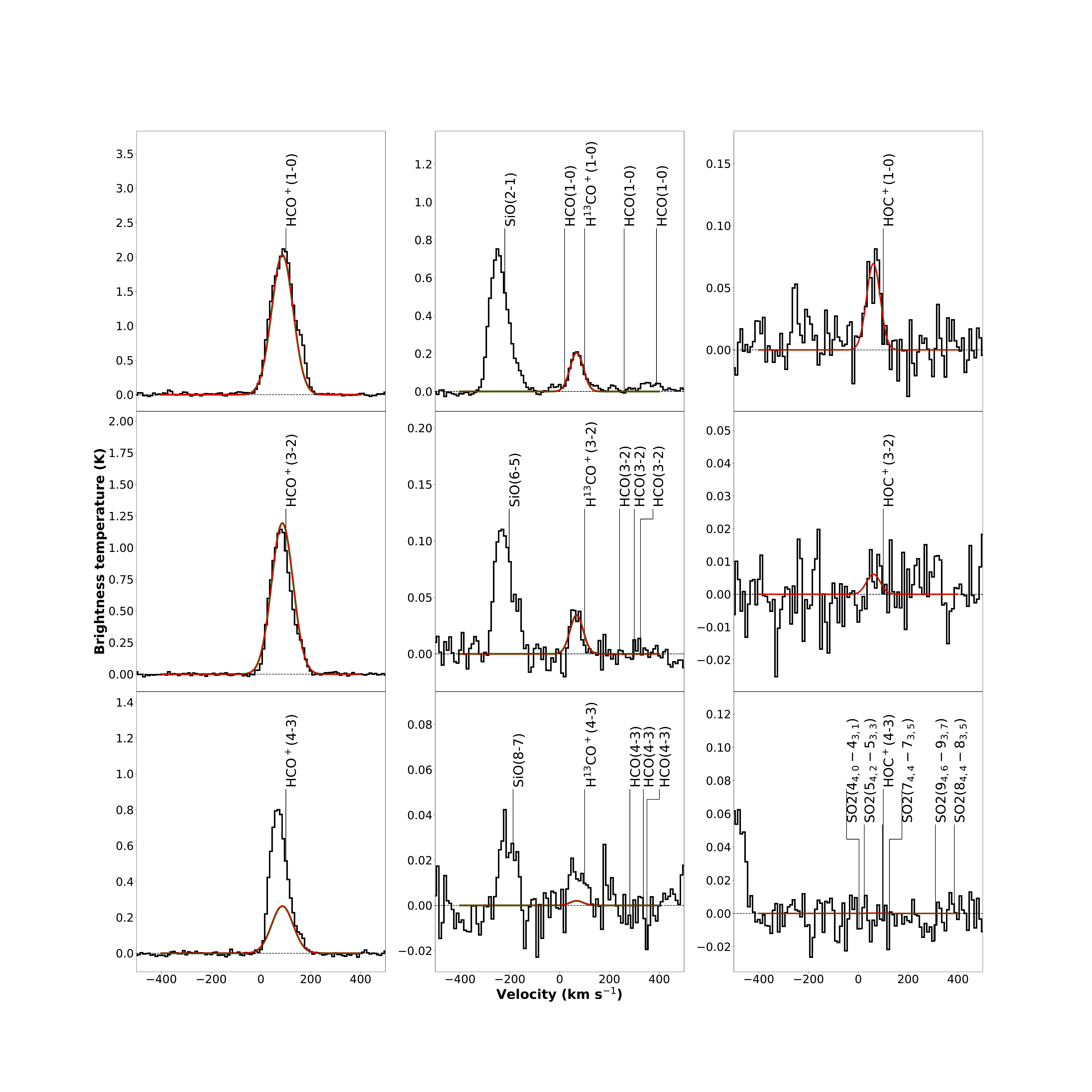}
}
\caption{Spectra at position M3. \label{fig:spec_m3}}
\end{figure*}

\begin{figure*}
\centering{
\includegraphics[width=0.99\textwidth,trim= 0 0 0 0]{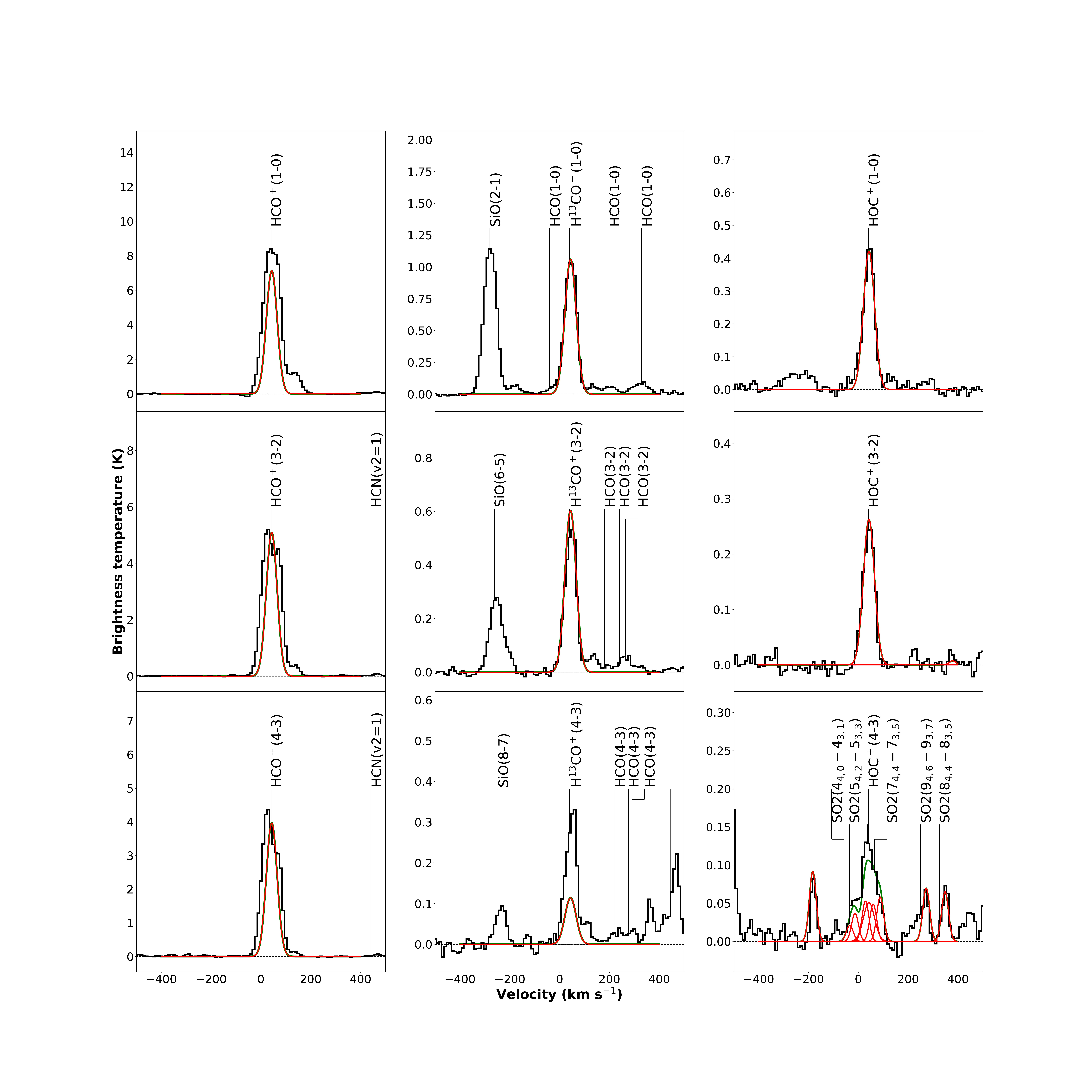}
}
\caption{Spectra at position M4. \label{fig:spec_m4}}
\end{figure*}

\begin{figure*}
\centering{
\includegraphics[width=0.99\textwidth,trim= 0 0 0 0]{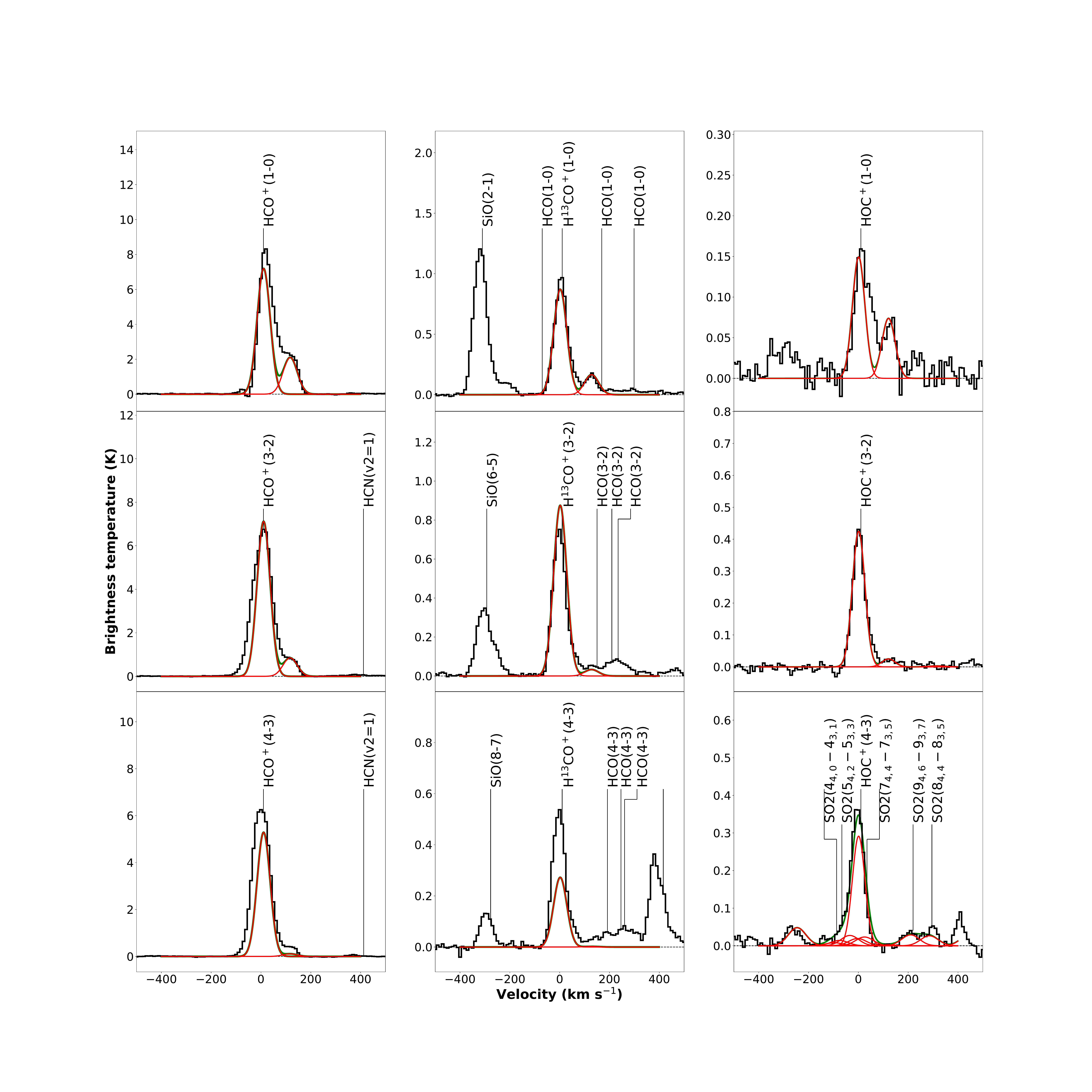}
}
\caption{Spectra at position M5. \label{fig:spec_m5}}
\end{figure*}

\begin{figure*}
\centering{
\includegraphics[width=0.99\textwidth,trim= 0 0 0 0]{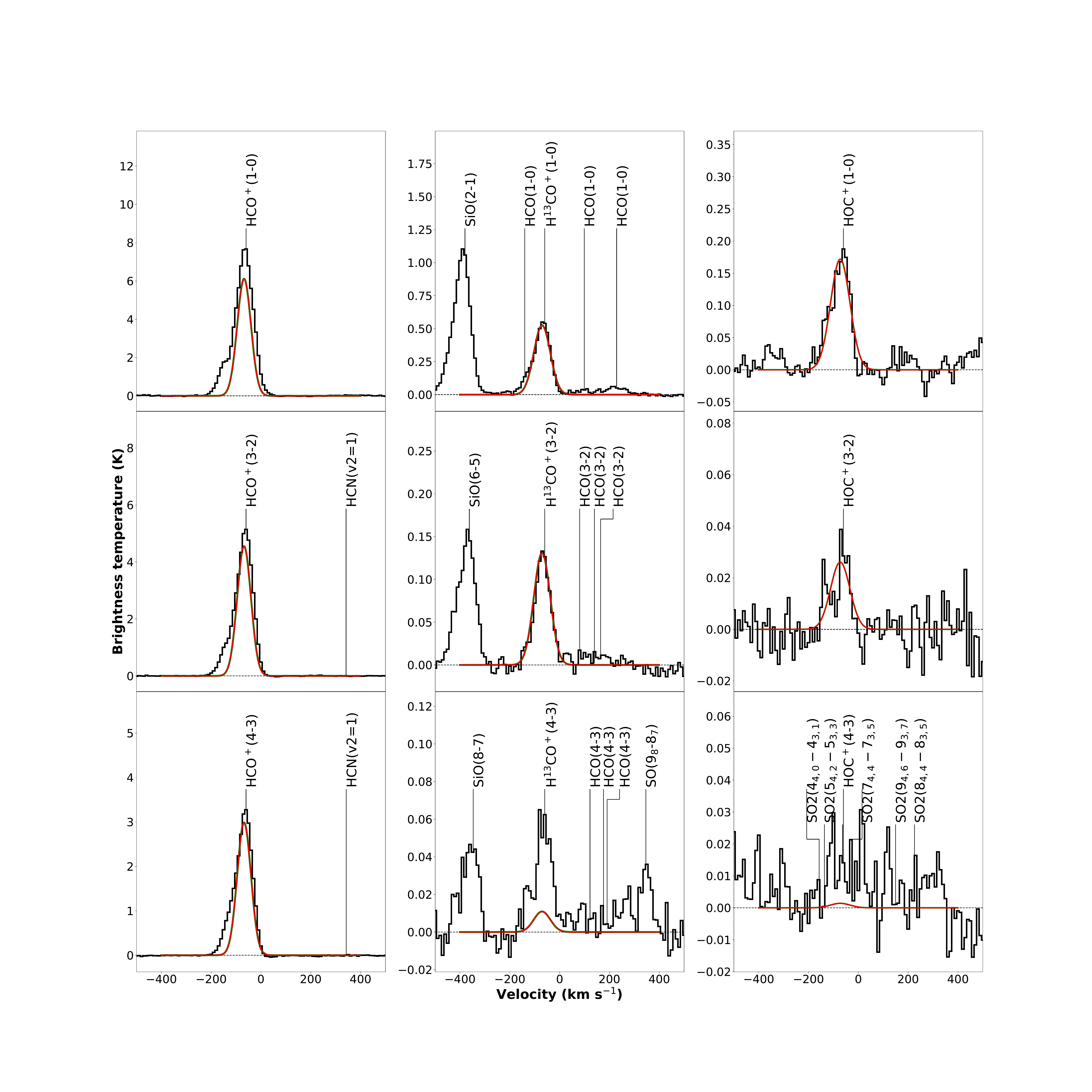}
}
\caption{Spectra at position M8. \label{fig:spec_m8}}
\end{figure*}

\begin{figure*}
\centering{
\includegraphics[width=0.99\textwidth,trim= 0 0 0 0]{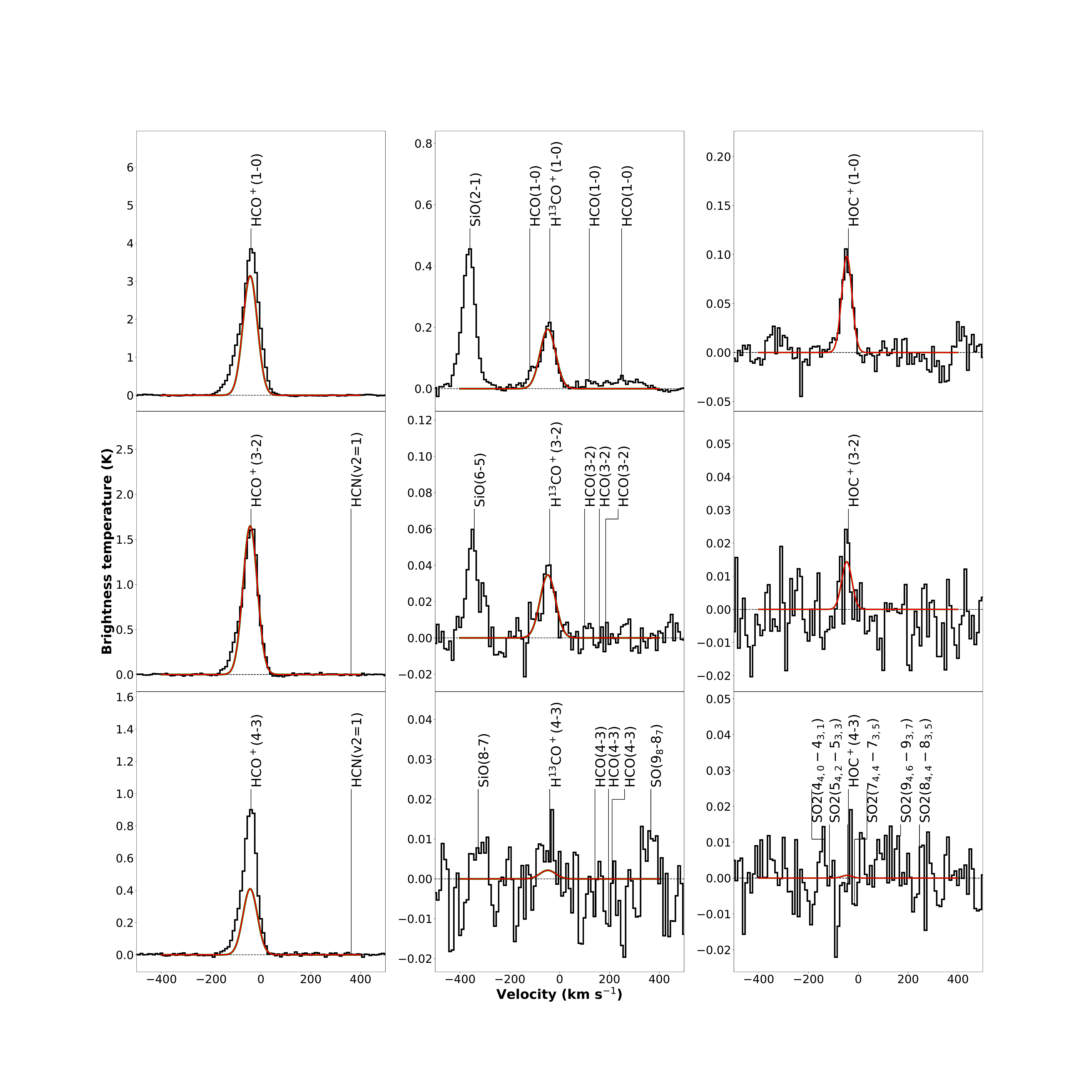}
}
\caption{Spectra at position M9. \label{fig:spec_m9}}
\end{figure*}

\begin{figure*}
\centering{
\includegraphics[width=0.99\textwidth,trim= 0 0 0 0]{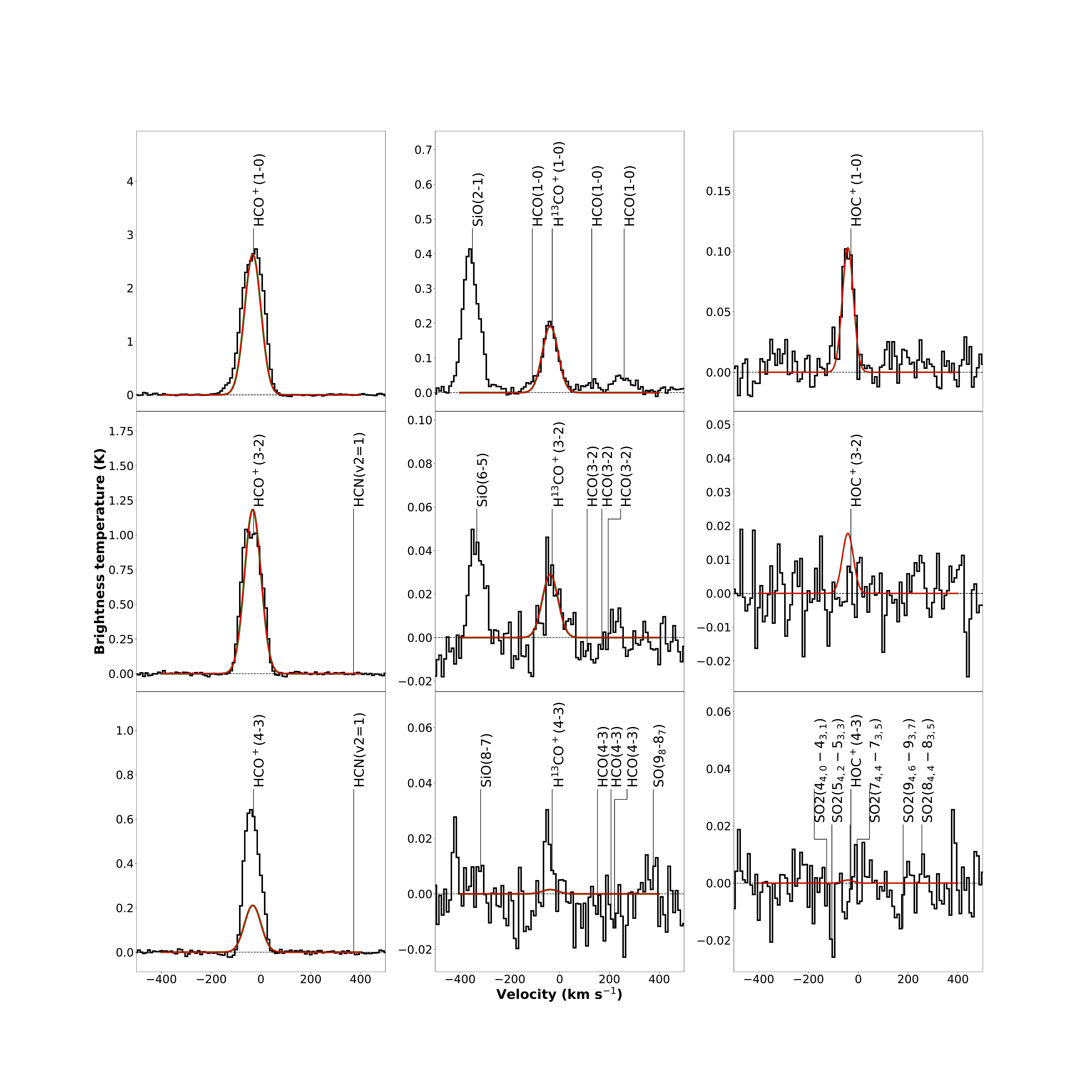}
}
\caption{Spectra at position M10. \label{fig:spec_m10}}
\end{figure*}

\begin{figure*}
\centering{
\includegraphics[width=0.99\textwidth,trim= 0 0 0 0]{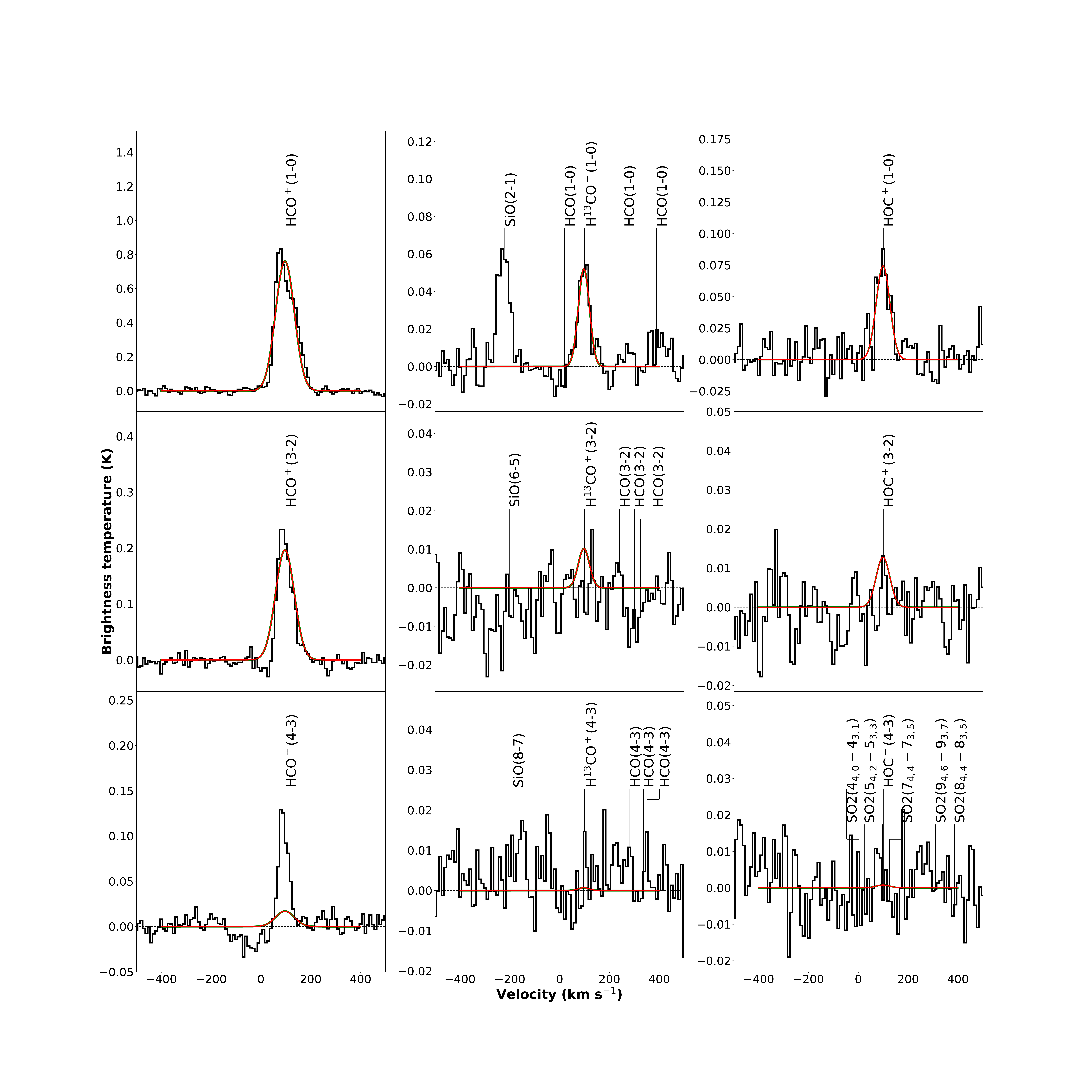}
}
\caption{Spectra at position A1. \label{fig:spec_a1}}
\end{figure*}

\begin{figure*}
\centering{
\includegraphics[width=0.99\textwidth,trim= 0 0 0 0]{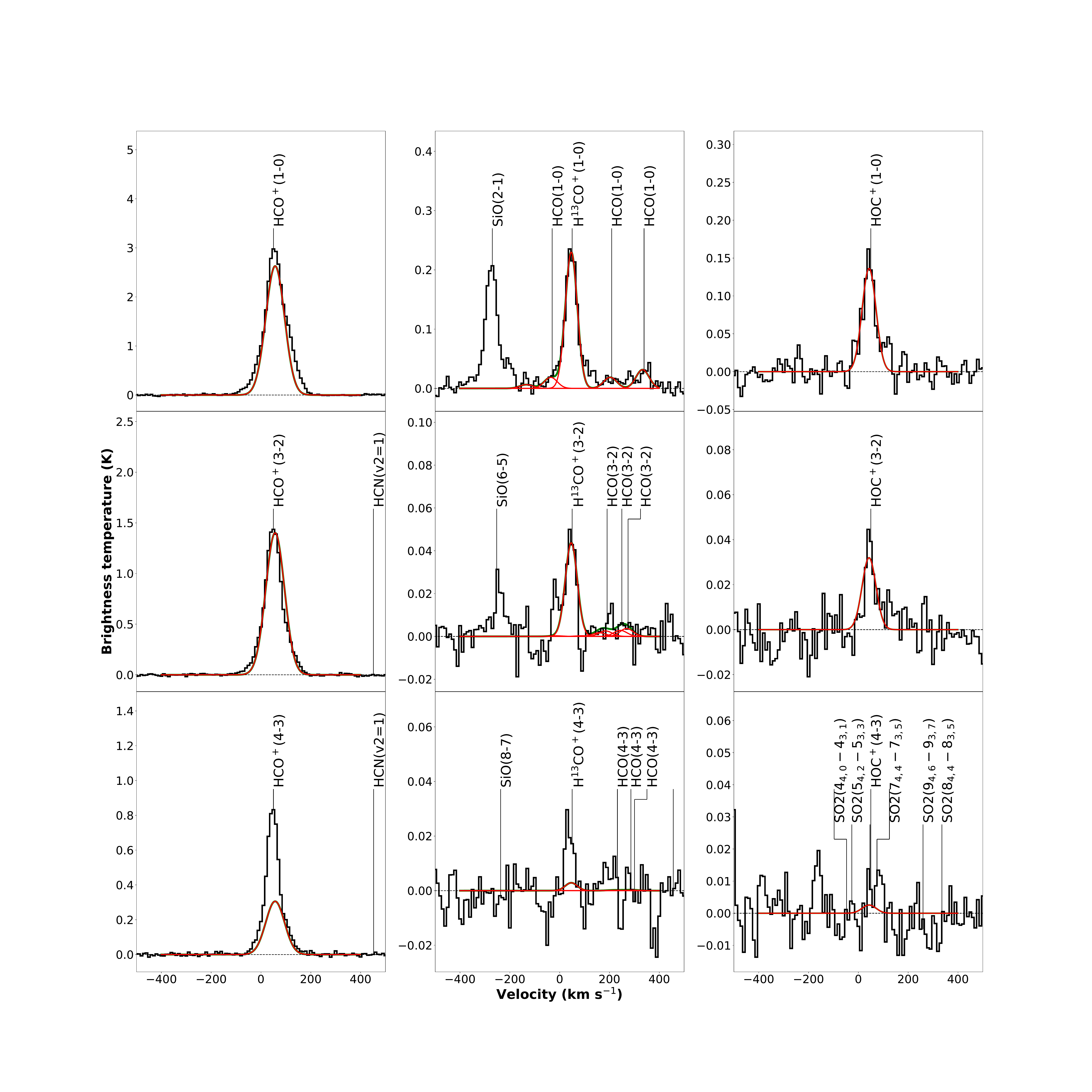}
}
\caption{Spectra at position A2. \label{fig:spec_a2}}
\end{figure*}

\begin{figure*}
\centering{
\includegraphics[width=0.99\textwidth,trim= 0 0 0 0]{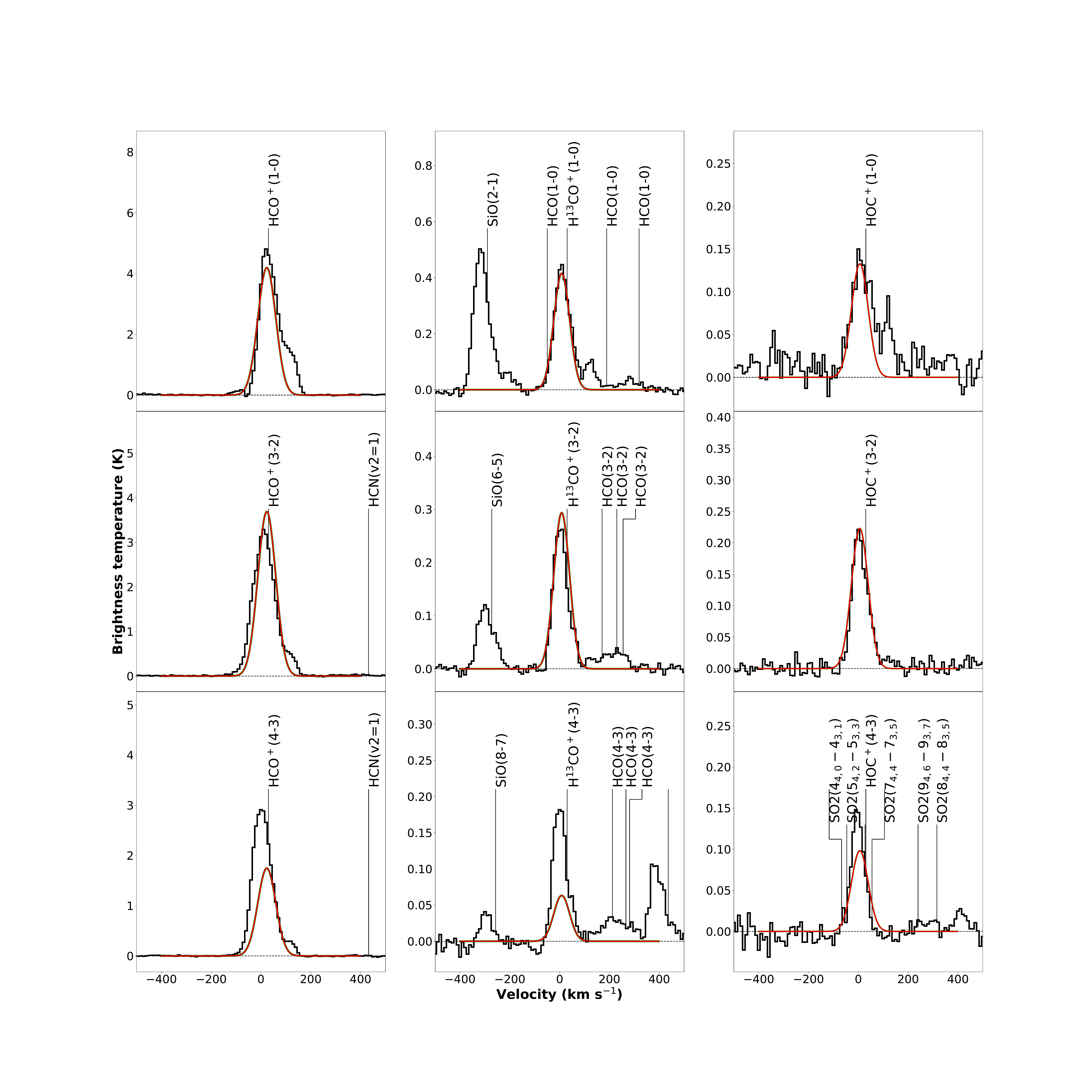}
}
\caption{Spectra at position A3. \label{fig:spec_a3}}
\end{figure*}

\begin{figure*}
\centering{
\includegraphics[width=0.99\textwidth,trim= 0 0 0 0]{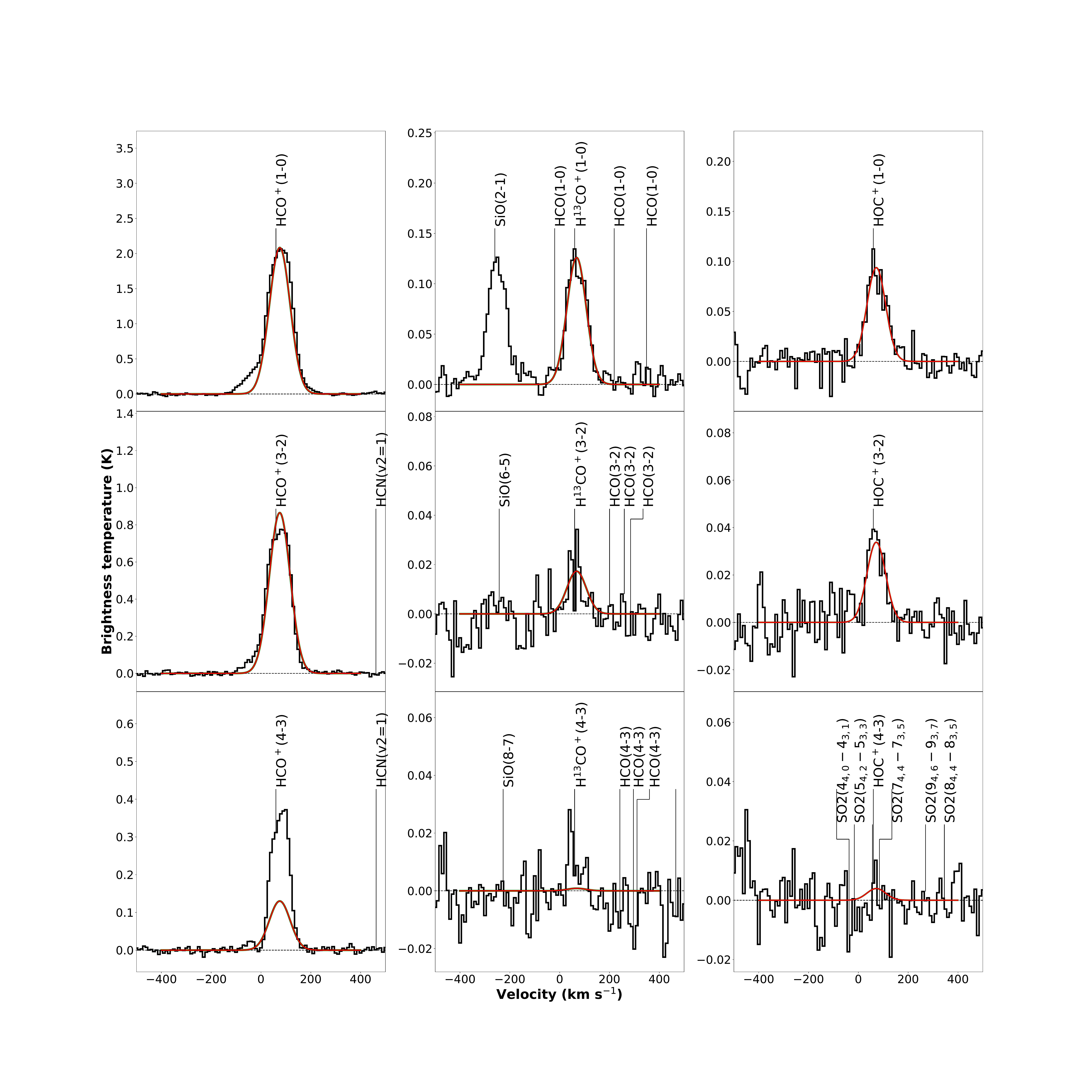}
}
\caption{Spectra at position A4. \label{fig:spec_a4}}
\end{figure*}

\begin{figure*}
\centering{
\includegraphics[width=0.99\textwidth,trim= 0 0 0 0]{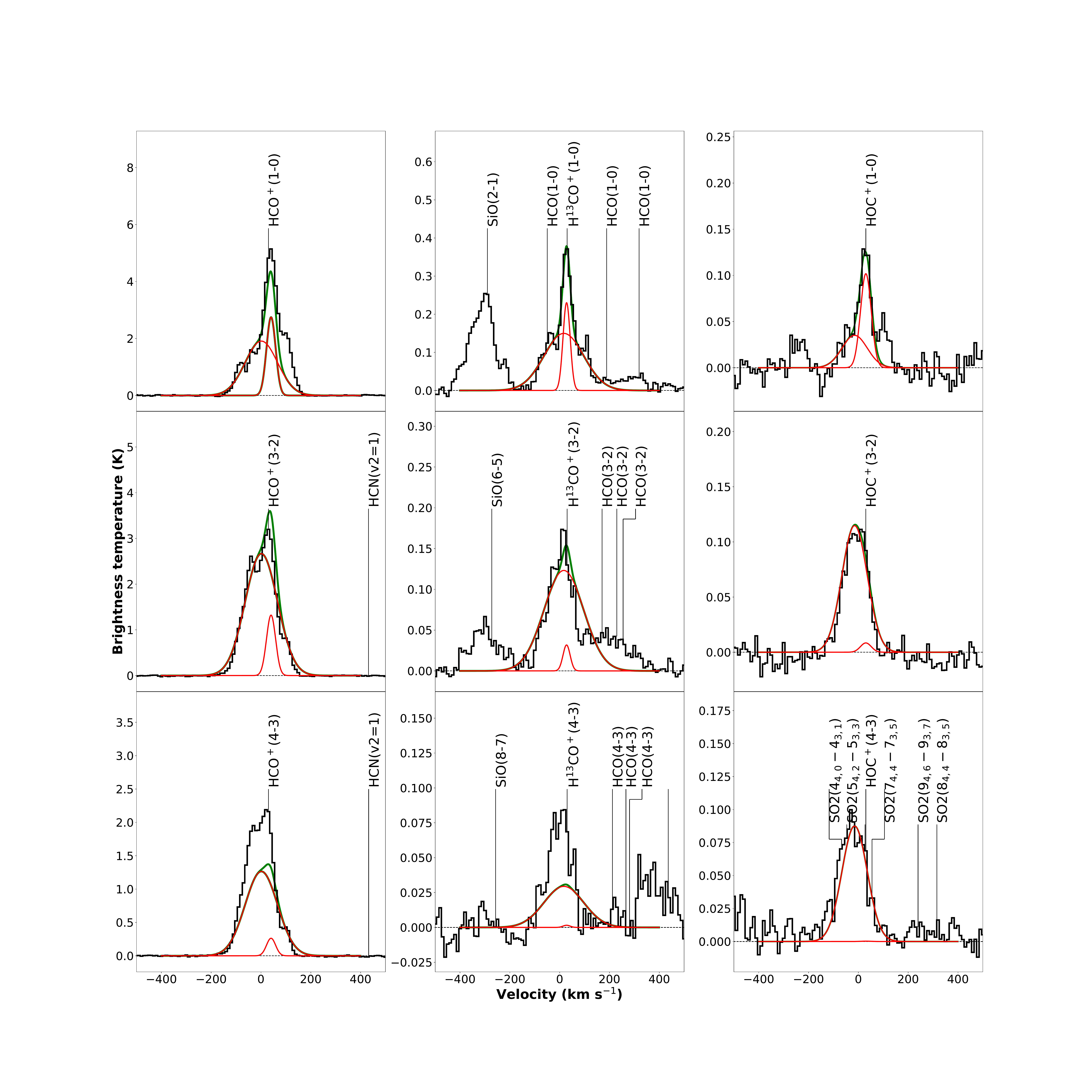}
}
\caption{Spectra at position A5. \label{fig:spec_a5}}
\end{figure*}

\begin{figure*}
\centering{
\includegraphics[width=0.99\textwidth,trim= 0 0 0 0]{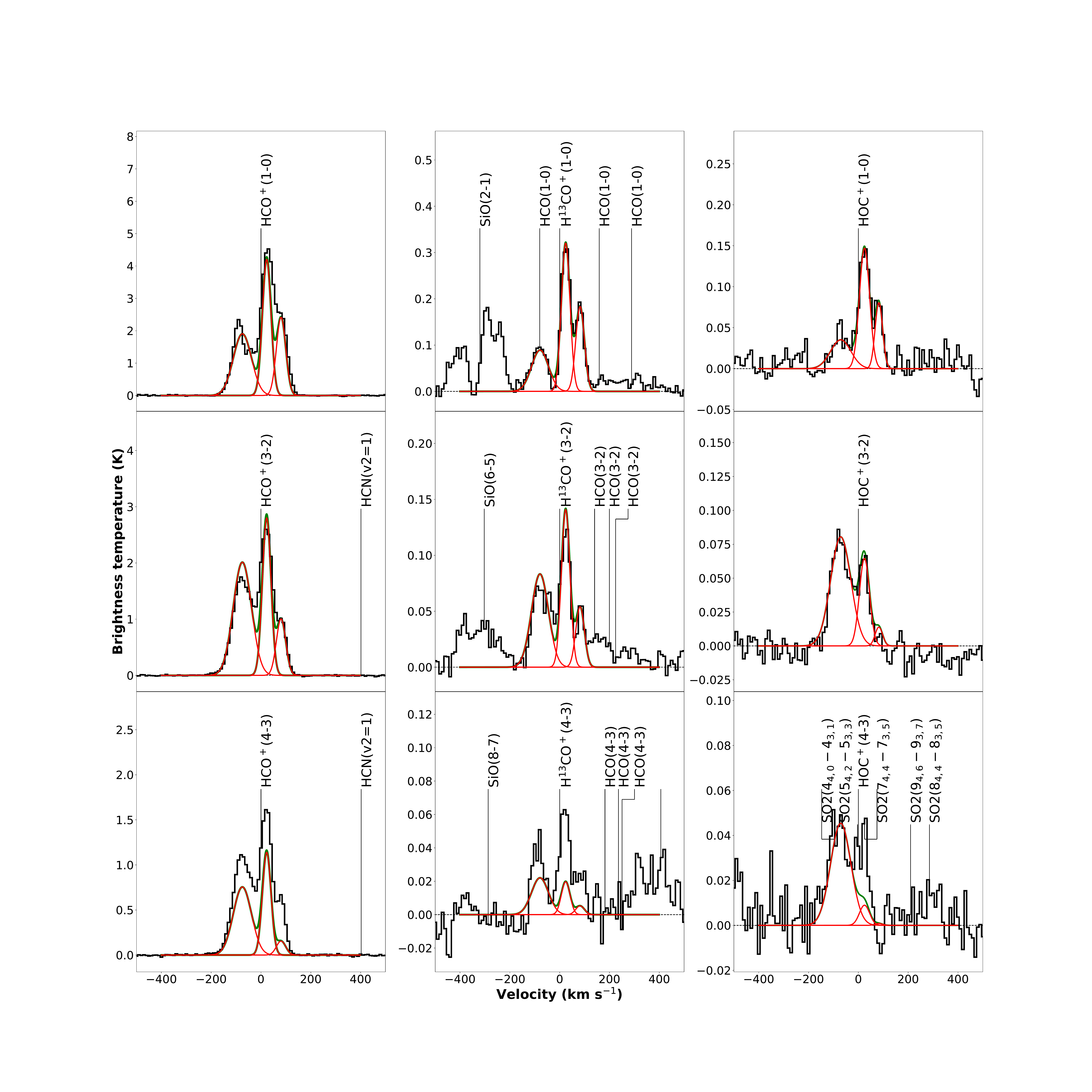}
}
\caption{Spectra at position A6. \label{fig:spec_a6}}
\end{figure*}

\begin{figure*}
\centering{
\includegraphics[width=0.99\textwidth,trim= 0 0 0 0]{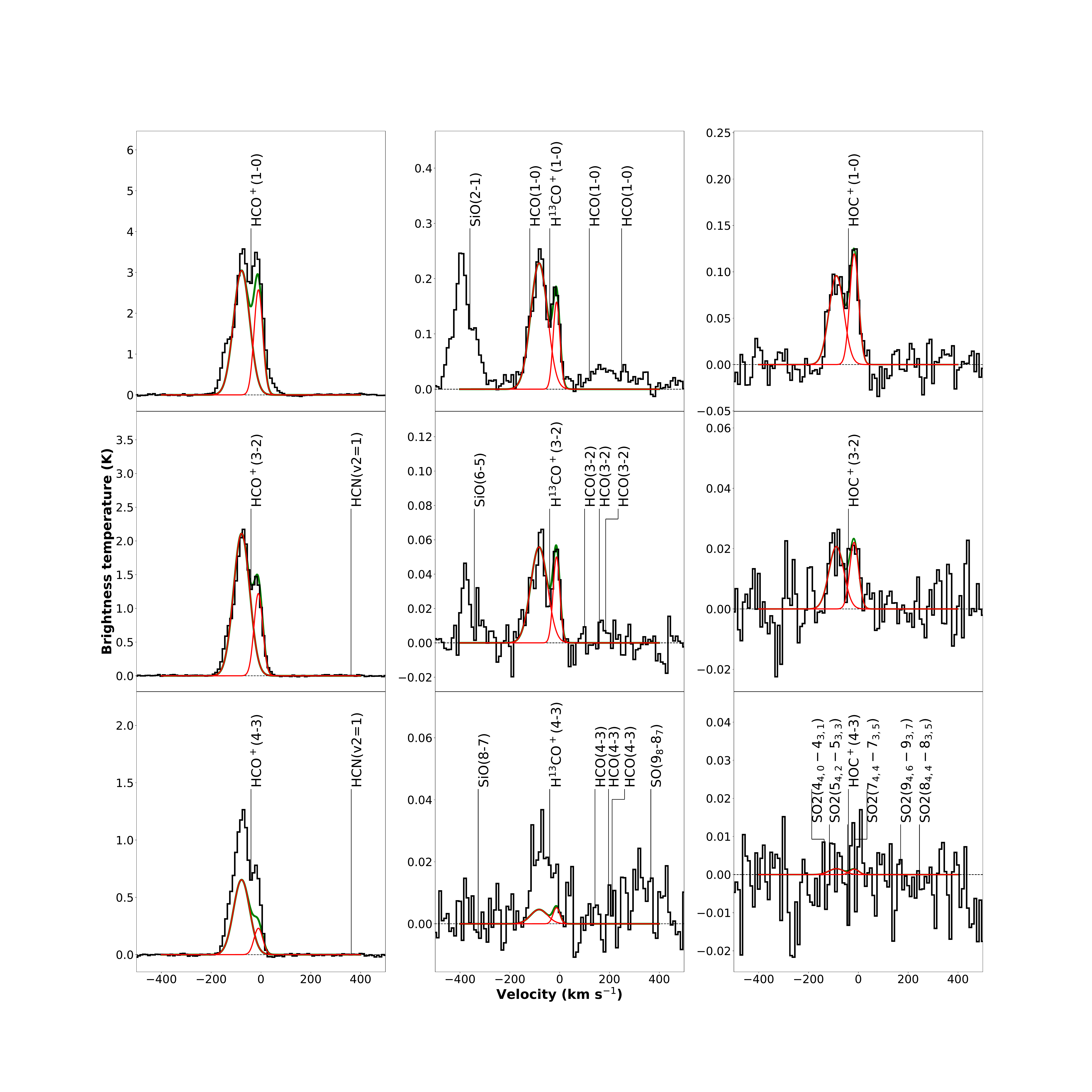}
}
\caption{Spectra at position A7. \label{fig:spec_a7}}
\end{figure*}

\begin{figure*}
\centering{
\includegraphics[width=0.99\textwidth,trim= 0 0 0 0]{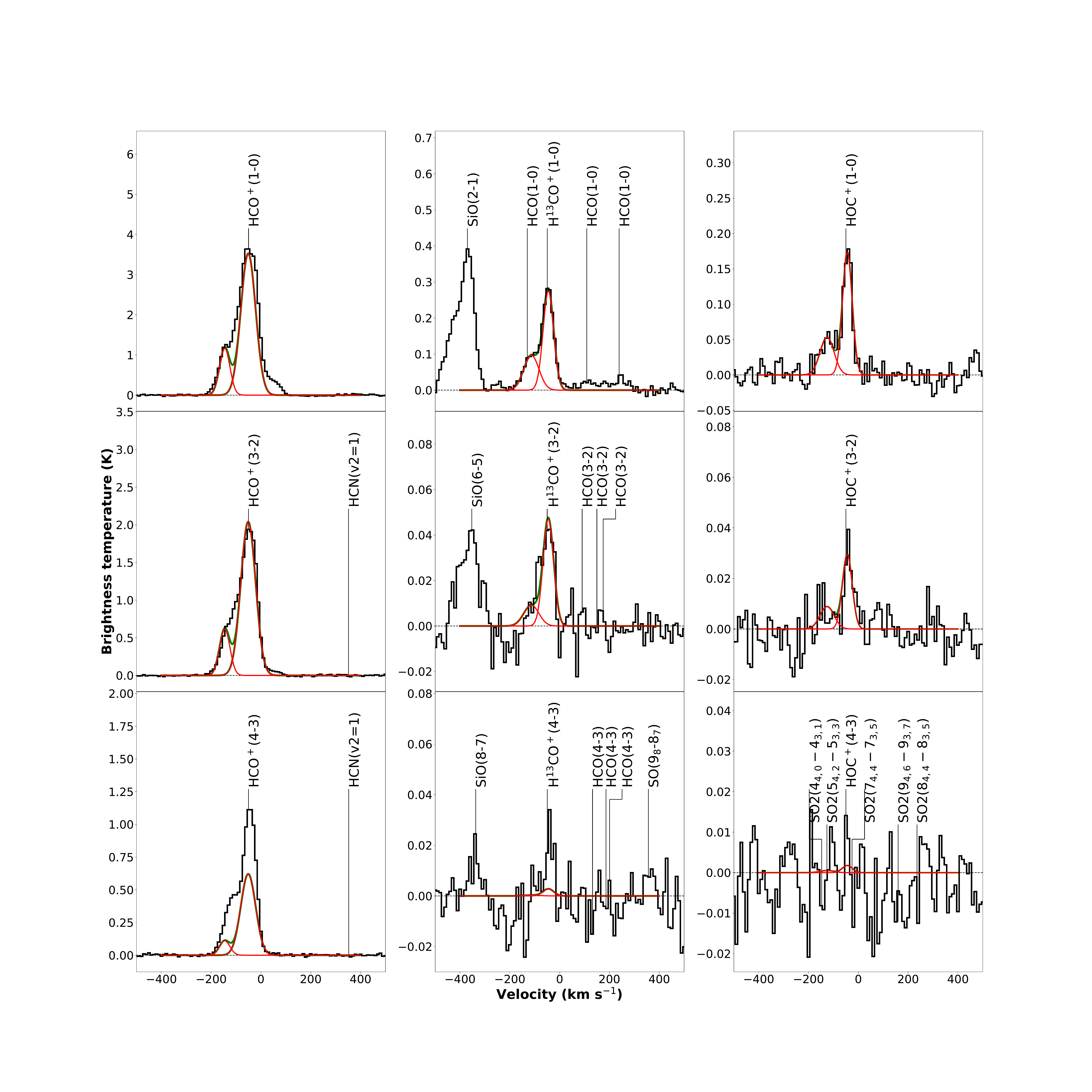}
}
\caption{Spectra at position A8. \label{fig:spec_a8}}
\end{figure*}

In Figures \ref{fig:spec_m2}-\ref{fig:spec_a8}, spectra of HCO$^+$, H$^{13}$CO$^+$, HOC$^+$ at $J=1-0$, $3-2$, and $4-3$ transitions are
shown for positions that were not shown in Figures \ref{fig:spectra} or \ref{fig:spectraM6}.

%\clearpage

\section{Details of Chemical Models}\label{sec:app_chem_model}

In this section we explain the detailed parameters used for our chemical models. As noted in Section \ref{sec:model}, we ran Meudon PDR code  to obtain the gas and dust temperatures in the PDRs \citep[][http://ism.obspm.fr]{2006ApJS..164..506L}.
We also assume $T_{\rm gas} =10\,$K if the calculated temperature is below 10 K because most molecular clouds have higher temperatures than 10 K.
We ran the Meudon code with the following parameters: $A_{\rm V, max}=5$, 
turbulent velocity $v=10$ km s$^{-1}$, maximum and minimum dust sizes of 0.3$\mu m$, and 3 nm with MRN distributions \citep{1977ApJ...217..425M}.
The current version of the Meudon code has several values of dust temperature corresponding to different dust number densities, 
which also translates to different dust sizes. 
We used the dust temperature when the dust number density is $n_{\rm dust} = 1.1\times 10^{-7}$ cm$^{-3}$.

With the temperature obtained with the Meudon code, we ran the time-dependent gas-grain code based on Nautilus. In this paper we show the results at $t  = t_{\rm dyn}$ where $t_{\rm dyn}= 3\times 10^7/\sqrt{n/cm^{-3}}$ yr. 
We note that species of interest in this paper did not have significant 
differences between the steady-state abundances and abundances at 
the dynamical time.
We do not include reactions with vibrationally excited molecular hydrogen as 
its effect on the chemistry we analyzed seems limited in most cases \citep[e.g., ][]{2010ApJ...713..662A}.

\subsection{PDR Models }

Because the Meudon PDR code requires radiation on both sides,
we set the radiation field of one side to be dominant, and we varied
this radiation field strength to be $G_0 = 1-10^5$. For the back side of the plane-parallel slab, 
we set the radiation field to be negligibly small, with $G_0=10^{-3}$.

We modified the H$_2$ formation in the PDR region due to the high 
grain temperature. The original implementation of H$_2$ formation
using Lanmuir-Hinschelwood mechanism among physisorbed (weakly bound) atoms cannot form molecular
hydrogen efficiently when
the dust temperature is $\gtrsim 30\,$K. However, molecular hydrogen has been
observed in PDR regions even with the dust temperatures of 50 K or so.
The alternative mechanisms have been proposed to form H$_2$ in warm dust temperatures such as 
Eley-Rideal reactions with chemisorbed (strongly bound) atoms.
\citet{2020A&A...634A..42T} included chemisorption and the diffusion of 
chemisorbed atoms for their formula of H$_2$ formation. 
Their results showed that the formation rate is similar to the standard values in 
the literature of $R= (1-3)\times 10^{-17}$ cm s$^{-1}$ \citep{1975ApJ...197..575J}  where R is the rate of H$_2$ formation, regardless of the dust temperature. 
Therefore, we used this rate without treating the rate with Lanmuir-Hinschelwood mechanism among 
the physisorbed atoms.

\subsection{CRDR Models}

For CRDR models, we set the interstellar radiation field 
of Meudon code to be $G_0 =1$, and took the temperature at
$A_{\rm V} = 4$, where the effects of the radiation on the temperature becomes negligible. 
We note that the Meudon code encounters bistability at certain values of visual extinction ($A_{\rm V} \sim 2-3$) with a relatively 
high cosmic-ray ionization rate of $\zeta \sim 10^{-15}$ s$^{-1}$.
This bistability did not affect the parameter range we took results from.

\begin{deluxetable}{cc} 
\tablecolumns{2} 
\tablewidth{0pc} 
\tabletypesize{\scriptsize} 
\tablecaption{Elemental abundances used for chemical model} \label{tab:elem_abun} 
\tablehead{\colhead{Element} &\colhead{X/H$_{\rm total}$}}
\startdata 
He &0.14  \\
N    &2.14(-05) \\
O           &1.76(-04) \\
C$^+$          &7.3(-05)\\
S$^+$          &8.00(-08) \\
Si$^+$        & 8.00(-09) \\
Fe$^+$        & 3.00(-09)\\
Na$^+$        & 2.00(-09)\\
Mg$^+$        & 7.00(-09) \\
P$^+$         & 2.00(-10) \\
Cl$^+$      & 1.00(-09)\\
F       & 6.68(-09)\\
\enddata 
\tablecomments{a(-b) means $a \times 10^b$.} 
\end{deluxetable} 

\section{Continuum Images}
\label{sec:app_cont}
\begin{figure*}
\centering{
\includegraphics[width=1.\textwidth]{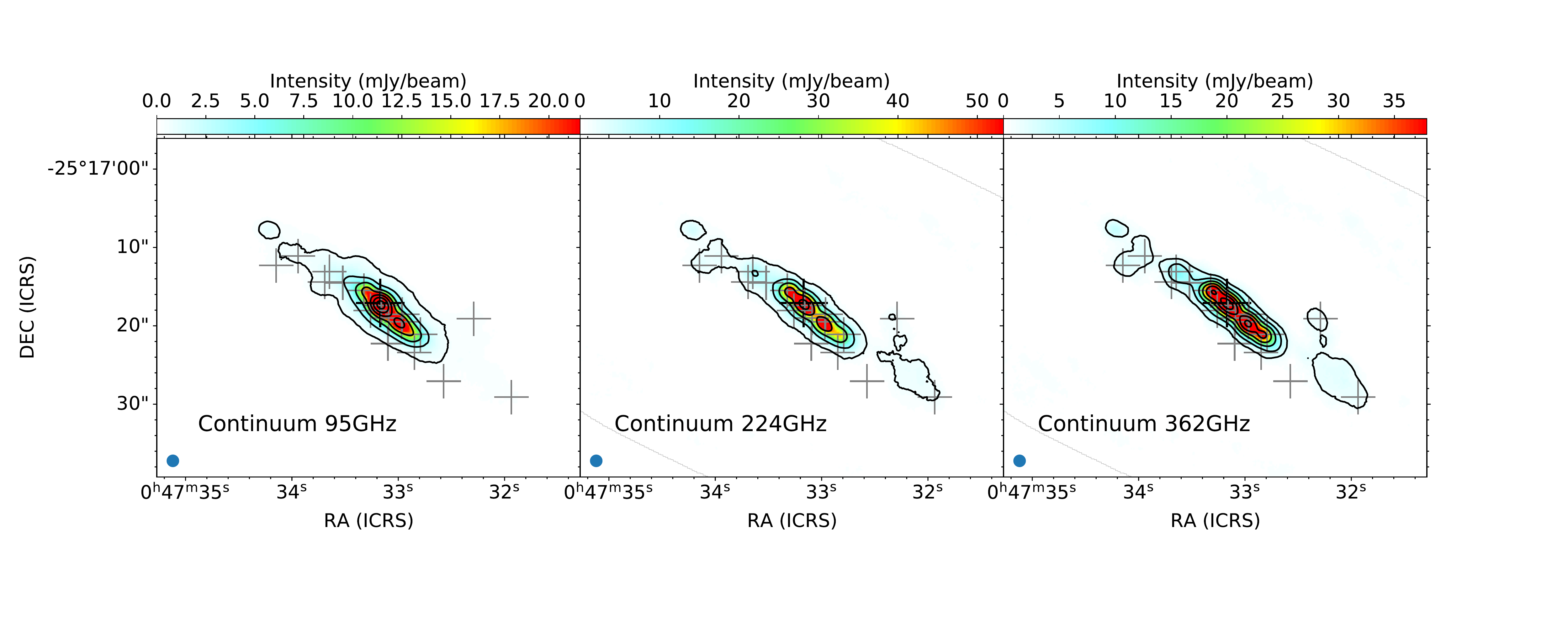}
}
\caption{Continuum images at (left) 95 GHz, (middle) 224 GHz, and (right) 362 GHz. These frequencies were chosen 
to select spectral windows with large numbers of continuum channels.
Contour levels are 5$\sigma$n$^{2.5}$ (n=1, 2, 3, 4, ...) for 95 and 224 GHz, and 5$\sigma$n$^{2}$ for 362 GHz images.
The RMS noise levels are 0.1, 0.3, and 1.5 mJy/beam for 95, 224, and 362 GHz, respectively.\label{fig:cont}}
\end{figure*}

As described in Martin et al., the continuum images were obtained from STATCONT. Continuum images at 3 different frequencies 95, 224, and 362 GHz are shown in Fig. \ref{fig:cont}. 
The emission in these continuum images are due to a combination of synchrotron, free-free, and dust emission.
At 95 GHz, free-free emission is likely to be the dominant source of emission (Figure 5 of Mart\'in et al. submitted),
while dust should dominate at higher frequencies. 
Therefore, we use the 224 and 362 GHz continuum to calculate the H$_2$ column densities.
We assume 95 GHz emission as an indicator of star formation in the discussion.
However, to obtain the exact contribution from each emission mechanism requires the fit of many frequencies.

\section{Obtaining Total Hydrogen Column Densities}\label{sec:app_coltot} 
To obtain the total hydrogen column densities ($N_{H2}$ or $N_H$),
we used two different methods. One is from the dust continuum
emission using the method proposed by \citet{1983QJRAS..24..267H}. 
Following the notation by \citet{mangum_fire_2019}, 
the molecular hydrogen column densities are estimated to be
\begin{equation}
N(H_2) = 7\times 10^{22} R_{dg} \left( \frac{\lambda(mm)}{0.4} \right) ^{\beta} \left( \frac{T_R}{T_d}\right) (cm^{-2}),
\end{equation}
where $R_{dg}$ is the dust-to-gas mass ratio, $\lambda$ is the wavelength, 
$\beta$ is the dust emissivity power law, $T_R$ is the brightness temperature of 
the continuum flux,
and $T_d$ is the dust temperature.
We used $R_{dg} = 150$, $\beta = 1.5$, and $T_d  = 30$\,K. The values for $R_{dg}$ and $\beta$ are taken
from \citet{mangum_fire_2019}, while we use a slightly lower value for the dust temperature, 30\,K, than that assumed by \citet{mangum_fire_2019} (35\,K).
Column densities are calculated using the continuum images
around 223 and 361 GHz that are relatively line-free.
An alternative method for calculating $N_{H_2}$ utilizes the C$^{18}$O column densities assuming 
CO/H$_2$ = $10^{-4}$ and $^{16}$O/$^{18}$O = 150 \citep{2009ApJ...706.1323M}. To obtain the column densities of C$^{18}$O, we used a MADCUBA fit to the C$^{18}$O emission within the ALCHEMI survey, which included the $J=1-0$, $2-1$, and $3-2$ transitions.
The values obtained from these two methods are 
consistent with each other on some cases, 
but not in other cases (see Table \ref{tab:tot_columns}).
The discrepancy can be as large as a factor of 4.

From our $H_2$ column densities derived from C$^{18}$O column densities,
continuum at 223 GHz, and continuum at 361 GHz, we used \begin{equation}
\log_{10}N = \frac{1}{2} (\log_{10}N_{max} + \log_{10}N_{min})
\end{equation}
where $N_{max}$, $N_{min}$ are maximum and minimum values of column densities among three values.

\begin{deluxetable}{cccc}
\tablecolumns{4}
\tablewidth{0pc}
\tabletypesize{\scriptsize}
\tablecaption{Total molecular hydrogen column densities.} \label{tab:tot_columns}
\tablehead{\colhead{Region} &\colhead{$N$(H$_2$) (cont 223GHz)}  &\colhead{$N$(H$_2$) (cont 361GHz)}    &\colhead{$N$(H$_2$) (C$^{18}$O)}\\
\colhead{} &\colhead{10$^{22}$(cm$^{-2}$)}  &\colhead{$10^{22}$(cm$^{-2}$)}   &\colhead{$10^{22}$(cm$^{-2}$)}
}
\startdata
M2 & 2.76 & 4.42&10.9\\
M3 &3.74 &5.18&7.47\\
M4 &71.2 &69.8&25.9\\
M5 &139 &109& 29.1\\
M7 &114 &95.4&31.0\\
M8 &17.6 &21.5& 18.1\\
M9 & 4.96 &5.43&7.82\\
M10 &5.29 &6.53&8.84\\
 A1 & $< 1.85$ &$< 1.72$& 4.96\\
A2 & 5.02 &5.42&6.07\\
A3 & 57.6 &50.3&17.8\\
A4 & $< 1.85$ &$< 1.72$&5.13\\
A5 &73.8 &53.2&17.5\\
A6 &49.6 &33.4&14.6\\
A7 & 12.9 &12.0&13.2\\
A8 & 7.23 &7.62&11.1\\
\enddata
\tablecomments{The abundance ratio of C$^{18}$O/H$_2 = 150$ is used to obtain the estimate from C$^{18}$O.}
\end{deluxetable}

\end{document}